\documentclass[12pt]{article}
\usepackage{amsfonts,amsmath,amssymb}
\usepackage{graphicx}

\newfont{\twelvemsb}{msbm10 scaled\magstep1}
\newfont{\eightmsb}{msbm8}
\newfam\msbfam
\textfont\msbfam=\twelvemsb \scriptfont\msbfam=\eightmsb
\catcode`\@=11
\def\Bbb{\ifmmode\let\next\Bbb@\else
\def\next{\errmessage{Use \string\Bbb\space only in math mode}}\fi\next}
\def\Bbb@#1{{\fam\msbfam{{#1}}}}

\newcommand{\be}{\begin{equation}}
\newcommand{\ee}{\end{equation}}
\newcommand{\ba}{\begin{eqnarray}}
\newcommand{\ea}{\end{eqnarray}}

\newcommand{\m}{\mathcal}

\newcommand{\nn}{\nonumber}
\newcommand{\q}{\theta}

\allowdisplaybreaks

\usepackage[usenames,dvipsnames]{color}

\begin{document}

\sloppy
\renewcommand{\thefootnote}{\fnsymbol{footnote}}
\newpage
\setcounter{page}{1} \vspace{0.7cm}
\vspace*{1cm}
\begin{center}
{\bf The contribution of scalars to ${\cal N}=4$ SYM amplitudes II: Young tableaux,
asymptotic factorisation and strong coupling} \\
\vspace{.6cm} {Alfredo Bonini$^a$, Davide Fioravanti $^a$, Simone Piscaglia $^{b}$, Marco Rossi $^c$}
\footnote{E-mail: bonini@bo.infn.it, fioravanti@bo.infn.it, piscaglia@th.phys.titech.ac.jp, rossi@cs.infn.it} \\
\vspace{.3cm} $^a$ {\em Sezione INFN di Bologna, Dipartimento di Fisica e Astronomia,
Universit\`a di Bologna} \\
{\em Via Irnerio 46, 40126 Bologna, Italy}\\


\vspace{.3cm} $^b$ {\em Department of Physics,
Tokyo Institute of Technology,
Tokyo 152-8551, Japan}

\vspace{.3cm} $^c${\em Dipartimento di Fisica dell'Universit\`a
della Calabria and INFN, Gruppo collegato di Cosenza} \\
{\em 87036 Arcavacata di Rende, Cosenza, Italy} \\
\end{center}
\renewcommand{\thefootnote}{\arabic{footnote}}
\setcounter{footnote}{0}
\begin{abstract}
{\noindent We disentangle the contribution of scalars to the OPE series of null polygonal Wilson loops/MHV gluon scattering amplitudes in multicolour ${\cal N}=4$ SYM. In specific, we develop a systematic computation of the $SO(6)$ matrix part of the Wilson loop by means of Young tableaux (with several examples too). Then, we use a peculiar factorisation property (when a group of rapidities becomes large) to deduce an explicit polar form. Furthermore, we emphasise the advantages of expanding the logarithm of the Wilson loop in terms of 'connected functions' as we apply this procedure to find an explicit strong coupling expansion (definitively proving that the leading order can prevail on the classical $AdS_5$ string contribution).}
\end{abstract}
\vspace{4cm}

\begin{flushright}
{\it Dedicated to the memory of Yassen Stanev}
\end{flushright}

\newpage


\tableofcontents

\index{}
\newpage

\section{Introduction and summary}
\label{intro}
\setcounter{equation}{0}
In the last years there has been much interest in $SU(N_c)$  ${\cal N}=4$ Super Yang-Mills (SYM) theory, especially in the so-called planar limit $N_c\rightarrow \infty$, $ g_{YM} \rightarrow 0$, and fixed 't Hooft coupling
\be
\lambda\equiv N_c g_{YM}^2 =16 \pi ^2 g^2 \, .
\ee
It cannot be clearer that we are interested in gauge theories for phenomenological reasons, nevertheless there are at least two other valid motivations. On the one hand there is the AdS/CFT correspondence \cite{MGKPW1,MGKPW2,MGKPW3}, namely a duality between type IIB superstring theory on $AdS_5 \times S^5$ and ${\cal N}=4$ SYM on the boundary of $AdS_5$. On the other hand there was the appearance of surprising connections with $1+1$ dimensional integrable models \cite{MZ, BS1,BS2,BS3,BS4,BS5,BES,TBA1,TBA2,TBA3,QSC1,QSC2}, which also have allowed a better comprehension and partial proof of the correspondence itself: in fact, being the latter a weak/strong coupling duality, only non-perturbative methodologies (like those of integrability theory) could really test it. After the computational successes for the spectrum of local operators, more recently, many ideas from the realm of integrable models have been adapted and used for exact computations of 4D scattering amplitudes in ${\cal N}=4$ SYM or, which is the same in the planar limit \cite{AM-amp, DKS, BHT}, vacuum expectation values (vevs) of null polygonal Wilson loops (WLs). These operators are among the simplest ones of non-local nature.

An efficient way to compute expectation values of WLs, valid in any QFT with conformal invariance, is the Operator Product Expansion (OPE) \cite{Anope}. Furthermore, quantum integrability theory gives this OPE the interpretation of a Form Factor (FF) (Infra-Red (IR)) spectral series of the many points correlation function of some peculiar {\it twist} field \cite{Knizhnik:1987xp, CCAD, BSV1}. Although, so far, the OPE series terms cannot be directly derived from the gauge or string theory, yet the integrability of the flux-tube dynamics spanned by the the Gubser, Klebanov and Polyakov (GKP) string \cite{GKP} has given many ideas on their properties. In fact, an integrable spin chain view has yielded some preliminary ideas \cite{Bel-Ope,Bel-Qua}, then expanded very efficiently in a beautiful axiom system \cite{BSV1,BSV2,BSV3,BSV5,BCCSV1}. Also, the interpretation as an integrable FF series of two (or more) point correlation function has helped the determination of the single terms and their re-summation \cite{FPR2,BFPR1,BFPR2} ({\it cf.} also the re-summations at weak coupling, {\it e.g.} \cite{Cordova,DP,Dixon}). Yet, checks, investigations and re-summations are still very needed. In a nutshell, the proposal was to write the expectation values of WLs as an infinite sum over intermediate excitations of the string GKP \cite{GKP} quantum vacuum \cite{Basso, FPR1}. In detail, these excitations are gluons with their bound states, fermions and antifermions and scalars which scatter in a non-trivial way \cite{Basso, FPR1}. Consequently, in order to pursue the OPE project, we need to know the dispersion laws of the GKP string excitations \cite{Basso} and the 2D scattering factors between them \cite{FPR1,FRO6, Basso-Rej}. And, although with more ambitious plans in mind, we shall first scrutinise regimes where explicit computations are possible. Naturally these are the weak ({\it cf.}, for instance, \cite{BSV2,P1,P2,DP}) and the strong coupling  ({\it cf.}, for instance, \cite{BSV3,FPR2,BFPR1,BSV4,Bel1509}) limits, where comparisons with gauge and string theory outcomes, respectively, are possible and have been successfully made. For what concerns the strong coupling limit, string theory has so far given only the (classical) leading order (LO) by means of a rather sophisticated mathematical minimisation of the bubble in ($AdS_3$ and eventually in) $AdS_5$ insisting on the boundary polygon \cite{Anope,TBuA,YSA, Hatsuda:2010cc}: the final outcome is (in both cases) a Thermodynamic Bethe Ansatz (TBA) system. We have already reproduced it by re-summing the contributions of gluons and fermions/antifermions bound states to the OPE series at LO \cite{FPR2, BFPR1}. This kind of computation is of very different nature (w.r.t. string calculations), and besides it is a genuine re-summation (among very rare cases) of a FF series which, very surprisingly and interestingly, generates a fully different integrability scenario, namely a TBA set-up (this happens for the first time, to our knowledge).

However, despite being worth further investigation, this is only part of the story concerning the strong coupling regime. After considering also scalars (which from the string side correspond to fluctuations on the five sphere $S^5$), the situation is even more intriguing. In fact, for very large coupling $\lambda$ the scalars decouple in a $O(6)$ Non Linear Sigma Model (NLSM) with an exponentially small (dynamically generated) mass $m_{gap}\sim e^{-\frac{\sqrt{\lambda}}{4}}$ \cite{AM, FGR, BK, FGR1,FB}, so that the theory is {\it almost} conformal. Therefore, the scalar contribution to the WLs can be guessed to be given in the limit by the conformal correlation functions of some pentagonal twist field. This makes concrete calculations possible and a surprising contribution proportional to $\sqrt {\lambda}$ has been found for the logarithm of the WL \cite {BSV4}. This result needed a corroboration by Monte Carlo simulations on the few-particles terms of the series \cite{BSV4}, which was bolstered by \cite{BEL} (see also \cite{BEL-last} where equivalence of sigma models to twisted parafermions is proposed as a calculation tool). Nevertheless, this behaviour was asking for a definitive and stringent proof directly from the OPE series because many subtleties are to be considered. Moreover, this contribution can be dominant on the $AdS_5$ string action, which gives $W_{AdS}\simeq C_{AdS} e^{-\sqrt{\lambda}\frac{A_6}{2\pi}}$, decaying with the hexagon area $A_6$. In fact, the latter is exponentially small $A_6 \sim O(e^{-\sqrt{2}\tau})$ at large $\tau$ (collinear limit) and thus the only multiplicative contribution to the WL is coming from the five sphere $S^5$, and it is actually exponentially large $W_{O(6)} \sim e^{\frac{J}{4} \sqrt{\lambda}}$ at least for small $m_{gap} \tau \sim e^{-\frac{\sqrt{\lambda}}{4}}\tau$ \footnote{So that we can conjecture that it dominates or contributes still at finite $m_{gap} \tau$ as long as $e^{-m_{gap} \tau}$ is not too small.} (see (\ref{fin-scal}) which is valid for $z\sim e^{-\frac{\sqrt{\lambda}}{4}}\tau \ll 1$). Similar considerations hold if we include string one loop corrections (whose details are still unknown but, relying on \cite{FPR2, BFPR1}, may have an analogue in the one-loop for in $\mathcal{N}=2$ partition function \cite{BoF,BoF2}): in the collinear limit the scalar contribution (given for any $\tau$ and $\sigma$ by (\ref{Wlambda})) is actually the dominating one. Of course, being a purely quantum effect (moreover in another sector of the theory), the leading term from scalars is missed by the classical minimisation of \cite{Anope,TBuA,YSA}.

Interestingly, we have found that a simple analytic derivation of this behaviour may be possible \cite {BFPR2}, although in that letter some issues could not receive the attention they deserve: here this lack will be solved along with new results. In fact, the idea anticipated in \cite{BFPR2} is that of passing from the OPE series for the expectation value of the WL, $W$, to the series for $\ln W$: this change corresponds, as for the generic term of the series, to passing from the (non-connected) multiparticle $2n$ function of $W$ to the $2n$ 'connected' multiparticle function of $\ln W$. Upon integrating {\bf each term} on one of the $2n$ rapidities and expanding for large $\lambda$ {\it inside} the multi-integral (this is a very delicate point as will be shown below in Subsection \ref {smallmass} in the part between formula (\ref{I2n}) and (\ref {fin-scal})), the $\sqrt{\lambda}$ factorises in front of the LO (of {\bf each term} of the series of $\ln W$, {\it cf.} below formula (\ref {fin-scal})). Despite the simplicity of this idea, a very essential point for its effectiveness is the convergence of the remaining integrals over the other $2n-1$ differences of rapidities, which is discussed in Subsection \ref {conn-func}, by using asymptotic properties\footnote{A secondary subtlety concerns the convergence of the particle series (\ref {fin-scal}) which gives the prefactor of $\sqrt{\lambda}$: unfortunately we can have only rough numerical and analytical ideas about it, but in the literature this series has been always found converging very fast to the conformal data.} studied in Section \ref {asy}: the connected functions belong to $L^{1}(\mathbb{R}^{2n-1})$ and, though we cannot compute explicitly the value of each integral contained in formula (\ref {fin-scal}), this result gives the definitive proof of the rightness of the procedure and hence of the diverging $\sqrt{\lambda}$ behaviour of the whole series of $\ln W$. Otherwise, in case of divergence of the integrals we could not trust the result as it might have been an artefact of the exchange of limits: in fact the range of the exponential damping of the rapidity is very large as it is roughly the inverse of the scalar mass gap ({\it cf.} below formula (\ref {m-lambda})). Besides, we will show that the general form ({\it i.e.} the $\lambda $ dependence) of the expansion (not only of the LO) stays the same upon summing further connected functions: in other words summing more connected functions simply improves the accuracy of the numerical coefficients. More in detail, the next orders require the use of a cut-off, which produces a subleading logarithmic behaviour in $\lambda$ and a constant term containing the cross ratios. Eventually, we shall not omit that our numerical calculations here on the connected functions are much easier that those on the original ones (non-connected) \cite{BSV4,BEL}.
Actually, this idea is not entirely new, since it is an extension to asymptotically free theories (here the O(6) NLSM) of a method already used in FF theory \cite{Smirnov}, thus reinforcing the idea that the OPE series is a FF one\footnote{If the hexagon gives rise to a two point correlation function, the one very well studied so far in FF theory, adding an additional side to the polygon corresponds to adding a field in the correlation function. The coordinates of the fields are indeed the conformal ratios fixing the polygon.}. In other words, physically the entire procedure corresponds to the non-perturbative problem of reproducing the ultra-violet (UV) data (critical exponents) from the IR ones (FFs). However, in usual theories like \cite{Smirnov} the connected functions enjoy an exponential fall-off, whilst asymptotically free theories are endowed with a softer power-like decay at infinity: this makes the discussion more delicate because of the need of cut-offs in the next-to-leading orders (and the peculiar appearance of terms $\sim \ln\lambda$, {\it cf. infra} formul\ae \, (\ref {I2n-bis}-\ref{conflog})).

This is the plan of the paper. In Section \ref {scalar} we analyse the contribution of scalars to the hexagonal WL and compute the $SO(6)$ matrix factor: by summing on Young tableaux we will obtain an explicit expression, formula (\ref{MatYoung}) below, for the (general) $2n$-scalar matrix factor. In Section \ref {asy} the multiparticle functions of scalars are considered and their factorisation properties\footnote{The latter assume a slightly more general form than that in \cite{Smirnov}.} found for large values of {\it some} rapidities. The polar structure of the matrix factor will be proven thanks to this feature. The factorisation is also crucial to elucidate, in Section \ref {min-area}, the finiteness of the integral of the connected functions despite their mild fall-off at infinity. Thus we can prove that, when the coupling grows to infinity, driving the mass to zero, the scalar contribution to the logarithm of the amplitude is proportional to $\sqrt{\lambda}$. Actually, we provide a systematic approach to the expansion in this regime and analyse in detail the two, four and $2n$ particle contributions to the Wilson loop, obtaining analytic formul{\ae} for the leading, subleading and constant terms.
This article closes with a summary of main results and perspectives (Section \ref {conc}). Several technical aspects will be developed in final appendices.

\section{The matrix part of scalars}
\label{scalar}
\setcounter{equation}{0}

We start this section by giving the notations for the whole paper. We want to analyse the bosonic null hexagonal Wilson loop, which is dual to the MHV six-gluon scattering amplitude. We will use the OPE of the null polygonal Wilson loop \cite{Anope} with the pentagon building blocks \cite{BSV1,BSV2}; namely the $N-$gon WL is represented in full generality as a sum over all the possible excitations of the (dual) flux-tube ({\it i.e.} the GKP string). What we have in the sum are the (free) propagation phases and the pentagonal probability transitions between two different states. For sake of simplicity and concreteness, we focus our attention in this section on the scalar sector and we will see other sectors in the following; in any case the form of the OPE is always the same (changing only the form of the different quantities involved).

Since we deal with the hexagonal bosonic WL, we need to consider, as intermediate states, only those which are singlet under the $SU(4)$ (residual) R-symmetry. For scalars this means that we need to sum only on even numbers of them\footnote{They form an antisymmetric representation $\boldsymbol{6}$ and produce a singlet $\boldsymbol{1}$ and the other two (irreducible) representations in the decomposition of the product of two of them $\boldsymbol{6}\otimes\boldsymbol{6}= \boldsymbol{1} + \boldsymbol{20} + \boldsymbol{15}$, so that only the product of an even number of them produces again a singlet.}, namely we decompose the WL expectation value as
\be\label{Wilson}
W=\sum_{n=0}^{\infty} \frac{1}{(2n)!}\int \left [ \prod_{i=1}^{2n}\frac{du_i}{2\pi}e^{\displaystyle - \tau E(u_i)-i\sigma p(u_i)} \right ] \,G^{(2n)}(u_1,\ldots,u_{2n}) \equiv \sum_{n=0}^{\infty}W^{(2n)} \, ,
\ee
where each term $W^{(2n)}$ denotes the contribution of $2n$ scalars with rapidities $\{ u_i \}$.

The exponential of the energy $E(u_i)$ and momentum $p(u_i)$ represents the free propagation in the Wick's rotated 2D space \cite{Anope}, while $G^{(2n)}$ correspond to the probability (modulus square) transitions from the vacuum to a ($2n$ scalars) state of the flux-tube \cite{BSV1}. The cross ratios $\tau ,\sigma$ determine the conformal geometry of the polygon (of the WL)\footnote{There is another cross ratio necessary to fix the hexagon, $\phi$, but it does not appear in the scalar contribution.} and thus are related to the dual momenta of the scattering gluons (in the amplitude).

\medskip

Focusing on the functions $G^{(2n)}$, they are conveniently factorised into a $\lambda$-dependent dynamical part $\Pi_{dyn}^{(2n)}$ and a $\lambda$-independent factor $\Pi_{mat}^{(2n)}$ reflecting the matrix structure of scalars under the internal $SO(6)$ symmetry \cite{BSV4}:
\be\label{Gi2n}
G^{(2n)}(u_1,\ldots,u_{2n})=\Pi_{dyn}^{(2n)}(u_1,\ldots,u_{2n})\,\Pi_{mat}^{(2n)}(u_1,\ldots,u_{2n}) \, .
\ee
The dynamical part in turn, can be further factorised in terms of functions involving just two particles at a time:
we have
\be\label{dynamical}
\Pi_{dyn}^{(2n)}(u_1,\ldots,u_{2n})=\prod_{k=1}^{2n}\mu_s(u_k)\prod_{i<j}^{2n}\frac{1}{P_{ss}(u_i|u_j)P_{ss}(u_j|u_i)} \ ,
\ee
being $P_{ss}$ the pentagonal amplitude and $\mu_s$ the measure for scalars.

On the other hand, the factor accounting for the matrix structure under $SO(6)$ does not depend on the coupling constant $\lambda$ and involves integrations over $n$ auxiliary roots of type $a$, $2n$ of type $b$, $n$ of type $c$:
\ba\label{Pi_mat}
\Pi_{mat}^{(2n)}(u_1,\ldots,u_{2n}) &=& \frac{1}{(2n)!(n!)^2}\int_{-\infty}^{+\infty}
\prod_{k=1}^{n}\frac{da_k}{2\pi}
\prod_{k=1}^{2n}\frac{db_k}{2\pi}\prod_{k=1}^{n}\frac{dc_k}{2\pi} \times \\
&\times& \frac{\displaystyle\prod_{i<j}^{n} g(a_i-a_j) \prod_{i<j}^{2n} g(b_i-b_j) \prod_{i<j}^{n} g(c_i-c_j)}
{\displaystyle \prod_{j=1}^{2n} \left(\prod_{i=1}^{n} f(a_i-b_j) \prod_{k=1}^{n} f(c_k-b_j)
\prod_{l=1}^{2n} f\left (u_l-b_j \right)\right)} \ , \nn
\ea
where the functions $f(x)$ and $g(x)$ are defined as
\be
f(x)=x^2+\frac{1}{4} \, , \quad g(x)=x^2(x^2+1) \, . \label {f-g}
\ee

\subsection{A Young tableaux approach}
\label{young}

Now we provide a way to compute explicitly the matrix factor (\ref{Pi_mat}) by residues, eventually based on Young tableaux. No need to say that the matrix factor does not depend on the coupling constant $\lambda$, so that the results apply at any coupling.

The variables $a,c$ in (\ref{Pi_mat}) do not couple to each other, are symmetric and can be integrated over to give us the same contribution
\ba\label{Pin_mat}
\Pi_{mat}^{(2n)}(u_1,\ldots,u_{2n}) =
\frac{1}{(2n)!(n!)^2}\int \prod_{k=1}^{2n}\frac{db_k}{2\pi}\,\left[\mathcal{D}_{2n}(b_1,\ldots,b_{2n})\right]^2\,
\frac{\displaystyle \prod_{i<j}^{2n} g(b_i-b_j) }
{\displaystyle \prod_{k=1}^{2n} \prod_{l=1}^{2n} f(u_l-b_k)} \, ,
\ea
where the result of the integrations over the $a_k$ (and identically for the $c_k$) is the symmetric function
\be\label{int_a}
\mathcal{D}_{2n}(b_1,\ldots,b_{2n})\equiv
\int_{-\infty}^{\infty}
\prod_{k=1}^{n}\frac{da_k}{2\pi}
\frac{\displaystyle\prod_{i<j}^{n} g(a_i-a_j)}
{\displaystyle \prod_{j=1}^{2n} \prod_{i=1}^{n} f(a_i-b_j)} \ .
\ee
The integrations in the auxiliary variables $a_1,\ldots , a_n$ can be evaluated by residues
\be\label{int_a2}
\mathcal{D}_{2n}(b_1,\ldots,b_{2n})=
\sum_{\alpha_1=1}^{2n}\ldots\sum_{\alpha_n=1}^{2n}
\frac{\displaystyle\prod_{i<j}^n g(b_{\alpha_i}-b_{\alpha_j})}
{\displaystyle\prod_{k=1}^n \prod^{2n}_{\gamma_k=1 \, ,\,\gamma_k\neq\alpha_k} f(b_{\alpha_k}-b_{\gamma_k}+\frac{i}{2})} \, .
\ee
Each term in the multiple sum depends on a partition of labels $\alpha_k$ (with $\alpha_k\in \{1,\ldots,2n\}$), which we indicate as
$S_{\vec{\alpha}}=\{\alpha_1,\ldots,\alpha_n\}$ (making use of the shorthand notation $\vec{\alpha}=\alpha_1,\ldots,\alpha_n$). It is convenient to introduce also the complementary set
$\bar S_{\vec{\alpha}}=\{1,\ldots,2n\} - \{\alpha_1,\ldots,\alpha_n\}$. Equipped with these notations, we rewrite (\ref{int_a2}) as
\be
\mathcal{D}_{2n}(b_1,\ldots,b_{2n})=2n \frac{\delta _{2n}(b_1,\ldots,b_{2n})}{\prod \limits_{\stackrel {i,j=1}{i<j}} ^{2n} [ (b_i-b_j)^2+1]} \, ,
\ee
where we introduced
\be
\delta _{2n}(b_1,\ldots,b_{2n}) \equiv \frac{n!}{2n} \sum _{\alpha _1<\alpha _2< \dots  < \alpha _{n}=1}^{2n} \left (  \prod _{\stackrel {i\in S_{\vec{\alpha}},
j\in S_{\vec{\alpha}}, i<j} {i\in \bar S_{\vec{\alpha}},
j\in \bar S_{\vec{\alpha}}, i<j}} [ (b_i-b_j)^2+1] \right ) \prod _{k=1}^n \prod _{\beta \in  \bar S_{\vec{\alpha}}} \frac{b_{\alpha _k}-b_{\beta}-i}{b_{\alpha _k}-b_{\beta}} \, .
\label{def-delta}
\ee
Thanks to the symmetry under permutation of rapidities the function defined above is a polynomial, since a single pole for $b_i=b_j$ would spoil this symmetry and double poles do not appear. Properties of the polynomials $\delta_{2n}$ are discussed in Appendix \ref {delta}. What is relevant to us now is that the matrix factor is expressed as
\be\label{Nekr-scal}
\Pi_{mat}^{(2n)}(u_1,\ldots,u_{2n})=\frac{4n^2}{(2n)!(n!)^2}\int \displaystyle\prod_{i=1}^{2n}\frac{db_i}{2\pi}\frac{[\delta_{2n}(b_1,\ldots,b_{2n})]^2}{\displaystyle\prod_{i,j}^{2n}f(u_i-b_j)}\displaystyle\prod_{i<j}\frac{b_{ij}^2}{(b_{ij}^2+1)}
\, , \quad b_{ij}\equiv b_i -b_j \, ,
\ee
which shows many similarities with the Nekrasov instanton partition function in $\mathcal{N}=2$ theories \cite{NEK}. More precisely, (\ref{Nekr-scal}) can be compared to $Z_{U(2n)}^{(2n)}$, the $2n$-instanton contribution to the partition function of a $U(2n)$ theory, where the physical rapidities $u_i$ play the role of the vevs $a_i$ of the scalar fields and the instanton positions $\phi_i$ are represented by the isotopic roots $b_i$. For these integrals, is well-known an evaluation by residues which results in a sum over Young tableaux configurations \cite{NOK}.  This observation allows us to put forward the proposal to compute $\Pi_{mat}^{(2n)}$ by residues and classify the contributions in Young tableaux, along with further diagrams obtained upon performing permutations on the column index. We must highlight, though, some differences with respect to the Nekrasov partition function. In our case the polar part is somehow simpler, since we do not have two deformation parameters $\epsilon_1,\epsilon_2$ as in $Z_{U(2n)}^{(2n)}$, but only one\footnote{In the Nekrasov function there is a Young tableaux associated to each vev $a_i$, while here we have a column associated to any $u_i$ and the Young tableaux description appears from the symmetrization.}, which is fixed to $\pm i$. The polynomial $\delta_{2n}$, on the contrary, is absent in the Nekrasov function and brings some important effects on the computation. However, the key features of the multiple integrals, which allow us to employ this method, are shared by $\Pi_{mat}^{(2n)}$, $Z_{U(2n)}^{(2n)}$ and are basically three:
\begin{itemize}
\item The poles of the type $b_i=u_j+\frac{i}{2}$, which relate the residues positions to the physical rapidities;
\item The double zeroes $b_{ij}^2$ that cancel the contributions in which two or more residues are evaluated at the same point: as an example, if we take the first residue in $b_1=u_k+\frac{i}{2}$, the poles in $b_{j\neq 1}=u_k+\frac{i}{2}$ disappear and we do not have to consider them when we integrate over the other variables $b_j$;
\item The polar part $\frac{1}{b_{ij}^2+1}$, whose effect is to arrange the residues in strings in the complex plane, displaced by $+i$: considering the example before, the first residue in $b_1=u_k+\frac{i}{2}$ generates poles in $b_{j\neq 1}=u_{k}+\frac{3i}{2}$.
\end{itemize}
Eventually, a particular residue configuration is represented by their $2n$ coordinates, which are all different and arranged in strings in the complex plane starting from $u_i+\frac{i}{2}$ and displaced by $+i$. The examples $n=1,2$ discussed later will clarify the procedure.

For the most general case of $2n$ scalars, this procedure leads to the formula
\be\label{PiMat-Young}
\Pi_{mat}^{(2n)}(u_1,\ldots,u_{2n})=\sum_{l_1+\ldots +l_{2n}=2n, l_i<3,l_{i+1}\leq l_i}(l_1,\ldots,l_{2n})_s=\sum_{|Y|=2n,l_i<3}(Y)_s \, .
\ee
Formula (\ref {PiMat-Young}) is our sum over Young tableaux. Some explanations are needed. The symbol $(l_1,\ldots,l_{2n})$ represents the contribution of a particular residue pattern: we have $l_i$ residues with real coordinate $u_i$ arranged in a string in the upper half plane, as described before. The constraint $\sum_{i=1}^{2n}l_i=2n$ follows from the fact that we have $2n$ integrations, while $l_i<3$ is due to a particular feature of the polynomials $\delta_{2n}$, see Appendix A.
The lower index $s$ in $(l_1,\ldots,l_{2n})_s$ means sum over permutation (of inequivalent
rapidities) of a single residue configuration:
\be
(l_1,\ldots,l_{2n})_s\equiv (l_1,\ldots,l_{2n}) + \textrm{permutations of $l_1,\ldots,l_{2n}$} \, .
\ee
Finally, the symbol $Y$ (with $|Y|=\sum _i l_i$) is a shorthand for $(l_1,\ldots ,l_{2n})$ and will be often used in the following to shorten various expressions.\\
The building block of (\ref {PiMat-Young}) is the contribution of the diagram $(l_1,\ldots,l_{2n})$. Its evaluation gives\footnote{The multiplicity $(2n)!$ due to the permutation of the variables $b_i$ is taken into account.}
\ba\label{diagr}
&&(l_1,\ldots,l_{2n})= \frac{4}{[(n-1)!]^2}\frac{1}{\displaystyle\prod_i^{2n}(l_i!)^2}\frac{1}{\displaystyle\prod_{k=1}^{2n}\prod_{j\neq k}\prod_{m=1}^{l_k}(u_k-u_j+(m-1)i)(u_k-u_j+mi)}\cdot\nn\\
&\cdot &\displaystyle\prod_{i<j}^{2n}\prod_{m=1}^{l_i}\prod_{k=1}^{l_j}\frac{(u_i-u_j+(m-k)i)^2}{(u_i-u_j+(m-k)i)^2+1}\delta ^2_{2n}(Y)\equiv \frac{4}{[(n-1)!]^2}\frac{1}{\displaystyle\prod_i^{2n}(l_i!)^2}\delta ^2_{2n}(Y)[l_1,\ldots ,l_{2n}] \, .
\ea
We split (\ref {diagr}), except for a factor, in two parts, one coming from $\delta_{2n}(Y)$ and the rest, called $[l_1,\ldots ,l_{2n}]$.
The notation $\delta_{2n} (Y)$ means that the arguments of $\delta _{2n}$ are the coordinates of the residues described by the pattern $Y=(l_1,\ldots ,l_{2n})$.\\
Formula (\ref{diagr}) can be specialized for our actual diagrams, with $l_i\leq 2$: the most generic contribution contains $k$ columns with  $l_i=2$, $2(n-k)$ with $l_i=1$ and $k$ with $l_i=0$, thus the total number of different Young tableaux is $n+1$.\\
To obtain a more compact expression of (\ref{diagr}) we start with the simplest case
\be
(1,1,\ldots,1,1)_{2n}=\frac{4}{[(n-1)!]^2}\delta_{2n}^2(u_1,\cdots ,u_{2n})[1,1,\ldots,1,1]_{2n} \, ,
\ee
which, using
\be
[1,1,\ldots ,1,1]_{2n}=\displaystyle\prod_{i<j}^{2n}\frac{1}{(u_{ij}^2+1)^2} \, ,
\ee
and the Pfaffian representation \cite{DID} of $\delta_{2n}$
\be
\delta_{2n}(b_1,\ldots ,b_{2n})=
\frac{n!}{2n}2^n\displaystyle\prod_{i<j}\frac{b_{ij}^2+1}{b_{ij}}\textrm{Pf} \, D \, , \quad D_{ij}=\left(\frac{b_{ij}}{b_{ij}^2+1}\right) \, ,
\ee
discussed in Appendix \ref {delta},\footnote{We are very grateful to Ivan Kostov and Didina Serban for the various discussions on the subject and for pointing out this specific representation of $\delta_{2n}$.} can be recast in the compact form
\be\label{1111}
(1,1, \ldots ,1,1)_{2n}=2^{2n}\textrm{Det}\left(\frac{u_{ij}}{u_{ij}^2+1}\right)\displaystyle\prod_{i<j}^{2n}\frac{1}{u_{ij}^2} =
\frac{4n^2}{(n!)^2} \prod _{i<j}^{2n} \frac{1}{(1+u_{ij}^2)^2} \, \delta^2_{2n}(u_1,\ldots ,u_{2n})
\, .
\ee
We introduced the subscript $2n$ in $(1,1,\ldots,1,1)_{2n}$ to highlight the fact that there are $2n$ variables.
At this stage this subscript is redundant, but it will be necessary in the following. The Det above is the determinant
of a matrix whose element $i,j$ is $u_{ij}/(u_{ij}^2+1)$. On the other hand, the formula (\ref{delta20}) gives our polynomial computed in a configuration of the type\footnote{We permuted the rapidities in order to get all the columns with two boxes on the left, to get $(2,\ldots ,2,0,\ldots ,0)$ in place of $(2,0,\ldots ,2,0)$.} $(2,\ldots ,2,0,\ldots ,0)$, which, combined with (always from (\ref{diagr}))
\be
[2,\ldots ,2,0,\ldots ,0]_{2n}=\displaystyle\prod_{i<j}^{n}\frac{1}{(u_{ij}^2+1)^2(u_{ij}^2+4)^2}\frac{1}{\displaystyle\prod_{i=1}^{n}\displaystyle\prod_{j=n+1}^{2n}u_{ij}
(u_{ij}+i)^2(u_{ij}+2i)} \, ,
\ee
yields the residues contribution for the other special case
\be\label{2200}
(2, \ldots ,2,0, \ldots , 0)_{2n}=\frac{1}{\displaystyle\prod_{i=1}^{n}\displaystyle\prod_{j=n+1}^{2n}u_{ij}(u_{ij}+i)^2(u_{ij}+2i)} \, .
\ee
The most general configuration, up to a permutation of the rapidities, contains $k$ columns of height two, $k$ of height  zero and $2n-2k$ with height one.

Relation (\ref{delta-fact}) in Appendix \ref {delta} gives the polynomial $\delta _{2n}$ corresponding to $(2, \ldots ,2,0, \ldots , 0_{2k},1, \ldots ,1)_{2n}$ in terms of the two special cases described before. The intermediate subscript $2k$ means that the first $2k$ columns are of the type $(2,\ldots, 2,0,\ldots ,0)_{2k}$, while the remaining $2n-2k$ contains $1$.
To get an explicit formula of the general contribution we also need the property, again from (\ref{diagr})
\ba
&&[2, \ldots ,2,0, \ldots , 0_{2k},1, \ldots ,1]_{2n}=[2, \ldots ,2,0, \ldots , 0]_{2k}\cdot [1_{2k+1},1, \ldots ,1,1]_{2n} \cdot \nn\\
&& \cdot \displaystyle\prod_{j=2k+1}^{2n}\displaystyle\prod_{i=1}^{k}
\frac{1}{u_{ij}(u_{ij}-i)(u_{ij}^2+1)(u_{ij}+2i)^2}\displaystyle\prod_{l=k+1}^{2k}\frac{1}{u_{lj}(u_{lj}-i)} \, .
\ea
Combining all the pieces, we write down the most general contribution as follows
\ba\label{general}
&&(2, \ldots ,2,0, \ldots , 0_{2k},1, \ldots ,1)_{2n}=(2, \ldots ,2,0, \ldots , 0)_{2k}\cdot (1_{2k+1},1, \ldots ,1,1)_{2n} \cdot \nn\\
&& \cdot \displaystyle\prod_{j=2k+1}^{2n}\displaystyle\prod_{i=1}^{k}\frac{1}{u_{ij}(u_{ij}+i)}\displaystyle\prod_{l=k+1}^{2k}\frac{1}{u_{lj}(u_{lj}-i)}= \frac{1}{\displaystyle\prod_{i=1}^{k}\displaystyle\prod_{j=k+1}^{2k}u_{ij}(u_{ij}+i)^2(u_{ij}+2i)}\cdot \nn\\
&& \cdot 2^{2n-2k}\textrm{Det}_{(i,j)=2k+1}^{2n}\left(\frac{u_{ij}}{u_{ij}^2+1}\right)\displaystyle\prod_{i<j=2k+1}^{2n}\frac{1}{u_{ij}^2} \displaystyle\prod_{j=2k+1}^{2n}\displaystyle\prod_{i=1}^{k}\frac{1}{u_{ij}(u_{ij}+i)}\displaystyle\prod_{l=k+1}^{2k}\frac{1}{u_{lj}(u_{lj}-i)} \, ,
\ea
where with $(1_{2k+1},1,\ldots,1,1)_{2n}$ it is intended the contribution of the type (\ref{1111}) restricted to the variables $u_{2k+1},\ldots,u_{2n}$. Analogously, the determinant in (\ref {general}) concerns a matrix whose elements
are $u_{ij}/(u_{ij}^2+1)$, with $2k+1\leq i,j\leq 2n$.

Recalling (\ref{PiMat-Young}), the matrix part is a sum over the Young tableaux configurations, which are in turn given by all the permutations of inequivalent rapidities of (\ref{general}). We perform the sum over $(Y)_s$ considering all the cases $k=0,\ldots,n$ in (\ref{general}); to symmetrize the Young tableaux, we sum over the $(2n)!$ permutations $P$ and divide by the overcounting factor $(k!)^2(2n-2k)!$, to get the final expression for the matrix part
\ba
&&\Pi_{mat}^{(2n)}(u_1, \ldots , u_{2n})=\sum_{k=0}^{n}\frac{2^{2(n-k)}}{(2n-2k)!(k!)^2}\sum_{P}\frac{1}{\displaystyle\prod_{i=1}^{k}\displaystyle\prod_{j=k+1}^{2k}u_{P_i P_j}(u_{P_i P_j}+i)^2(u_{P_i P_j}+2i)}\cdot \nn\\
&& \label{MatYoung} \\
&& \cdot \textrm{Det}_{(i,j)=2k+1}^{2n}\left(\frac{u_{P_i P_j}}{u_{P_i P_j}^2+1}\right)\displaystyle\prod_{i<j=2k+1}^{2n}\frac{1}{u_{P_i P_j}^2}\displaystyle\prod_{j=2k+1}^{2n}\displaystyle\prod_{i=1}^{k}\frac{1}{u_{P_i P_j}(u_{P_i P_j}+i)}\displaystyle\prod_{l=k+1}^{2k}\frac{1}{u_{P_l P_j}(u_{P_l P_j}-i)} \nn \, .
\ea
Formula (\ref{MatYoung}) is the main result of this section: it represents the matrix factor as a finite sum of rational functions. However, the polar structure of $\Pi_{mat}^{(2n)}$ remains somehow hidden in that representation, since many poles appearing in the sum cancel once we consider all the terms. The exact polar structure of $\Pi_{mat}^{(2n)}$ will be analysed in next section using another feature, the asymptotic factorisation.

\medskip
In order to elucidate the method of the Young tableaux, below we outline the computations for the simplest cases.

\medskip

\noindent\textbf{$\bullet$\ Two scalars ($n=1$):}

\medskip

The simplest case ($n=1$) involves just a couple of scalars, $u_1,\,u_2$. From (\ref{Pin_mat}) we have
\ba
\Pi_{mat}^{(2)}(u_1,u_{2})=\frac{1}{2}\int\frac{da\,dc}{(2\pi)^2}\,\frac{db_1\,db_2}{(2\pi)^2}\,
\frac{g(b_1-b_2)}{f(u_1-b_1)f(u_1-b_2)f(u_2-b_1)f(u_2-b_2)}\cdot \nn\\
\cdot \frac{1}{f(a-b_1)f(a-b_2)f(c-b_1)f(c-b_2)} \ ,
\ea
which, upon performing the integrations over the variables $a$ and $c$, turns to
\be
\Pi_{mat}^{(2)}(u_1,u_{2})=2\int\frac{db_1\,db_2}{(2\pi)^2}\,
\frac{1}{f(u_1-b_1)f(u_1-b_2)f(u_2-b_1)f(u_2-b_2)}\,\frac{(b_1-b_2)^2}{(b_1-b_2)^2+1} \ .
\ee
This result means that our polynomial is trivial for two particles, {\it i.e.} $\delta_2=1$. The contour integrals over $b_1$, $b_2$ are easy to perform without any Young tableaux technique (we have just $3\times 2=6$ residues to evaluate) and we finally get
\be\label{2scalars}
\Pi_{mat}^{(2)}(u_1,u_{2})=
\frac{6}{[(u_1-u_2)^2+1][(u_1-u_2)^2+4]} \ .
\ee
Even though the answer is already known, it is meaningful to solve the $n=1$ case within the Young tableaux framework, in order to give a simple illustration of how it works. Afterwards, we will address the first non trivial case, $n=2$.\\
We start from the double integral over $b_1$ and $b_2$,
\be
\Pi_{mat}^{(2)}(u_1,u_{2})=2\int\frac{db_1\,db_2}{(2\pi)^2}\,
\frac{1}{f(u_1-b_1)f(u_1-b_2)f(u_2-b_1)f(u_2-b_2)}\,\frac{(b_1-b_2)^2}{(b_1-b_2)^2+1} \, .
\ee
We perform the integration on the real axis closing the contour in the upper half plane. Therefore the integral over $b_1$ gets contributions from the poles $b_2+i$, $u_1+i/2$ and $u_2+i/2$, leading to the following expression
\be
\Pi_{mat}^{(2)}(u_1,u_{2})=\int\frac{db_2}{2\pi}\frac{A+B+C}{(b_2-u_1-i/2)(b_2-u_1+i/2)(b_2-u_2-i/2)(b_2-u_2+i/2)}  \, ,
\ee
where
\ba
A&&=\frac{-1}{(b_2-u_1+3i/2)(b_2-u_1+i/2)(b_2-u_2+3i/2)(b_2-u_2+i/2)} \nn\\
B&&=\frac{2}{(u_1-u_2)(u_1-u_2+i)}\frac{(b_1-u_1-i/2)^2}{(b_1-u_1-3i/2)(b_1-u_1+i/2)} \nn\\
C&&=\frac{2}{(u_2-u_1)(u_2-u_1+i)}\frac{(b_1-u_2-i/2)^2}{(b_1-u_2-3i/2)(b_1-u_2+i/2)}
\ea
are the contributions respectively for $b_1=b_2+i$, $u_1+i/2$, $u_2+i/2$. As for the integral over $b_2$, we see that each term contains two poles. Therefore, in total we have $3\times 2=6$ residues. We can represent the various contributions by the position of the poles of the isotopic roots $(b_1,b_2)$: it is easy to check that they are $(u_1+i/2,u_1+3i/2)$, $(u_1+3i/2,u_1+i/2)$, $(u_1+i/2,u_2+i/2)$, $(u_2+i/2,u_1+i/2)$, $(u_2+i/2,u_2+3i/2)$ and $(u_2+3i/2,u_2+i/2)$. The key property that allows us a quick evaluation of the integrals is that the residues are invariant under exchange of isotopic roots, then only three terms are truly different and can be represented by an array of two numbers $(l_1,l_2)$ with $l_1+l_2=1$, where $l_i$ labels the number of roots with real coordinate $u_i$. Therefore, we define the three residues configurations
\ba
(2,0)&&\equiv(u_1+i/2,u_1+3i/2) + (u_1+3i/2,u_1+i/2)= 2\times (u_1+i/2,u_1+3i/2) \nn\\
(0,2)&&\equiv(u_2+i/2,u_2+3i/2) + (u_2+3i/2,u_2+i/2)= 2\times (u_2+i/2,u_2+3i/2)\nn\\
(1,1)&&\equiv(u_1+i/2,u_2+i/2) + (u_2+i/2,u_1+i/2) = 2\times (u_1+i/2,u_2+i/2)\, ,
\ea
which are the $n=1$ version of $(l_1,\ldots, l_{2n})$. Eventually, the total matrix part amounts to
\be
\Pi_{mat}^{(2)}(u_1,u_2)=(1,1)+(2,0)+(0,2)
\ee
with
\ba
(1,1)&&=\frac{4}{[(u_1-u_2)^2+1]^2}\nn\\
(2,0)&&=\frac{1}{(u_1-u_2)(u_1-u_2+i)^2(u_1-u_2+2i)}\nn\\
(0,2)&&=\frac{1}{(u_2-u_1)(u_2-u_1+i)^2(u_2-u_1+2i)} \, ,
\ea
which is in agreement with (\ref{2scalars}).
As a last step, we note that $(2,0)$ and $(0,2)$ are related by the symmetry $u_1\leftrightarrow u_2$ and thus we define the symmetric $(2,0)_s=(2,0)+(0,2)$ which we call Young tableaux, to get our final result
\be
\Pi_{mat}^{(2)}(u_1,u_2)=(1,1)_s+(2,0)_s \, ,
\ee
where $(1,1)_s\equiv (1,1)$, as it is already symmetric.

\medskip

\noindent\textbf{$\bullet$\ Four scalars ($n=2$):}

\medskip

When dealing with four scalars ($n=2$), the $\delta$-polynomial (\ref{def-delta}) reads
\ba
\delta_4(b_1,\ldots,b_4) &=&
14 + \frac{1}{2}(b_1 - b_2)^2 [(b_4 - b_3)^2+4] +
 \frac{1}{2}(b_1 - b_3)^2 [(b_2 - b_4)^2+4] + \nn\\
& +& \frac{1}{2}(b_1 - b_4)^2 [(b_2 - b_3)^2+4] +
 \frac{1}{2}(b_2 - b_3)^2 [(b_1 - b_4)^2+4] + \nn\\
&+& \frac{1}{2}(b_2 - b_4)^2 [(b_1 - b_3)^2+4] + \frac{1}{2} (b_3 - b_4)^2 [(b_1 - b_2)^2+4] = \\
&=& 2+[(b_1-b_2)^2+2][(b_3-b_4)^2+2] +[(b_1-b_3)^2+2][(b_2-b_4)^2+2]+ \nn \\
&+& [(b_1-b_4)^2+2][(b_2-b_3)^2+2] \, . \nn
\ea
Hence, for $n=2$, formula (\ref{Pin_mat}) becomes:
\be
\Pi_{mat}^{(4)}(u_1,\ldots,u_{4})= \frac{1}{6}\int
\frac{db_1 db_2 db_3 db_4}{(2\pi)^4}\,
\frac{[\delta_4(b_1,\ldots,b_4)]^2}
{\displaystyle \prod_{i,j=1}^{4}f(u_l-b_j)}
\,\prod_{i<j}\frac{(b_i-b_j)^2}{(b_i-b_j)^2+1} \, .
\ee
A standard evaluation by residues would be very long since there are $7\times 6\times 5\times 4=820$ contributions (each integration lowers the number of residues by one).\\
The Young tableaux expansion employs two symmetries: under permutations of isotopic rapidities $b_i$ (which brings a factor $4!=24$) and under permutations of $u_i$ (which is responsible for the subscript $_s$) to get only 5 different Young tableaux: $(1,1,1,1)_s$, $(2,1,1,0)_s$, $(2,2,0,0)_s$, $(3,1,0,0)_s$ and $(4,0,0,0)_s$. Each of them is a sum over the permutations of rapidities of terms $(l_1,\ldots,l_4)$, they are respectively $1,12,6,12,4$.
As a check,  the total number of residues is $(1+12+6+12+4)\times 24=840$
as stated before, but our method tells us that many of them are either equal (by permutations of $b_i$) or related by permutations of $u_i$.
The latter two diagrams vanish as a result of the property of the $\delta$-polynomial already stated: $\delta_4(u_1,u_1+i,u_1+2i,u_1+3i)=\delta_4(u_1,u_1+i,u_1+2i,u_2)=0$.
To sum up, the four scalars matrix part is given by
\be
\Pi_{mat}^{(4)}(u_1,u_2,u_3,u_4)=(1,1,1,1)_s + (2,1,1,0)_s + (2,2,0,0)_s \, ,
\ee
where the building blocks are, according to (\ref{general})
\ba
(1,1,1,1)_s&=&(1,1,1,1)=16 \, \textrm{Det}\left(\frac{u_{ij}}{u_{ij}^2+1}\right)\displaystyle\prod_{i<j}^{4}\frac{1}{u_{ij}^2} \, ,  \\
(2,2,0,0)&=&\frac{1}{\displaystyle\prod_{i=1}^2\prod_{j=3}^4 u_{ij}(u_{ij}+i)^2(u_{ij}+2i)} \, , \\
(2,0,1,1)&=&\frac{1}{u_{12}(u_{12}+i)^2(u_{12}+2i)}
\frac{4}{(u_{34}^2+1)^2}\frac{1}{\displaystyle\prod_{j=3}^{4}u_{1j}(u_{1j}+i)u_{2j}(u_{2j}-i)} \,
\ea
and the symmetrisation is explicitly obtained through
\ba
(2,2,0,0)_s &=& (2,2,0,0) + (2,0,2,0) + (2,0,0,2) + (0,2,2,0) + (0,2,0,2) + (0,0,2,2)\nn\\
(2,1,1,0)_s&=& (2,1,1,0) + (2,1,0,1)+ (2,0,1,1)+(1,2,1,0)+(1,1,2,0)+(1,0,1,2) + \nn\\
&+& (1,2,0,1) + (1,0,2,1) + (1,1,0,2) + (0,2,1,1) + (0,1,1,2) + (0,1,2,1) \, .
\ea

As an application of the method outlined in this section, we compute the residue of $\Pi_{mat}^{(2n)}$ in $u_i=u_j+2i$, which can be nicely expressed in terms of the matrix part with $2$ scalars less
\be\label{ResPimat}
-2i \textit{Res}_{u_2=u_1+2i} \Pi_{mat}^{(2n)}(u_1,\cdots ,u_{2n})= \frac{\Pi_{mat}^{(2n-2)}(u_3,\cdots ,u_{2n})}{\displaystyle\prod_{j=3}^{2n}u_{1j}(u_{1j}+i)^2 (u_{1j}+2i)} \, .
\ee
In order to prove (\ref{ResPimat}), we start from the sum over Young configurations (\ref{PiMat-Young}) and note that the pole in $u_2=u_1+2i$ appears only in the terms belonging to the type $(2,0,l_3,\cdots ,l_{2n})$, with $\sum_{i=3}^{2n}l_i=2n-2$. The sum on the RHS reduces then to that of $\Pi_{mat}^{(2n-2)}(u_3,\cdots ,u_{2n})$. To go further we need to work out the expression of $(2,0,l_3,\cdots ,l_{2n})$ and split it in three different contributions
\be\label{split}
(2,0,l_3,\cdots ,l_{2n})=(2,0)\cdot (l_3,\cdots ,l_{2n})M_{\lbrace l_i \rbrace}
\ee
where the mixed term $M_{\lbrace l_i \rbrace}$ depends on the specific configuration ${\lbrace l_i \rbrace}$ and on all the variables $u_i$. We stress that (\ref{split}) is a different and more complicated split than (\ref{general}).
The pole for $u_2=u_1 +2i$ is contained only in $(2,0)$ with residue $i/2$ and, as we will show, when $M_{\lbrace l_i \rbrace}$ is evaluated for $u_2 = u_1 + 2i$, which we call $M^*$, the dependence on ${\lbrace l_i \rbrace}$ drops out, thus we get a prefactor multiplying the sum and the matrix part for fewer scalars is recovered
\be
\textit{Res}_{u_2=u_1+2i}\Pi_{mat}^{(2n)}(u_1,\cdots ,u_{2n}) = -\frac{1}{2i} M^*(u_1,u_3,\cdots ,u_{2n}) \Pi_{mat}^{(2n-2)}(u_3,\cdots ,u_{2n})
\ee
By means of (\ref{2200}) and (\ref{general}), the mixed contribution can be specified for the configuration $(2_3,\cdots ,2_{k+1},0,\cdots ,0_{2k},1,\cdots ,1)$
\be
\frac{1}{\displaystyle\prod_{j=3}^{k+1}u_{1j}(u_{1j}+i)^2(u_{1j}+2i)\prod_{j=k+2}^{2k}u_{2j}(u_{2j}-i)^2(u_{2j}-2i)\prod_{j=2k+1}^{2n}u_{1j}(u_{1j}+i)u_{2j}(u_{2j}-i)}
\ee
the other being obtained by a suitable permutation of the variables. By the identification $u_2=u_1 +2i$ we find
\be
M^*(u_1,u_3,\cdots ,u_{2n})=\frac{1}{\displaystyle\prod_{j=3}^{2n}u_{1j}(u_{1j}+i)^2(u_{1j}+2i)}
\ee
and finally prove the claim.
We remark that a residue formula like (\ref{ResPimat}) was expected for physical reasons, as $\Pi_{mat}^{(2n)}$ is a part of the squared form factor of a specific operator, which must satisfy certain axioms. Among them, one concerns the residues of its kinematic poles and relate them to the form factor with two particles less. The kinematic poles are those in $u_i=u_j+2i$, thus (\ref{ResPimat}) is nothing but a consequence of the form factor interpretation of the pentagonal transitions.

\subsection{Other representations}

The matrix part, starting from its expression (\ref{Nekr-scal}) as an integral over the isotopic roots $b_i$, enjoys other alternative representations. They are originated form the properties of the $\delta_{2n}$ polynomials.

\subsubsection{The matrix factor as an integral of a determinant}
\label{mat-det}

As previously anticipated, the $\delta$-polynomials enjoy a nice expression in terms of the Pfaffian of a skew-symmetric matrix \cite{DID}, which implies a determinant representation for its square
\be
\delta^2_{2n}(b_1,\ldots ,b_{2n})=
\frac{(n!)^2}{4n^2}2^{2n}\displaystyle\prod_{i<j}\frac{(b_{ij}^2+1)^2}{b^2_{ij}}\textrm{Det} \, D \, , \quad D_{ij}=\left(\frac{b_{ij}}{b_{ij}^2+1}\right) \, .
\ee
It then follows, from (\ref {Nekr-scal}), a determinant representation for the (integrand of) matrix part
\be\label{Pimat-detD}
\Pi_{mat}^{(2n)}(u_1,\cdots ,u_{2n})=
\frac{2^{2n}}{(2n)!}\int\displaystyle\prod_{i=1}^{2n}
\frac{db_i}{2\pi}\frac{\displaystyle\prod_{i<j}(b_{ij}^2+1)}{\displaystyle\prod_{i,j=1}^{2n}f(u_i-b_j)}\textrm{Det} \, D \, .
\ee
We can put the functions $f(u_i-b_j)$ inside and define the new $2n\times 2n$ matrix $A$
\be
\Pi_{mat}^{(2n)}=
\frac{1}{(2n)!}\int\displaystyle\prod_{i=1}^{2n}
\frac{db_i}{2\pi}\displaystyle\prod_{i<j}(b_{ij}^2+1)\textrm{Det} \, A \, , \quad A_{ij}=\frac{2b_{ij}}{b_{ij}^2+1}\frac{1}{\displaystyle\prod_{k=1}^{2n}f(u_k-b_i)} \, .
\label {Pimat-det}
\ee
We can go further by using the Vandermonde formula
\be\label{Vandermonde}
\displaystyle\prod_{i<j}^{2n }b_{ij}=(-1)^n \textrm{Det} \, B \, , \quad B_{ij}=\left(b_i^{j-1}\right) \, ,
\ee
combined with the well-known Cauchy identity
\be\label{Cauchy}
\displaystyle\prod_{i<j}^{2n}\frac{b_{ij}^2}{b_{ij}^2 + 1} = \textrm{Det} \, C \, , \quad C_{ij}=\left(\frac{i}{b_{ij}+i}\right) \, ,
\ee
to represent the integrand of the matrix part completely as a determinant.
Starting from (\ref {Pimat-det}) and using (\ref{Vandermonde}) together with (\ref{Cauchy}), it is straightforward
to recast the matrix part in the form\footnote {For the square of the delta polynomials we have also a
representation in terms of determinants:
\be
\delta_{2n}^2(b_1,\ldots ,b_{2n})=\frac{(n!)^2 2^{2n}}{4n^2}\left[\textrm{Det}\, B\right]^2\left[\textrm{Det}\,
C\right]^{-2}\textrm{Det} \, D .
\ee
}
\be
\Pi_{mat}^{(2n)}(u_1,\cdots, u_{2n})=
\frac{1}{(2n)!}\int\displaystyle\prod_{i=1}^{2n}
\frac{db_i}{2\pi} \textrm{Det} \, A [\textrm{Det} \, B]^2 [ \textrm{Det} \, C]^{-1}
\label {Pi2n-det} \, .
\ee
or, in terms of the matrix $D$
\be
\Pi_{mat}^{(2n)}(u_1,\cdots, u_{2n})=
\frac{2^{2n}}{(2n)!}\int\displaystyle\prod_{i=1}^{2n}
\frac{db_i}{2\pi}\frac{\textrm{Det} \, D [\textrm{Det} \, B]^2 [ \textrm{Det} \, C]^{-1}}{\displaystyle\prod_{i,j=1}^{2n}f(u_i-b_j)} \, .
\ee
and also
\be
\Pi_{mat}^{(2n)}(u_1,\cdots, u_{2n})=
\frac{2^{2n}}{(2n)!}\frac{1}{\displaystyle\prod_{i<j}^{2n}u_{ij}^2}
\int\displaystyle\prod_{i=1}^{2n}\frac{db_i}{2\pi}\,\textrm{Det}\,D
\frac{\textrm{Det} \left(\frac{1}{u_i-b_j+\frac{i}{2}}\right)\textrm{Det} \left(\frac{1}{u_i-b_j-\frac{i}{2}}\right)}
{\textrm{Det} \left(\frac{i}{b_{ij}+i}\right)} \ .
\ee

\subsubsection{The matrix factor as a scalar product}
\label{sca-pro}

Another nice representation of the matrix factor is as a scalar product between two symmetric multiparticle wave functions. These are defined by
\be
\psi^{(2n)}(b_1,\ldots,b_{2n})\equiv
\frac{2n}{n!}\,\frac{\delta_{2n}(b_1,\ldots,b_{2n})}
{\displaystyle\prod_{i,j=1}^{2n} \left ( u_i-b_j+\frac{i}{2}\right ) }  \,
\prod_{i<j}^{2n}\frac{b_{ij}}{b_{ij}+i} \, .
\ee
In terms of $\psi^{(2n)}$ the matrix factor (\ref{Nekr-scal}) can be recast as
\be\label{Pi_PP}
\Pi_{mat}^{(2n)}(u_1,\ldots,u_{2n})=
\frac{1}{(2n)!}\int\prod_{i=1}^{2n}\frac{db_i}{2\pi}\,\left[\psi^{(2n)}(b_1,\ldots,b_{2n})\right]^\ast
\psi^{(2n)}(b_1,\ldots,b_{2n}) \,,
\ee
denoting with $[\psi^{(2n)}]^\ast$ the complex conjugate of $\psi^{(2n)}$ \footnote{In fact, the splitting of the integrand in (\ref{Nekr-scal}) is not univocally determined; we chose to define $\psi^{(2n)}$ such that it be regular on the points $u_i-b_j=i/2$.}. The recursive behaviour of the $\delta $ polynomials (\ref {delta-rec}) also affects the
functions $\psi^{(2n)}$: indeed, they enjoy the following recursive relation
\ba\label{recursive_psi}
\psi^{(2n)}(b_1,\ldots,b_{2n}) &=&
2\sum_{\stackrel{l=1}{l\neq k}}^{2n}\frac{\psi^{(2n-2)}(b_1,\ldots,\underline{b_k},\underline{b_l}\ldots,b_{2n})}
{\displaystyle\prod_{i,j\in\{k,l\}}(u_i-b_j+\frac{i}{2})}
\frac{1}{\displaystyle\prod_{\stackrel{i\in\{k,l\}}{j\notin\{k,l\}}}(u_i-b_j+\frac{i}{2})(u_j-b_i+\frac{i}{2})}\cdot\nn\\
&\cdot & \left(\frac{b_{kl}}{b_{kl}+i(-1)^{\Theta(k-l)}}\right)
\prod^{2n}_{\stackrel{j=1}{j\neq k,l}}(b_{jk}-i(-1)^{\Theta(j-k)})
\prod^{2n}_{\stackrel{j=1}{j\neq k,l}}(b_{jl}-i(-1)^{\Theta(j-l)}) \quad
\ea
where $\Theta(j)$ stands for the Heaviside step function; the index $k$ can be arbitrarily chosen, then held fixed, with no summation involved. Upon introducing the set
$$
\m{T}_{kl}=\left\{(i,j)\in\{1,\ldots,2n\}\times\{1,\ldots,2n\}|\, i\in\{k,l\}\vee j\in\{k,l\}\right\} \, ,
$$
whose elements are couples of natural number (from $1$ to $2n$), of whom at least one is equal to $k$ or $l$, the relation above finds a more compact expression ($k$ arbitrarily chosen):
\ba
\psi^{(2n)}(b_1,\ldots,b_{2n}) &=&
2\sum_{\stackrel{l=1}{l\neq k}}^{2n}\frac{\psi^{(2n-2)}(b_1,\ldots,\underline{b_k},\underline{b_l}\ldots,b_{2n})}
{\displaystyle\prod_{(i,j)\in\m{T}_{kl}}\left(u_i-b_j+\frac{i}{2}\right)}
\left(\frac{b_{kl}}{b_{kl}+i(-1)^{\Theta(k-l)}}\right) \cdot\nn\\
&\cdot & \prod_{h\in\{k,l\}}\prod^{2n}_{\stackrel{j=1}{j\notin \{k,l\}}}(b_{jh}-i(-1)^{\Theta(j-h)}) \ .
\ea
For clarity, the very first $\psi^{(2n)}$ are listed, making use of the short-hand notation $\omega_{ij}\equiv u_i-b_j+i/2$\,:
\ba
\psi^{(2)}(b_1,b_2) &=& \frac{2}{\omega_{11}\omega_{12}\omega_{21}\omega_{22}}
\frac{b_{12}}{b_{12}+i} \ , \\
\psi^{(4)}(b_1,\ldots,b_4) &=& \frac{2\psi^{(2)}(b_3,b_4)}{\omega_{11}\omega_{12}\omega_{21}\omega_{22}}
  \frac{(b_{13}-i)(b_{14}-i)(b_{23}-i)(b_{24}-i)}{\omega_{23}\omega_{32}\omega_{24}\omega_{42}\omega_{13}\omega_{31}\omega_{14}\omega_{41}}
  \frac{b_{12}}{b_{12}+i} + \nn\\
&+& \frac{2\psi^{(2)}(b_1,b_4)}{\omega_{22}\omega_{23}\omega_{32}\omega_{33}}
  \frac{(b_{12}-i)(b_{13}-i)(b_{24}-i)(b_{34}-i)}{\omega_{21}\omega_{12}\omega_{24}\omega_{42}\omega_{13}\omega_{31}\omega_{34}\omega_{43}}
  \frac{b_{23}}{b_{23}+i} + \nn\\	
&+& \frac{(-1)2\psi^{(2)}(b_1,b_3)}{\omega_{44}\omega_{42}\omega_{24}\omega_{22}}
  \frac{(b_{12}-i)(b_{14}-i)(b_{23}-i)(b_{34}-i)}{\omega_{21}\omega_{12}\omega_{23}\omega_{32}\omega_{14}\omega_{41}\omega_{34}\omega_{43}}
  \frac{b_{24}}{b_{24}+i} \nn
\ea
(in reference to (\ref{recursive_psi}), we set the fixed index $k=2$ to write $\psi^{(4)}$, as an example).

\section{Asymptotic factorisation}
\label{asy}
\setcounter{equation}{0}

The main aim of this article is to arrive at definite expressions for the contributions of scalars to the expectation values of hexagonal Wilson loop in the strong coupling limit, which is expected to be exponential in  $\sqrt{\lambda}$. In the framework of the OPE discussed before,
this entails going through integrations involving the functions $G^{(2n)}$. The different terms of the series $W^{(2n)}$ turn out to be of different orders in $\sqrt{\lambda}$, in particular $W^{(2n)} \sim (\sqrt{\lambda})^{n}$. Given the exponential behaviour of $W$, what we have in mind is to study its logarithm, whose terms of the series expansion contain the connected counterparts $g^{(2n)}$ of the $G^{(2n)}$ and their order is expected to be $\sqrt{\lambda}$: we will describe the procedure in details in Section \ref{min-area}. This part is devoted to prove some fundamental properties of the function $G^{(2n)}$, which in turn will apply on the connected parts enabling us to effectively employ the series of the logarithm.

In order to understand the properties of the expansions of $W$ and $\ln W$, it is necessary to analyse the behaviour of these functions when some of their arguments, the rapidities $u_i$, go to infinity. More in detail, we are going to show a factorisation property for any $G^{(2n)}$ when we shift an even number $m$ of its rapidities by large amounts $\Lambda _i$, while holding fixed the remaining $2n-m$. To be specific, we will prove that factorisation arises when considering
\be
G^{(2n)}(u_1+\Lambda _1, u_2 +\Lambda _2,\ldots , u_m+\Lambda _m, u_{m+1},\ldots ,u_{2n}) \, ,
\ee
where each of the $\Lambda _i$ is parametrised as $\Lambda _i=c_i R +O(R^0)$, with $c_i$ constants and $R\rightarrow \infty$. The procedure is an extension of the one discussed in the letter \cite{BFPR2}, where we proved the
factorisation $G^{(2n)}\to G^{(2k)}G^{(2n-2k)}$ when the $2n$ particles are split in two groups composed respectively by $2k$ and $2n-2k$ particles, separated by the large parameter $\Lambda$ (in this particular case all the $c_i$ above are equal).
More in general the discussion we are going to expose now are extensions to asymptotically free theories of analogous results \cite {Smirnov} found in Form Factor computations. While in \cite {Smirnov} the corrections to factorisation are exponentially suppressed, in our case they enjoy a simple power-like decay as a consequence of asymptotic freedom. This very important fact is the core of this section since, as it will be extensively clarified throughout the paper, the integral of the connected functions $g^{(2n)}$ over the $2n-1$ variables on which they depend must be finite. In order to assure that, we need to unravel the properties of $G^{(2n)}$ when some rapidities go to infinity separately.

\medskip

We start our analysis from the dynamical factor (\ref{dynamical}) which, in the strong coupling (non-scaling) regime, can be given an explicit form through
\ba\label{dynamical2}
\frac{1}{P_{ss}(u_i|u_j)P_{ss}(u_j|u_i)}&=&
\frac{4(u_i-u_j)\tanh \left(\frac{\pi (u_i-u_j)}{4}\right)\Gamma \left (\frac{3}{4}+i\frac{u_i-u_j}{4} \right)
\Gamma \left (\frac{3}{4}-i\frac{u_i-u_j}{4} \right)}{\Gamma \left (\frac{1}{4}+i\frac{u_i-u_j}{4} \right) \Gamma \left (\frac{1}{4}-i\frac{u_i-u_j}{4} \right)}
\equiv \Pi(u_i-u_j)\ \, , \nn \\
\quad\quad \mu_s(u)&=&\sqrt{\pi}\frac{\Gamma\left(\frac{3}{4}\right)}{\Gamma\left(\frac{1}{4}\right)}\equiv \mu \ .
\ea
Hence, $\Pi_{dyn}^{(2n)}$ enjoys the behaviour (valid, indeed, even if $m$ is odd)
\ba
&& \Pi_{dyn}^{(2n)}(u_1+\Lambda _1,\ldots,u_{m}+\Lambda_m,u_{m+1},\ldots,u_{2n})\longrightarrow  \nonumber \\
&& \Pi_{dyn}^{(m)}(u_{1}+\Lambda _1,\ldots,u_{m}+\Lambda _m) \left ( \prod _{i=1}^{m} \Lambda _i ^2 \right ) ^{2n-m} \Pi_{dyn}^{(2n-m)}(u_{m+1},\ldots,u_{2n}) \label {as-dyn} \cdot \nonumber \\
&& \cdot \left [ 1+2 \sum _{i=1}^m\sum _{j=m+1}^{2n}\frac{u_i-u_j}{\Lambda _i} +O\left (R^{-2} \right ) \right ] \, ,
\ea
when the displacements $\Lambda _i$ are sent to infinity, as a consequence of the asymptotic behaviour
\be\label{asymptF}
u \rightarrow \infty \quad \Rightarrow \quad \Pi(u)= u^2-\frac{1}{2}-\frac{9}{8u^2}+O\left (u^{-4}\right ) \ .
\ee
We remark that $\Pi_{dyn}^{(m)}(u_{1}+\Lambda _1,\ldots,u_{m}+\Lambda _m)$ in fact is divergent if at least one of the
$\Lambda _i$ is different from the others, because in this case
\be
\Pi_{dyn}^{(m)}(u_{1}+\Lambda _1,\ldots,u_{m}+\Lambda _m) =
\mu ^m \prod _{\underset{i<j}{i,j=1}}^{m} (\Lambda _i-\Lambda _j)^2 + \ldots  \, ,
\ee
as a consequence of (\ref {asymptF}).

\medskip

In order to show how to work out the matrix part (\ref{Pi_mat}) instead, it is convenient to tackle the simplest non trivial case first, {\it i.e.} the asymptotic factorisation $\Pi_{mat}^{(4)}\rightarrow \Pi_{mat}^{(2)}\Pi_{mat}^{(2)}$: eventually, the procedure can be straightforwardly adapted to the general case
$\Pi_{mat}^{(2n)}\rightarrow \Pi_{mat}^{(2k)}\Pi_{mat}^{(2n-2k)}$\,.
To start with, we perform the shift on the rapidities $u_1\to u_1+\Lambda _1 ,\,u_2\to u_2+\Lambda _2$; then, for large $\Lambda_ 1$, $\Lambda _2$ the integrals (\ref{Pi_mat}) receive the main contribution from the region in which two roots $b$, one $a$ and one $c$ are comparable with $\Lambda _i$.  Therefore, we write (\ref{Pi_mat}) after shifting, for instance, $a_1$ by $\Lambda _1^{a}$, $c_1$ by $\Lambda _1^{c}$ and $b_1,b_2$ by $\Lambda _1^{b}$, $\Lambda _2^{b}$, respectively, where the large shifts of isotopic variables can be equal to $\Lambda _1$ or $\Lambda _2$.
Actually, we have to sum over all possible choices for shifts $\Lambda _i^{\alpha}$, $\alpha=a,b,c,$, which give the same result. We indicate shortly $\sum \limits_{\textrm{shifts}}$ this sum in formula (\ref {Pi_mat2}). Eventually, the resulting expression has to be multiplied by a multiplicity factor
\be
24= \binom{4}{2} \cdot 2 \cdot 2
\ee
taking into account the $\binom{4}{2}$ independent choices of a pair of $b_i$ out of four and the two choices for the $a_i$ and $c_i$, all giving the
same result. Details of the calculations are reported in Appendix \ref {calc-4}.
The final result is relation (\ref {pimat-final}), which reads
\ba\label{4-fact-PiMat}
&&\Pi_{mat}^{(4)}(u_1+\Lambda_1,u_2+\Lambda_2,u_3,u_4)=
\Lambda_1^{-4}\Lambda _2^{-4}\Bigl [1+2(u_3+u_4)\left (\frac{1}{\Lambda _1}+\frac{1}{\Lambda _2} \right )- \nn \\
&&- 4\left (\frac{u_1}{\Lambda _1}+\frac{u_2}{\Lambda _2} \right ) +O(R^{-2}) \Bigr ]  \Pi_{mat}^{(2)}(u_1+\Lambda _1,u_2+\Lambda _2)\Pi_{mat}^{(2)}(u_3,u_4) \ , \label {Pi4-fact}
\ea
where, if $c_1 \not= c_2$, $\Pi_{mat}^{(2)}(u_1+\Lambda _1,u_2+\Lambda _2) =6/\Lambda _{12}^4 + \ldots $, whilst if $c_1=c_2$ this function is finite.

On the other hand, for the dynamical parts we have
\ba
&& \Pi_{dyn}^{(4)}(u_1+\Lambda_1,u_2+\Lambda_2,u_3,u_4)\rightarrow \Pi_{dyn}^{(2)}(u_1+\Lambda_1,u_2+\Lambda_2)\Pi_{dyn}^{(2)}(u_3,u_4)
\Lambda _1^4 \Lambda _2^4 \cdot \nn \\
&& \cdot \left [1-2(u_3+u_4)\left (\frac{1}{\Lambda _1}+\frac{1}{\Lambda _2} \right )+4\left (\frac{u_1}{\Lambda _1}+\frac{u_2}{\Lambda _2} \right ) +O(R^{-2}) \right ]  \, ,
\ea
where, if $c_1 \not= c_2$, $\Pi_{dyn}^{(2)}(u_1+\Lambda _1,u_2+\Lambda _2) = \mu ^2 \Lambda _{12}^2 + \ldots $, otherwise if $c_1=c_2$ this function is finite.

Putting the dynamical and matrix parts together (\ref{Gi2n}), we get the following asymptotic factorisation when the first two rapidities get large:
\be
G^{(4)}(u_1+\Lambda_1,u_2+\Lambda_2,u_3,u_4)\ \overset{\Lambda_i\rightarrow\infty}{\longrightarrow}\ G^{(2)}(u_1+\Lambda _1,u_2+\Lambda _2)G^{(2)}(u_3,u_4) [1+O(R^{-2}) ]\, , \label{4-fact}
\ee
where, if $c_1 \not= c_2$, $G^{(2)}(u_1+\Lambda _1,u_2+\Lambda _2)=6\mu ^2 /\Lambda _{12}^2+\ldots $, if $c_1=c_2$, $G^{(2)}$ is finite.
Since $G^{(4)}(u_1,u_2,u_3,u_4)$ is a symmetric function of the four $u_i$, property (\ref {4-fact}) is indeed valid when any couple of rapidities is very large.

\medskip

We now consider a more general case: we shift an even number $m=2k$ of rapidities by amounts $\Lambda _i$. Because of the symmetry of the function $G$ we can stick to the case in which the first $m$ rapidities are shifted: $u_i\to u_i+\Lambda _i$ for $1\leq i\leq m$. In order to get the main contribution to the integrals for large $\Lambda _i$, we also shift $a_i\to a_i+\Lambda _i^a$ and $c_i\to c_i+\Lambda_i^c$ for $1\leq i\leq k$, along with $b_i\to b_i+\Lambda _i^b$ for $1\leq i\leq m$. Details of the calculations are reported in Appendix \ref {calc-2n}.

The final result is:
\ba
&& \Pi_{mat}^{(2n)}(u_1+\Lambda _1,\ldots,u_{2k}+\Lambda _{2k},u_{2k+1},\ldots,u_{2n}) \rightarrow
\frac{1}{\left ( \prod \limits _{i=1}^m \Lambda _i \right )^{4n-4k}}\Bigl [1+2 \sum _{i=1}^{2k}\frac{1}{\Lambda _i}\sum _{j=2k+1}^{2n}u_j- \nonumber \\
&-& 2(2n-2k)\sum _{i=1}^{2k}\frac{u_i}{\Lambda _i} + O\left (R^{-2}\right ) \Bigr ] \Pi_{mat}^{(2k)}(u_1+\Lambda _1,\ldots,u_{2k}+\Lambda_{2k})\Pi_{mat}^{(2n-2k)}(u_{2k+1},\ldots,u_{2n}) \, ,
\ea
which is an extension to the case in which the $\Lambda_i$ can be different to analogous results \cite{BFPR2} holding when all the $\Lambda_i$ are equal.
We remark that $\Pi_{mat}^{(2k)}(u_1+\Lambda _1,\ldots,u_{2k}+\Lambda_{2k})$ in the general case with different $\Lambda _i$ goes to zero like
\be
\Pi_{mat}^{(2k)}(u_1+\Lambda _1,\ldots,u_{2k}+\Lambda_{2k})\sim
\prod _{i<j=1}^{2k} (\Lambda _i-\Lambda _j)^{-2} \left [ \Lambda _{12}^{-2}\Lambda _{34}^{-2}\ldots \Lambda _{2k-1,2k}^{-2}+pairings \right ]
\, . \label{pimatzero}
\ee
If some of the $\Lambda_i$ are equal, the function $\Pi_{mat}^{(2k)}(u_1+\Lambda _1,\ldots,u_{2k}+\Lambda_{2k})$ goes to zero less rapidly than
(\ref {pimatzero}).

On the other side, the dynamical part enjoys the behaviour (\ref {as-dyn}). Putting dynamical and matrix part together, we get the sought factorisation property:
\ba\label{2n-fact}
&& G^{(2n)}(u_1+\Lambda _1,\ldots,u_{2k}+\Lambda _{2k},u_{2k+1},\ldots,u_{2n}) \rightarrow \nonumber \\
&& \rightarrow G^{(2k)}(u_1+\Lambda _1,\ldots,u_{2k}+\Lambda_{2k})G^{(2n-2k)}(u_{2k+1},\ldots,u_{2n})\Bigl [1+ O\left (R^{-2}\right ) \Bigr ] \, ,
\ea
which extends the known factorisation property of \cite{Smirnov}. We remark that the function $G^{(2k)}(u_1+\Lambda _1,\ldots,u_{2k}+\Lambda_{2k})$, when $\Lambda _i$ are all different, behaves like
\be
G^{(2k)}(u_1+\Lambda _1,\ldots,u_{2k}+\Lambda_{2k})
\sim \left [ \Lambda _{12}^{-2}\Lambda _{34}^{-2}\ldots \Lambda _{2k-1,2k}^{-2}+pairings \right ] \, .
\ee
Therefore, putting $\Lambda _i =c_i R +O(R^0)$, with different $c_i$ and $R\rightarrow +\infty$, the behaviour of $G^{(2k)}(u_1+\Lambda _1,\ldots,u_{2k}+\Lambda_{2k})$ and of $G^{(2n)}(u_1+\Lambda _1,\ldots,u_{2k}+\Lambda _{2k},u_{2k+1},\ldots,u_{2n})$ is
\be\label{G-even}
G^{(2k)}(u_1+\Lambda _1,\ldots,u_{2k}+\Lambda_{2k}) \sim R^{-2k} \, , \quad G^{(2n)}(u_1+\Lambda _1,\ldots,u_{2k}+\Lambda _{2k},u_{2k+1},\ldots,u_{2n})
\sim R^{-2k} \, .
\ee
If $p$ of the $c_i$ are equal, {\it i.e.} we have $\Lambda _1=\Lambda _2=\ldots =\Lambda _p \not = \Lambda _{p+1} \ldots \not= \Lambda _{2k}$, $G^{(2k)}$ and consequently $G^{(2n)}$ go to zero as $R^{-2k+2\left [\frac{p}{2}\right ]}$: some of these cases and more general configurations will be examined in Appendix \ref {int-con}.
\medskip

Now, we spend a few words to discuss the behaviour of $G^{(2n)}$ when we shift an odd number $m$ of rapidities.
This discussion is necessary in order to study the behaviour for large rapidities of the integrands appearing in the matrix part. Obviously, in this case there is no factorisation, since there are no functions $G$ with an odd number of arguments.

When $m$ is odd, we find convenient to define $m=2k-1$. Then, we shift $a_i\to a_i+\Lambda _i^a$ and $c_i\to c_i+\Lambda_i^c$ for $1\leq i\leq k$, along with $b_i\to b_i+\Lambda _i^b$ for $1\leq i\leq m$. With these positions, formula (\ref {Intpimat}) still holds, along with (\ref {r-int}). Sending $\Lambda _i \rightarrow +\infty$ inside the integrals, we get
\ba
&& \m{R}^{(2n,m)} \rightarrow  \frac{1}{\left ( \prod \limits _{i=1}^m \Lambda _i \right )^{4n-2m}}
\prod _{i=1}^m (\Lambda _i^b)^2 \prod _{i=1}^k (\Lambda _i^a \Lambda _i ^c )^{-2}
\Bigl \{ 1+2 \sum _{i=1}^{m}\frac{1}{\Lambda _i^b}\sum _{j=m+1}^{2n}u_j-2(2n-m)\sum _{i=1}^{m}\frac{u_i}{\Lambda _i} + \nonumber \\
&& + \sum _{j=m+1}^{2n} 2b_j \left [ \sum _{i=1}^{m}\frac{1}{\Lambda _i}-2\sum _{i=1}^{m} \frac{1}{\Lambda _i^b} + \sum _{i=1}^k\left (\frac{1}{\Lambda _i^a}+\frac{1}{\Lambda _i^c} \right ) \right ] +  \sum _{j=k+1}^{n} 2a_j \left ( \sum _{i=1}^{m}\frac{1}{\Lambda _i^b}-2 \sum _{i=1}^{k}\frac{1}{\Lambda _i^a} \right ) + \nn \\
&& +  \sum _{j=k+1}^{n} 2c_j \left ( \sum _{i=1}^{m}\frac{1}{\Lambda _i^b}-2 \sum _{i=1}^{k}\frac{1}{\Lambda _i^c} \right ) + 2 \sum _{i=1}^m \frac{b_i}{\Lambda _i^b}-2\sum _{i=1}^k \frac{a_i}{\Lambda _i^a}-2\sum _{i=1}^k \frac{c_i}{\Lambda _i^c}+ O\left (R^{-2} \right ) \Bigr \} \, . \label {R-int2}
\ea
Therefore, sticking only to the leading term, when $m=2k-1$ for the matrix part we have the behaviour for different $\Lambda _i$
\ba
&& \Pi_{mat}^{(2n)}(u_1+\Lambda _1,\ldots,u_{2k-1}+\Lambda _{2k-1},u_{2k},\ldots,u_{2n}) \sim
\frac{(\prod \limits _{i=1}^k \Lambda _i ^{-4}+perm.)}{\left ( \prod \limits _{i=1}^{2k-1} \Lambda _i \right )^{4n-4k}} \cdot\nn\\
&& \cdot\prod \limits _{\underset{i<j}{i,j=1}}^{2k-1} (\Lambda _i-\Lambda _j)^{-2}
\left [ \Lambda _{12}^{-2}\Lambda _{34}^{-2}\ldots \Lambda _{2k-3,2k-2}^{-2}+pairings \right ]
\cdot \textrm{finite function} (u_{2k},\ldots,u_{2n}) \, , \nonumber
\ea
where for 'finite function' we mean the third line of (\ref {Intpimat}) excluding the function $\m{R}^{(2n,m)}$.
We have to multiply the matrix part by the dynamical part which behaves - also for odd $m$ - as (\ref {as-dyn}).
Doing this we get
\ba
&& G^{(2n)}(u_1+\Lambda _1,\ldots,u_{2k-1}+\Lambda _{2k-1},u_{2k},\ldots,u_{2n}) \sim \nonumber \\
&& 
\left [ \Lambda _{12}^{-2}\Lambda _{34}^{-2}\ldots \Lambda _{2k-3,2k-2}^{-2}+pairings \right ]
\prod \limits _{i=1}^{2k-1}\Lambda _i ^2 \left ( \prod \limits _{i=1}^k \Lambda _i ^{-4} +perm. \right ) \cdot  \textrm{finite function} (u_{2k},\ldots,u_{2n}) \, . \nonumber
\ea
Therefore, we conclude that, if all the $\Lambda_i$ are different and of order $R$ as before, the behaviour of $G^{(2n)}(u_1+\Lambda _1,\ldots,u_{2k-1}+\Lambda _{2k-1},u_{2k},\ldots,u_{2n}) $  is
\be
G^{(2n)}(u_1+\Lambda _1,\ldots,u_{2k-1}+\Lambda _{2k-1},u_{2k},\ldots,u_{2n})  \sim R^{-2k}=R^{-m-1} \, . \label{G-odd}
\ee
If $p$ of the $\Lambda _i$ are equal, the function $G^{(2n)}$ vanishes as $R^{-2k+2\left [ \frac{p}{2}\right ]}$.

\vspace{0.3cm}

\vspace{0.3cm}

It is worth to point out that there exists a different method to prove the factorisation properties obtained so far, which makes use of the Young tableaux representation (\ref{MatYoung}).
To give a sketch, let us consider the $n=2$ split $4\to 2+2$, which corresponds to (\ref{4-fact-PiMat}) with equivalent shifts $\Lambda_1=\Lambda_2\equiv\Lambda$.
We thus observe that many diagrams of $\Pi_{mat}^{(4)}$ split into a product of two diagrams already encountered when computing $\Pi_{mat}^{(2)}$, weighted by a factor $\Lambda^{-8}$
\footnotesize
\ba
(1,1,1,1)&&\to \Lambda^{-8}(1,1)_{12}\times(1,1)_{34} \nn\\
(2,0,2,0)+(2,0,0,2)+(0,2,2,0)+(0,2,0,2)&&\to \Lambda^{-8}[(2,0)_{12}+(0,2)_{12}]\times[(2,0)_{34}+(0,2)_{34}]\nn\\
(1,1,2,0)+(1,1,0,2)+(2,0,1,1)+(0,2,1,1)&&\to \Lambda^{-8}(1,1)_{12}\times[(2,0)_{34}+(0,2)_{34}] + 12 \leftrightarrow 34
\ea
\normalsize
where the right hand side (RHS) members sum up to $\Lambda^{-8}\Pi_{mat}^{(2)}(u_1,u_2)\Pi_{mat}^{(2)}(u_3,u_4)$. We remark that other diagrams of $\Pi_{mat}^{(4)}$ are of subleading order $o(\Lambda^{-8})$ and thus they do not contribute to the factorisation. The procedure can be extended to the general split $2n\to 2(n-k) + 2k$, allowing also to have different shifts $\Lambda_i$.

\vspace{0.3cm}

To summarize, we addressed some\footnote{As discussed in the Appendix \ref{int-con}, there are other asymptotic regions to be analysed.} different splits of the rapidities and obtained the corresponding asymptotic behaviours of the $G^{(2n)}$. The main results are equations (\ref{2n-fact}), (\ref{G-even}) and (\ref{G-odd}), they will turn out to be useful in section \ref{min-area} where we study the connected functions $g^{(2n)}$.

\subsection{Polar structure of the matrix factor}

We are now coming to an important point of this article. Indeed,
making use of the asymptotic factorisation discussed in this section, we can now prove in general that the matrix part can be written as follows
\be\label{P2n}
\Pi_{mat}^{(2n)}(u_1,\cdots , u_{2n})=\frac{P_{2n}(u_1,\ldots,u_{2n})}{\displaystyle\prod_{i<j}^{2n}(u_{ij}^2+1)(u_{ij}^2+4)} \, ,
\ee
where $P_{2n}$ is a symmetric\footnote{{\it I.e.} invariant under permutations of the arguments} polynomial. Therefore, $\Pi_{mat}^{(2n)}$ has poles only when the difference of two rapidities equals $\pm i$ or $\pm 2i$.

Actually, for the present proof we need only to know the behaviour of $\Pi_{mat}^{(2n)}$ when two arbitrary rapidities $u_p$, $u_q$ get large in the same way, {\it i.e.} $c_p=c_q$ and $c_i=0$ for $i\not= p,q$:
\be\label{fact}
\Pi_{mat}^{(2n)}(u_1, \ldots, u_{p}+\Lambda,\ldots ,u_{q}+\Lambda, \ldots, u_{2n})\simeq \Lambda^{-8(n-1)}\Pi_{mat}^{(2)}(u_p,u_q)\Pi_{mat}^{(2n-2)}(u_1, \ldots, \underline{u_p},\ldots ,\underline{u_q}, \ldots, u_{2n}) \, ,
\ee
where the structure of the two point function is (\ref {2scalars}) ({\it i.e.} $\Pi_{mat}^{(2)}(u_1,u_{2})=6/\{[(u_1-u_2)^2+1][(u_1-u_2)^2+4]\}$) and the notation $\underline{u_k}$ means the omission of the rapidity $u_k$. Then, we remember that $\Pi_{mat}^{(2n)}(u_1,\ldots, u_{2n})$ depends only on the differences $u_{ij}$, with $i<j$, $i,j=1,\ldots, 2n$, and, consequently, may show singularities when $u_{ij}$ pick particular values. Of course, any singular values of $\Pi_{mat}^{(2n)}$ for the particular difference $u_{pq}$ are left unchanged by the shifts in the left hand side (LHS) of (\ref {fact}), whose RHS (in its second factor) tells us where they occur: $u_{pq}=\pm i, \pm 2i$. Repeating this reasoning for all the possible differences of rapidities, we obtain the structure (\ref {P2n}).

What is left unknown in (\ref {P2n}) are the polynomials $P_{2n}$. The simplest ones, corresponding to $n=1,2$, are reported in Appendix \ref {sca-pol} ; for $n\geq 3$ expressions for $P_{2n}$ get rapidly unwieldy. The residue formula (\ref{ResPimat}) allows us to relate the polynomial $P_{2n}$, evaluated in a specific configuration, to a smaller polynomial. Other general properties of these polynomials, which can be found without much ado, are their degree and their highest degree monomial. The highest degree monomial will be discussed in Appendix \ref {sca-pol}. Instead, the degree of the polynomial $P_{2n}(u_1,\ldots, u_{2n})$ may be found here by comparing (\ref{P2n}) to (\ref{Nekr-scal}). The degree of $\Pi_{mat}^{(2n)}(u_1,\ldots, u_{2n})$ is found to be equal to $-4n^2$ by using integral representation (\ref{Nekr-scal}) and the fact that the degree of $\delta _{2n}(u_1,\ldots, u_{2n})$ is $2n(n-1)$. It then follows that the degree of $P_{2n}(u_1,\ldots, u_{2n})$ is $-4n^2+4\frac{2n(2n-1)}{2}=4n(n-1)$. It is worth to remark that the two polynomials $P_{2n}$ and $\delta^2_{2n}$ have both degree $4n(n-1)$ and, as discussed in appendix \ref{sca-pol}, their highest degrees share the same structure.

\section{The expansion in the strong coupling regime}
\label{min-area}
\setcounter{equation}{0}

As a preliminary remark, we will show that, since scalars decouple from the rest of the particles for large values of $\lambda$, the hexagonal Wilson loop can be decomposed into the product of a factor accounting for the minimal surfaces on $AdS_5$, multiplied per an $O(6)$ factor ascribable to scalars. We will carefully show how the scalar factor can be isolated, by initially mixing scalars with a single kind of particles at once, then by considering all the species together.\\
As a notation remark, in what follows $W_{\alpha_1,\dots,\alpha_k}$ stands for the expectation value of a hexagonal Wilson loop, taking into account scalars, gluons, fermions and antifermions as excitations ($i.e.\ \alpha_1,\dots,\alpha_k\in\{s,g,f,\bar f\}$),
while $W^{(N_1\alpha_1+\dots+N_k\alpha_k)}$ denotes the contribution to $W_{\alpha_1,\dots,\alpha_k}$, brought by an intermediate state made of $N_1$ particles of type $\alpha_1$, $N_2$ of type $\alpha_2$, and so on. In order to keep in touch with the previous notation, the $2N$-scalar contribution to the hexagonal Wilson loop and the $2N$-scalar matrix factor are shortly denoted as $W^{(2N)}$ and $\Pi^{(2N)}_{mat}$ (instead of $W^{(2Ns)}$ and $\Pi^{(2Ns)}_{mat}$), while in the strong coupling limit $\mu_s(u)$ reduces to $\mu$ (\ref{dynamical2}).
\medskip

\noindent\textbf{$\bullet\ $ Scalars and gluons:}\\
When considering $2N$ scalars of rapidities $u_k$ along with $M$ gluons with rapidities $v_k$, their contribution to the hexagonal Wilson loop reads
\ba
W^{(2Ns+Mg)} &=& \frac{1}{(2N)!M!}\int\prod_{i=1}^{M}\frac{d v_i\, e^{-\tau E_g(v_i)-i\sigma p_g(v_i)}}{2\pi} \prod_{j=1}^{2N}\frac{d u_j\, e^{-\tau E(u_j)-i\sigma p(u_j)}}{2\pi} \cdot \nn\\
&\cdot & \Pi_{mat}^{(2Ns)}(u_1\ldots u_{2N})\,\Pi_{dyn}^{(2Ns+Mg)}(u_1,\ldots,v_M) \, .
\ea
Since gluons behave as singlets under $SU(4)$, the matrix factor $\Pi_{mat}^{(2Ns)}$ takes into account only the $2N$ scalars, arranged into the singlet configuration, whereas the dynamical factor enjoys a pairwise decomposition
\cite{Bel1501}
\be
\Pi_{dyn}^{(2Ns+Mg)}(u_1,\ldots,v_M)=\frac{\displaystyle \prod_{i=1}^{M} \mu_g(v_i) \prod_{j=1}^{2N}\mu_s(u_j)}
{\displaystyle\prod_{i<j}P_{ss}(u_i|u_j)P_{ss}(u_j|u_i)\prod_{i<j}P_{gg}(v_i|v_j)P_{gg}(v_j|v_i)
\prod_{i,j}P_{sg}(u_i|v_j)P_{gs}(v_j|u_j)} \,,
\ee
being $E_g,\,p_g,\,\mu_g$ the expressions for energy, momentum and measure of gluons, whereas $P_{gg},\,P_{sg}$ stand for the gluon-gluon and scalar-gluon amplitudes.
In the strong coupling regime considered, the gluon rapidities get rescaled, $i.e.\ v_k=\frac{\sqrt{\lambda}}{2\pi}\bar v_k$, while holding the scalar rapidities fixed, resulting in a decoupling at the level of pentagon amplitudes \cite{Bel1607}
\be
P_{sg}(u|v)=1+O(e^{-\sqrt{\lambda}/4}) \,,
\ee
and as a by-product the scalar and gluon contributions become clearly distinguishable
\ba
W^{(2Ns+Mg)} &=& 
\frac{1}{(2N)!M!}\int\prod_{i=1}^{M}\frac{d v_i\, e^{-\tau E_g(v_i)-i\sigma p_g(v_i)}}{2\pi} \prod_{j=1}^{2N}\frac{d u_j\, e^{-\tau E(u_j)-i\sigma p(u_j)}}{2\pi} \cdot \nn\\
&\cdot & \frac{\Pi_{mat}^{(2N)}(u_1\ldots u_{2N})} {\displaystyle\prod_{i<j}^{2N}P_{ss}(u_i|u_j)P_{ss}(u_j|u_i)}\,\frac{\displaystyle\prod_{i=1}^{2N}\prod_{j=1}^{M}[1+O(e^{-\pi g})]}{\displaystyle\prod_{i<j}^{M}P_{gg}(v_i|v_j)P_{gg}(v_j|v_i)}
\simeq W^{(2N)} W^{(Mg)} \, . \label{fatt_sg}
\ea
From the last line of (\ref{fatt_sg}) we can infer the strong coupling factorisation of the hexagonal Wilson loop into a scalar part $W$ (\ref{Wilson}) and a gluon part $W_{g}$:
\ba
W_{sg}=\sum_{N=0}^\infty\sum_{M=0}^\infty W^{(2Ns+Mg)}\simeq
\sum_{N=0}^\infty\sum_{M=0}^\infty W^{(2N)} W^{(Mg)}=
\sum_{N=0}^\infty W^{(2N)} \sum_{M=0}^\infty W^{(Mg)}=
W \,W_{g} \,.\
\ea

\medskip
\noindent
\textbf{$\bullet\ $Scalars, fermions and antifermions:}\\
When studying the contribution to the hexagonal Wilson loop from $M$ fermions (with rapidities $u_k$), $M$ antifermions ($v_k$) and $2N$ scalars ($w_k$),
\ba
W^{(2Ns+Mf+M\bar f)} &=& \int\prod_{i=1}^{M}\frac{d u_i\, d v_i\, e^{-\tau [E_f(u_i)+E_f(v_i)]-i\sigma [p_f(u_i)+p_f(v_i)]}}{4\pi^2} \prod_{j=1}^{2N}\frac{d w_j\, e^{-\tau E(w_j)-i\sigma p(w_j)}}{2\pi} \cdot \nn\\
&\cdot & \frac{\Pi_{mat}^{(2Ns+Mf+M\bar f)}(\{w_i,u_i,v_i\})\Pi_{dyn}^{(2Ns+Mf+M\bar f)}(\{w_i,u_i,v_i\})}{(2N)!(M!)^2}
\, ;
\ea
the computation is more involved, since the matrix factor needs to take into account that fermions and antifermions are not $SU(4)$-singlets,
\ba
&&\Pi_{mat}^{(2Ns+Mf+M\bar f)}(\{w_i,u_i,v_i\})=\frac{1}{(K_a !)^2 K_b!}\int\prod_{i=1}^{K_a}\frac{da_i dc_i}{(2\pi)^2}\prod_{i=1}^{K_b}\frac{db_i}{2\pi}\cdot \nn\\
&& \cdot\frac{\displaystyle\prod_{i<j}^{K_a}[g(a_i-a_j)g(c_i-c_j)]\prod_{i<j}^{K_a}g(b_i-b_j)}
{\displaystyle\prod_{i=1}^{K_a}\prod_{j=1}^{K_b}[f(a_i-b_j)f(c_i-b_j)]
\prod_{i=1}^{K_a}\prod_{j=1}^{M}[f(a_i-u_j)f(c_i-v_j)]
\prod_{i=1}^{K_b}\prod_{j=1}^{2N}f(b_i-w_j)}
\ea
where the number of $a$-roots $K_a$ equals the number of $c$-roots, the number of $b$-roots is $K_b$, and they are related to the number of particles by
\ba
K_a &=& M+N \\
K_b &=& M+2N \ . \nn
\ea
In the strong coupling limit, we take finite scalar rapidities, $w_k=O(1)$, whereas fermion/antifermion rapidities get rescaled, $u_k=\frac{\sqrt{\lambda}}{2\pi}\bar u_k$ with $u_k=O(1)$, hence $M$ roots of types $a_k,b_k,c_k$ shall be rescaled accordingly, so that $a,b,c=O(\sqrt{\lambda})$ for $k\in\{1,\ldots,M\}$.
In view of the asymptotic behaviour
\ba
&& \frac{\displaystyle\prod_{i=1}^{M}\left[\prod_{j=M+1}^{K_a}g(a_i-a_j)g(c_i-c_j)\prod_{j=M+1}^{K_b}g(b_i-b_j)\right]}
{\displaystyle\prod_{i=M+1}^{K_a}\prod_{j=1}^{M}f(a_i-u_j)f(c_i-v_j)\prod_{i=1}^{M}\prod_{j=1}^{2N}f(b_i-w_j)
\prod_{i=1}^{M}\prod_{j=M+1}^{K_b}f(a_i-b_j)f(c_i-b_j)} \cdot\nn\\
&& \cdot\frac{1}{\displaystyle\prod_{i=M+1}^{K_a}\prod_{j=1}^{M}f(a_i-b_j)f(c_i-b_j)}=
\prod_{i=1}^M \frac{(a_i c_i)^{4N}b_i^{8N}}{(u_i v_i)^{2N}(a_i c_i)^{4N}b_i^{8N}}[1+O(1/\sqrt{\lambda})]
=\prod_{i=1}^M \frac{1+O(1/\sqrt{\lambda})}{(u_i v_i)^{2N}} \,,\nn\\
\ea
the whole matrix factor $\Pi_{mat}^{(2Ns+Mf+M\bar f)}$, in the zero $SU(4)$-charge configuration, factorises into the product of scalar and fermion parts:
\be
\Pi_{mat}^{(2Ns+Mf+M\bar f)}(\{w_i,u_i,v_i\})\simeq\Pi_{mat}^{(Mf+M\bar f)}(\{u_i,v_i\})\Pi_{mat}^{(2N)}(\{w_i \})\prod_{i=1}^M \frac{1}{(u_i v_i)^{2N}} \,.
\ee
The pairwise decomposition of the dynamical factor  
\cite{Bel1607}\footnote{Here portrayed for scalars and fermions only, for simplicity.}, together with the strong coupling behaviour of the mixed pentagons \cite{BEL} 
\ba
\Pi_{dyn}^{(2Ns+Mf)}(w_1,\ldots,w_{2N},u_1,\ldots,u_M) &=&
\frac{\displaystyle \prod_{i=1}^{2N}\mu_s(w_i) \prod_{j=1}^{M}\mu_f(u_i)}
{\displaystyle\prod_{i<j}P_{ss}(w_i|w_j)P_{ss}(w_j|w_i)\prod_{i<j}P_{ff}(u_i|u_j)P_{ff}(u_j|u_i)}\cdot \nn\\
&\cdot & \frac{1}{\displaystyle\prod_{i,j}P_{sf}(w_i|u_j)P_{fs}(u_j|w_i)} \\
P_{sf}(w,u) &=& \sqrt{\frac{2\pi}{\lambda^{1/2}\bar u}}\,\exp\left\{\frac{2\pi w^-}{\sqrt{\lambda}\bar u}+\ldots\right\}
\ea
allow us to the separate the scalar contribution from the fermion contribution
\ba
&& W^{(2Ns+Mf+M\bar f)} \simeq W^{(2N)}W^{(Mf+M\bar f)} \nn\\
&& W_{sf\bar f} = \sum_{N=0}^\infty\sum_{M=0}^\infty W^{(2Ns+Mf+M\bar f)}\simeq
\sum_{N=0}^\infty W^{(2N)} \sum_{M=0}^\infty W^{(Mf+M\bar f)}=
W\, W_{f\bar f} \, .
\ea

\medskip
\noindent
\textbf{$\bullet\ $ Scalars, gluons, fermions and antifermions:}\\
Finally, we consider scalars, gluons, fermions and antifermions altogether, each kind of particle being labelled by a Greek index
$\alpha\in\{s,g,f,\bar f\}$ (for scalars, gluons, fermions and antifermions respectively), while the Latin label distinguishes the particle of a given type $\alpha$ ($i=1,\ldots,N_\alpha$); since the system carries no overall $SU(4)$-charge, we need $N_f=N_{\bar f}$. Once the multiparticle pentagon factorisation is assumed \cite{Bel1501}
\be
\Pi_{dyn}(\{u^\alpha_i\})=\prod_{\alpha}\frac{\displaystyle\prod_{i=1}^{N_\alpha}\mu_\alpha(u_i^\alpha)}
{\displaystyle\prod_{i<j}^{N_\alpha}P_{\alpha\alpha}(u_i^\alpha|u_j^\alpha)P_{\alpha\alpha}(u_j^\alpha|u_i^\alpha)}
\prod_{\alpha}\prod_{\beta\neq\alpha}\prod_{i=1}^{N_\alpha}\prod_{j=1}^{N_\beta}\frac{1}{P_{\alpha\beta}(u_i^\alpha|u_j^\beta)} \,,
\ee
the hexagonal Wilson loop can be split into the product of the minimal area contribution $W_{AdS}$ and a scalar contribution $W_{O(6)}=W$
(\ref{Wilson})
\ba
W &=& \sum_{\stackrel{N_f,N_{\bar f},}{N_s,N_g}}\int\prod_\alpha\left[\frac{1}{N_\alpha!}\prod_{i=1}^{N_\alpha}\frac{d u_i^\alpha\,e^{-\tau E_\alpha(u^\alpha_i)-i\sigma p_\alpha(u^\alpha_i)}}{2\pi}\right]
\,\Pi_{mat}(\{u^s_i,u^f_i,u^{\bar f}_i\})\Pi_{dyn}(\{u^s_i,u^g_i,u^f_i,u^{\bar f}_i\})\simeq \nn\\
&\simeq & \sum_{N_s}\frac{1}{N_s!}\int\prod_{i=1}^{N_s}\frac{d u_i^s\,e^{-\tau E(u^s_i)-i\sigma p(u^s_i)}}{2\pi}
\,\frac{\Pi_{mat}^{(N_s)}(\{u^s_i\})}{\displaystyle\prod_{i<j}^{N_s}P_{ss}(u^s_i|u^s_j)P_{ss}(u^s_j|u^s_i)} \cdot\nn\\
&\cdot & \sum_{N_g,N_f,N_{\bar f}}\int\prod_{\alpha\neq s}\left[\frac{1}{N_\alpha!}\prod_{i=1}^{N_\alpha}\frac{d u_i^\alpha\,e^{-\tau E_\alpha(u^\alpha_i)-i\sigma p_\alpha(u^\alpha_i)}}{2\pi}\right] \Pi_{mat}^{(N_f f+N_f \bar f)}(\{u^f_i,u^{\bar f}_i\}) \cdot \nn\\
&\cdot & \prod_{\alpha\neq s}\prod_{i<j}^{N_\alpha}\frac{1}{P_{\alpha\alpha}(u_i^\alpha|u_j^\alpha)P_{\alpha\alpha}(u_j^\alpha|u_i^\alpha)}
\prod_{\alpha\neq s}\prod_{\stackrel{\beta\neq s}{\beta\neq\alpha}}\prod_{i=1}^{N_\alpha}\prod_{j=1}^{N_\beta}\frac{1}{P_{\alpha\beta}(u_i^\alpha|u_j^\beta)}=\nn\\
&=& W\,W_{gf\bar f}
=W_{O(6)}\,W_{AdS} \ .
\ea
As a consequence, in order to compute the same-order correction to the minimal area contribution, it is sufficient to consider scalars alone, ignoring their interactions with other particles. It is worth remarking that the two contributions, $AdS_5$ and $S^5$, behave very differently in the collinear limit $\tau\to+\infty$. The classical contribution $W_{AdS}\simeq e^{-\frac{\sqrt{\lambda}}{2n}A_6}$ becomes trivial, since the area is exponentially suppressed $A_6 \sim O(e^{-\sqrt{2}\tau})$ in the regime considered. On the other hand, the effect of the five sphere $S^5$, or scalars from the OPE point of view, may be finite and remains the only contribution to the hexagonal Wilson loop in the strong coupling limit. As it will be clearer in the following, the necessary condition is that the combination $z\simeq e^{-\frac{\sqrt{\lambda}}{4}}\tau$ does not get too large, otherwise also the five sphere contribution would be trivial: $W_{O(6)}=1+O(e^{-2z})$.
With this in mind, in the next subsection we are going to compute the contribution of scalars in the strong coupling limit of the hexagonal Wilson loop.

\subsection{The importance of being connected}
\label{conn-func}

A {\it Leitmotiv} ({\it cf.} the explicit Young tableaux computation of subsection \ref{young} and the parallel with $\mathcal{N}=2$ theories \cite{NEK}) is to consider the WL (\ref{Wilson}), in general, as a partition function. Therefore, we find convenient to compute its logarithm in terms of the 'connected' functions $g^{(2n)}$:
\be\label{logW}
 {\cal F}  \equiv \ln W  =\sum_{n=1}^{\infty}\frac{1}{(2n)!}\int\prod_{i=1}^{2n}\frac{du_i}{2\pi}g^{(2n)}(u_1,\ldots,u_{2n})
\,e^{\displaystyle -\sum_{i=1}^{2n}\displaystyle  [\tau E(u_i)+i\sigma p(u_i)]} \equiv \sum_{n=1}^{\infty}{\cal F}^{(2n)} \,.
\ee
As well known the 'non-connected' function $G^{(2n)}$ can be expressed in terms of the connected $g^{(2l)}$, with $l\leq n$; here we list the first few examples, upon introducing the shorthand notation $g_{i_1 \ldots i_n}\equiv g^{(n)}(u_{i_1},\ldots,u_{i_n})$:
\ba\label{Gg}
&& G_{12}=g_{12} \nn\\
&& G_{1234}=g_{1234}+g_{12}g_{34}+g_{13}g_{24}+g_{14}g_{23}= g_{1234} + (g_{12}g_{34} + 2 \textit{ perm.}) \nn\\
&& G_{123456}=g_{123456} + (g_{12}g_{3456} + 14\textit{ perm.}) + (g_{12}g_{34}g_{56} +14 \textit{ perm.}) \, .
\ea
The relations above can be inverted to gain the desired $g^{(2n)}$ in terms of the $G^{(2l)}$, $l\leq n$:
\ba\label{gG}
&& g_{12}=G_{12} \nn\\
&& g_{1234}=G_{1234}-G_{12}G_{34}-G_{13}G_{24}-G_{14}G_{23} = G_{1234}-(G_{12}G_{34} + 2 \textit{ perm.}) \nn\\
&& g_{123456}=G_{123456} - (G_{12}G_{3456} + 14\textit{ perm.}) + 2(G_{12}G_{34}G_{56} +14 \textit{ perm.}) \, .
\ea
These formul{\ae} can easily be made fully general (arbitrary $n$) as explained in Appendix \ref{int-con}.

As well established in field theory, the connected functions $g^{(2n)}$ enjoy a plethora of computational advantages with respect to the $G^{(2m)}$ quite in general. In the present case, for instance, they make possible the large coupling expansion by allowing this limit inside the series $\mathcal{F}$ on the ${\cal F}^{(2n)}$: this exchange is not possible on the original (\ref{Wilson}) because of the asymptotic divergence of the $G^{(2m)}$. In physical words, the connected functions re-sum many infinities to finite contributions. Therefore, as we will extensively see later, it is crucial that the functions $g^{(2n)}$ (differently from the $G^{(2m)}$) are integrable over the $2n-1$ variables they depend on. To this aim, it is sufficient to prove that $g^{(2n)}$ belongs to the class $L^1(\mathbb{R}^{2n-1})$, which is a stronger condition since it involves the modulus $|g^{(2n)}|$ inside the integral. To ensure $g^{(2n)}\in L^1(\mathbb{R}^{2n-1})$ we need to address all the possible asymptotic behaviours in the integration space. The most general situation concerns $l$ subsets composed of $k_i \ (i=1,\ldots,l)$ variables going to infinity by the shifts $\Lambda_i=c_i R$, $i=1,\ldots ,l$, where $R$ is large and the coefficients $c_i\neq c_j$ ( $i\neq j$) are finite. Sufficient condition for a connected function to be integrable at infinity is the behaviour
\be\label{convg2n}
g^{(2n)}(u_1 + \Lambda_1, \ldots ,u_{k_1}+\Lambda_1, u_{k_1 +1} + \Lambda_2, \ldots ,u_{k_1 +k_2}+\Lambda_2 ,\ldots , u_{\sum_i k_i} +\Lambda_l, \ldots ,u_{2n} ) \simeq O(R^{a\leq -l-1}) \,.
\ee
This condition is the generalisation of the one dimensional case $R^{a\leq -2}$, once we take into account the integration volume which grows as $R^{l-1}$.
A rigorous proof of $g^{(2n)}\in L^1(\mathbb{R}^{2n-1})$ is not easy, as the number of regions grows very rapidly with $n$. However, there are many indications and explicit computations that all the functions belong to $L^1(\mathbb{R}^{2n-1})$. In particular, a thorough discussion of the condition (\ref{convg2n}) for the first cases $g^{(4)}$ and $g^{(6)}$ can be found in Appendix \ref{int-con}. In conclusion, we can assume that all the multi-integrals are finite. Eventually, we shall not forget that also numerical calculations here on the $g^{(2n)}$ are much easier that those on the $G^{(2m)}$ \cite{BSV4,BEL}.

\subsection{Small mass behaviour}
\label {smallmass}

In this subsection we provide analytical evidence that, when the strong coupling limit is considered, the scalar partition function (\ref{Wilson}) yields an exponentially large contribution to the Wilson loop, which happens to be of the same order as the one from the classical area \cite{AM}. In fact, the energy and momentum are in general complicated coupling dependent dispersion relations (in terms of the rapidity $u$) \cite{Basso}, but reduce to the relativistic ones in the non perturbative regime $\lambda\rightarrow +\infty$
\be\label{disp}
p(u)=m_{gap}(\lambda) \sinh \frac{\pi}{2}u \, , \quad E(u)=m_{gap}(\lambda) \cosh \frac{\pi}{2}u \, ,
\ee
with the characteristic, for the scalar sector, of the dynamically generated mass \cite{AM, FGR, BK, FGR1, FB}
\be\label{m-lambda}
m_{gap}(\lambda)=\frac{2^{1/4}}{\Gamma (5/4)} \lambda ^{1/8}e^{-\sqrt{\lambda}/4} \left [ 1+O(1/\sqrt{\lambda} ) \right ] \, .
\ee
Importantly, in this regime the limiting value of the function $G^{(2n)}(u_1,\ldots,u_{2n})$ does not contain the coupling constant and depends only on the differences $u_i-u_j$. This fact, combined with the dispersion relations (\ref{disp}), allows us to think of $W$ as a two-point function $\langle \hat{V}(z_1)\hat{V}(z_2)\rangle$ in the (Euclidean) relativistic invariant $O(6)$ NLSM of a specific twist operator $\hat{V}$. In this picture $G^{(2n)}$ is the square modulus, summed over the internal $O(6)$ indices, of the form factor $\langle 0|\hat{V}(0)|\Phi_{a_1}(u_1)\cdots \Phi_{a_{2n}}(u_{2n})\rangle$, where $\Phi_a(u)$ represents a scalar with rapidity $u$ and $a$ as $O(6)$ degree of freedom. The two cross ratios are just the coordinate of the difference $z_1-z_2\equiv z_{12}=(\tau,\sigma)$ and rotational invariance (we have rotated into the euclidean space) imposes that everything must depend only on the distance (modulus) $|z_{12}|\equiv \sqrt{\sigma ^2+\tau^2} $. In fact, we just need to insert the identities
\be
\tau=\sqrt{\sigma ^2+\tau^2} \cos \arctan \frac{\sigma}{\tau} \, , \quad \sigma = \sqrt{\sigma ^2+\tau^2} \sin \arctan \frac{\sigma}{\tau} \, ,
\ee
inside (\ref {Wilson}):
\be
W^{(2n)}=\frac{1}{(2n)!}\int \left [ \prod_{i=1}^{2n}\frac{du_i}{2\pi}e^{\displaystyle - m_{gap}(\lambda) \sqrt{\sigma ^2+\tau^2} \cosh \left ( \frac{\pi}{2}u_i+i\arctan \frac{\sigma}{\tau} \right ) }  \right ] \,G^{(2n)}(u_1,\ldots,u_{2n})\, ,
\ee
and then define the natural variable
\be
z=m_{gap}(\lambda)\sqrt{\tau^2+\sigma^2} \, , \label {z}
\ee
and shift the integration variables $u_i\longrightarrow u_i-\displaystyle\frac{2i}{\pi}\arctan \frac{\sigma}{\tau}$
\be
W^{(2n)}=\frac{1}{(2n)!}\int _{\textrm{Im}u_i=\frac{2\arctan \sigma /\tau}{\pi}}\left [ \prod_{i=1}^{2n}\frac{du_i}{2\pi}e^{\displaystyle - z \cosh \frac{\pi}{2}u_i }  \right ] \,G^{(2n)}(u_1,\ldots,u_{2n}) \, . \label {W2n-1}
\ee
as this does not affect the functions $G^{2n}(u_1,\ldots, u_{2n})$, depending only on the differences $u_i-u_j$. For the same reason we can safely perform a shift back of the contours to the real axis\footnote{As a proof, we can make a change of variables
 in (\ref {W2n-1}), integrating in $u_1$ on $\textrm{Im}u_1=\frac{2\arctan \sigma /\tau}{\pi}$ and in the differences $u_{i\neq 1}-u_1$ on the real line; since $G^{(2n)}$ depends only on $u_{i\neq 1}-u_1$, the shift of the $u_1$ contour to $\textrm{Im}u_1=0$ does not produce any additional terms, since $\exp\{\displaystyle - z \cosh \frac{\pi}{2}u_1 \}$ is analytic.}:
\be
W^{(2n)}= \frac{1}{(2n)!}\int _{\textrm{Im}u_i=0}\left [ \prod_{i=1}^{2n}\frac{du_i}{2\pi}e^{\displaystyle - z \cosh \frac{\pi}{2}u_i }  \right ] \,G^{(2n)}(u_1,\ldots,u_{2n}) \, . \label {W2n-z}
\ee
The final expression (\ref {W2n-z}) depends on the cross ratios (better: on the modulus) and on $\lambda$ only through the 'adimensional' variable (\ref{z}) $z=m_{gap}(\lambda)|z_{12}|$  .

In this (non-scaling) regime the pentagonal amplitude $P_{ss}(u_i|u_j)$ depends at LO only on $u_i - u_j$ and not on the coupling, while the measure becomes a constant (\ref{dynamical2}).
The function $\Pi (u)$ has simple poles when $u=\pm (4m+2)i$, $u=\pm (4m+3)i$, with $m$ a positive or null integer. In addition, the function
$\Pi(u)$ (\ref{dynamical2}) has a double zero for $u=0$ and simple zeroes when $u=\pm (4m+1)i$, $u=\pm (4m+4)i$ with $m$ a positive or null integer.

The asymptotic behaviour of the connected functions assumes a paramount importance when studying the logarithm of the Wilson loop at strong coupling, or small $z$, as conceived by \cite {Smirnov}, from which we generalise to the asymptotically free case. The physical reason for their efficiency resides in the fact that they re-sum an infinite number of particle contributions from the original series (\ref{Wilson}).
In order to highlight how the connected functions depend only on $2n-1$ independent differences of the rapidities, let the reader allow us for a slight abuse of notation: upon introducing the rescaled rapidities $\theta _i=\frac{\pi}{2}u_i$, we denote by $g^{(2n)}(\theta _2-\theta _1, \theta _3-\theta _1, \ldots ,\theta _{2n}-\theta _1)$ the function $(2/\pi) ^{2n}g^{(2n)}(u_1, u_2, \ldots ,u_{2n})$. Hence, the generic term ${\cal F}^{(2n)}$ of the series (\ref {logW}) for the logarithm of the Wilson loop reads
\be\label{Wconn}
{\cal F}^{(2n)} = \frac{1}{(2n)! (2\pi)^{2n}}I^{(2n)} \, , \quad I^{(2n)} =\int d\theta _1 \ldots d\theta _{2n} e^{\displaystyle -z \sum _{i=1}^{2n}\cosh \theta _i} g^{(2n)}(\theta _2-\theta _1, \theta _3-\theta _1, \ldots ,\theta _{2n}-\theta _1) \, .
\ee
The new set of variables
$\alpha _i=\theta _{i+1} -\theta _1$ for $i=1,\ldots , 2n-1$,
allows us to recast the integral $I^{(2n)}$
\ba
I^{(2n)}=\int d\theta _1 \prod _{i=1}^{2n-1}d \alpha _i \exp \left[ -z \cosh \theta _1 -z \sum _{i=2}^{2n}\cosh (\theta _1 + \alpha _{i-1} )  \right]  g^{(2n)}(\alpha _1, \ldots , \alpha _{2n-1}) = \nn\\
=\int d\theta _1 \prod _{i=1}^{2n-1}d \alpha _i \exp \left [ -z \cosh \theta _1 -z \sum _{i=2}^{2n} \left ( \cosh \theta _1 \cosh \alpha _{i-1} + \sinh \theta _1 \sinh \alpha _{i-1} \right )  \right]  g^{(2n)}(\alpha _1,.., \alpha_{2n-1}) \,.\nn
\ea
It turns out convenient to define
$
a=1+\displaystyle\sum _{i=2}^{2n} \cosh  \alpha _{i-1}
$
and
$
b= \displaystyle\sum _{i=2}^{2n} \sinh  \alpha _{i-1}
$,
satisfying the relation
\ba\label{xi}
a^2-b^2=2n+2  \sum _{i=2}^{2n} \cosh  \alpha _{i-1} + 2 \sum _{i=2}^{2n} \sum _{j=i+1}^{2n} \cosh
(\alpha _{i-1} - \alpha _{j-1})= \xi ^2 >0 \ :
\ea
therefore $a$ and $b$ enjoy the parametrisation
\be
a=\xi \cosh \eta \, , \quad b=\xi \sinh \eta \, ,
\ee
in terms of a real parameter $\eta$, depending on the $\alpha _i$ but not on $\theta _1$, which can be thus integrated away:
\ba
I^{(2n)}&=&\int  \prod _{i=1}^{2n-1}d \alpha _i   g^{(2n)}(\alpha _1, \ldots , \alpha _{2n-1}) \int d\theta _1 \exp \Bigl [ -z \xi \Bigl ( \cosh \theta _1\cosh \eta + \sinh \theta _1 \sinh \eta \Bigr ) \Bigr ]  =\nonumber \\
&=& \int \prod _{i=1}^{2n-1}d \alpha _i g^{(2n)}(\alpha _1, \ldots , \alpha _{2n-1})  \int d\theta _1 \exp \Bigl [ -z \xi \cosh (\theta _1 + \eta )  \Bigr ] = \nonumber \\
&=& 2 \int \prod _{i=1}^{2n-1}d \alpha _i g^{(2n)}(\alpha _1, \ldots , \alpha _{2n-1})  K_0 (z \xi) \,.\label{I2n}
\ea
We stress that the result (\ref{I2n}) for $I^{(2n)}$ holds for any $z$ and does not rely on any properties of the functions $g^{(2n)}$, but their dependence on the rapidities only through their differences $\theta_{ij}$. 
Motivated by the findings from Section \ref{conn-func}, we claim that the integral (\ref {I2n}) is finite regardless of the damping factor $K_0(z \xi)$.
On the contrary, the functions $G^{(2n)}$ are not integrable with respect to the $2n-1$ variables $\alpha_i$:
indeed, when an even number $m<2n$ of $\theta _i$ get shifted by the same quantity $\Lambda\gg 1$
($i.e.\ \q_i\rightarrow\q_i+\Lambda\ \forall i\leq m $), the factorisation $G^{(2n)}\longrightarrow G^{(2n-m)}G^{(m)}$ prevents
$G^{(2n)}$ from decreasing to zero, since both the factors are order $O(1)$ as a result of their dependence on differences of rapidities.
As already observed, the connected functions $g^{(2n)}$ (contrarily to the $G^{(2n)}$) are of class $L^1(\mathbb{R}^{2n-1})$ and this fact allow us to expand the Bessel function (inside the integral (\ref{I2n})) at any order of small $z\ll 1$:
\be\label{Bessel}
K_0(z \xi)=-\ln z -\ln\xi +\ln 2 -\gamma +O(z^2\ln z)
\,,
\ee
($\gamma$ being the Euler-Mascheroni constant). Thus, the important leading order emerges as\footnote {We can touch by hand here as this expansion is not allowed in the original multi-integral (\ref{W2n-z}) with the functions $G^{(2n)}$: the function $K_0$ must be kept in the integration and the final result is proportional to $(\ln z)^{n}$.}
\be
I^{(2n)}= -2 \ln z  \int \prod _{i=1}^{2n-1}d \alpha _i g^{(2n)}(\alpha _1, \ldots , \alpha _{2n-1})
+ O\left(\ln\ln (1/z)\right) \, .
\ee
This provides us with two key issues about the logarithm of the Wilson loop (\ref {logW}): first of all an analytic form of its expansion for $z\ll 1$, {\it e.g.} large $\lambda$, valid and staying the same at any value of $n$ from the smallest one, then exact expressions for the coefficients of this expansion. Indeed, the leading coefficient takes up the form
\ba\label{fin-scal}
\ln W &=& -\frac{\ln z}{\pi} \sum _{n=1}^{+\infty} \frac{1}{(2n)!}\int \prod _{i=1}^{2n-1} \frac{d\alpha _i}{2\pi}
g^{(2n)}(\alpha _1, \ldots , \alpha _{2n-1}) + O(\ln \ln (1/z)) = \,  \\
&=& \frac{\sqrt{\lambda }}{4\pi} \sum _{n=1}^{+\infty} \frac{1}{(2n)!}\int \prod _{i=1}^{2n-1} \frac{d\alpha _i}{2\pi}
g^{(2n)}(\alpha _1, \ldots , \alpha _{2n-1}) + O(\ln \lambda) \, ,\nonumber
\ea
where the second equality follows by expanding the definition (\ref {z}) at large coupling $\ln z = - \frac{\sqrt{\lambda}}{4}+O(\ln \lambda)$. A caveat arises, though, when putting forward a systematic expansion: in fact, the term $\ln\xi$ in the asymptotic series (\ref{Bessel})  grows linearly for large rapidities, making the integral diverging at infinity. We can overcome this difficulty by introducing a cutoff $z\xi<1$ and splitting (\ref {I2n}) accordingly
\be
I^{(2n)}=2 \int _{z\xi <1}\prod _{i=1}^{2n-1}d \alpha _i g^{(2n)}(\alpha _1, \ldots , \alpha _{2n-1})  K_0 (z \xi)
+ 2 \int _{z\xi >1}\prod _{i=1}^{2n-1}d \alpha _i g^{(2n)}(\alpha _1, \ldots , \alpha _{2n-1})  K_0 (z \xi)
\label {I2n-bis} \, ,
\ee
so that the second term 
vanishes as $z\rightarrow 0$\footnote{$K_0(x>1)$ is bounded from above in the region $z\xi>1$, thus we have un upper bound for the integral which decreases to zero in the limit $z\to 0$.}.
Now, we may again expand the Bessel function within the integral,
\ba\label {I2n-quater}
I^{(2n)}&\simeq&2 \left [ \ln (1/z) + (\ln 2 -\gamma ) \right ]\int _{z\xi <1}\prod _{i=1}^{2n-1}d \alpha _i g^{(2n)}(\alpha _1, \ldots , \alpha _{2n-1})  -    \\
&-& 2\int _{z\xi <1}\prod _{i=1}^{2n-1}d \alpha _i g^{(2n)}(\alpha _1, \ldots , \alpha _{2n-1}) \ln \xi 
\, . \nonumber
\ea
While the cutoff $z\xi <1$ can be safely removed from the first line of (\ref{I2n-quater})\footnote{The price to pay is an $O(1)$ term, as we will discuss in details for the cases $n=1,2$.} , the second line needs a regularisation, which entails the peculiar form $\ln\ln (1/z)$ of the subleading term. In conclusion, the small $z$ ({\it e.g.} strong coupling) expansion of the logarithm of the hexagonal WL (\ref{logW}) enjoys the form
\be\label{conflog}
\mathcal{F}(z)\simeq J\ln(1/z) + s\ln\ln(1/z) + t
\ee
which reveals its peculiar double logarithmic behaviour, ascribable to the asymptotic freedom of the $O(6)$ NLSM. Correspondingly, the Wilson loop
\be\label{conf-corr}
W(z)\simeq c\frac{\ln^s(1/z)}{z^J} \qquad \mbox{where}\ c\equiv e^t \, ,
\ee
can be rewritten by means of (\ref{m-lambda}) and (\ref{z}), so to highlight its dependence on the coupling $\lambda$ and the cross ratios
\be
W_{O(6)}\simeq C(\tau,\sigma) \lambda^B e^{\sqrt{\lambda}A}, \quad A=\frac{J}{4}, \quad  B=\frac{s}{2}-\frac{J}{8}, \quad C(\tau ,\sigma)=\frac{c}{4^s}\left[\frac{\Gamma(5/4)}{2^{1/4}\sqrt{\tau^2+\sigma^2}}\right]^{J} \ . \label{Wlambda}
\ee
Now, it is evident that the leading term does not depend on the cross ratios and can be comparable to (if not bigger than) the classical minimal area contribution \cite{AM-amp} $W_{AdS}\simeq C_{AdS} e^{-\sqrt{\lambda}\frac{A_6}{2\pi}}$ (arising from the contribution of gluons and fermions in \cite{FPR2}). Moreover, from (\ref {Wlambda}) one can directly read the subleading correction $\lambda^B$, brought by scalars: it is the only contribution of that type, as the one-loop string corrections give a constant contribution $C_{AdS}(\tau,\sigma,\phi)$ of the same kind of $C(\tau,\sigma)$. In the collinear limit $\tau\to +\infty$, though, the one-loop corrections become negligible and the prefactor is fully given by $C(\tau,\sigma)$.

The small $z$ expression (\ref{conf-corr}) proves the proposal of \cite{BSV4} coming from associating the pentagonal amplitudes to $O(6)$ twist fields, whose scaling properties suggest the values $J=1/36$ and $s=-1/24$ \cite{Knizhnik:1987xp}  (also confirmed by numerical computations \cite{BSV4,BEL}, which are much easier here thanks to the employment of the connected $g^{(2n)}$). Differently, we obtained formula (\ref{conf-corr}) directly from the OPE series and thus can fruitfully decompose
\be\label{Jst2n}
J=J^{(2)}+\sum _{n=2}^{\infty} \delta J^{(2n)} \, , \quad s=s^{(2)}+\sum _{n=2}^{\infty} \delta s^{(2n)} \, , \quad
t=t^{(2)}+\sum _{n=2}^{\infty} \delta t^{(2n)} \, ,
\ee
so that the $2n$-particle connected contribution to $\mathcal{F}$ is parametrised, in the limit $z\to 0$, according to \footnote{In the following notation, $\delta J^{(2)}\equiv J^{(2)}$, $\delta s^{(2)}\equiv s^{(2)}$ and $\delta t^{(2)}\equiv t^{(2)}$.}
\be\label{F2npar}
\mathcal{F}^{(2n)}\simeq \delta J^{(2n)}\ln(1/z) + \delta s^{(2n)}\ln\ln(1/z) + \delta t^{(2n)} \, .
\ee
For later convenience we also introduce the partial sums according to (\ref{Jst2n})
\be\label{Jst2ntot}
J^{(2n)}=\sum_{k=1}^{n}\delta J^{(2k)}, \quad s^{(2n)}=\sum_{k=1}^{n}\delta s^{(2k)}, \quad t^{(2n)}=\sum_{k=1}^{n}\delta t^{(2k)} \,,
\ee
where clearly $\displaystyle J=\lim_{n\rightarrow\infty}J^{(2n)}$, $\displaystyle s=\lim_{n\rightarrow\infty}s^{(2n)}$, $\displaystyle t=\lim_{n\rightarrow\infty}t^{(2n)}$.
In fact, we wish to determine analytically these coefficients for $n=1,2$ and provide some considerations for arbitrary $n$.

We emphasise that this is not the original number of particles contributing to $W$ (\ref{W2n-z}), as any connected function re-sums (in $\ln W$) an infinite number of particle contributions from the non-connected ones in $W$ (\ref{W2n-z}). This simple fact entails a great improvement in the accuracy; in fact, on one side, the functional form of the expansion is the right one even for the lowest $n=1$ ({\it cf.} also below), on the other the numerical values of the coefficients are quite precise yet for lower $n$.

\medskip
\noindent\textbf{$\bullet $\ Two scalars:}\\
When considering (\ref{I2n}) for $n=1$, there is only one integration variable $\alpha_1\equiv \alpha$, while $\xi=2\cosh\frac{\alpha}{2}$
\be
\mathcal{F}^{(2)}=\frac{1}{(2\pi)^2}\int d\alpha g^{(2)}(\alpha)K_0\left (2z\cosh\frac{\alpha}{2}\right)=\frac{2}{(2\pi)^2}\int_0^{\infty} d\alpha g^{(2)}(\alpha)K_0\left (2z\cosh\frac{\alpha}{2}\right) \, ,
\ee
where the rescaled function $g^{(2)}(\alpha)=\frac{4}{\pi^2}g^{(2)}(u_1,u_2)$ reads
\be
g^{(2)}(\alpha)=\frac{\Gamma^2(3/4)}{\Gamma^2(1/4)}\frac{\alpha\tanh(\alpha/2)\Gamma\left (\frac{3}{4}-\frac{i\alpha }{2\pi} \right)\Gamma\left (\frac{3}{4}+\frac{i\alpha }{2\pi} \right)}{\Gamma\left (\frac{1}{4}-\frac{i\alpha }{2\pi} \right)\Gamma\left (\frac{1}{4}+\frac{i\alpha }{2\pi} \right)}\frac{12\pi^2}{\left(\alpha^2 + \frac{\pi^2}{4} \right)\left(\alpha^2 + \pi^2\right)}
\ee
and enjoys the asymptotic behaviour $g(\alpha)=C\alpha^{-2} + O(\alpha^{-4})$ with $C=6\pi\frac{\Gamma^2(3/4)}{\Gamma^2(1/4)}$.
We want to determine the coefficients in the expansion (\ref{F2npar})
\be\label{F2}
\mathcal{F}^{(2)}= J^{(2)}\ln(1/z) + s^{(2)}\ln\ln(1/z) + t^{(2)} + O\left(\frac{1}{\ln z}\right) \, .
\ee
According to (\ref{I2n-bis}), we divide the integral in two parts
\be
\mathcal{F}^{(2)}=\int_0^{2\ln(1/z)} \frac{d\alpha}{2\pi^2} g^{(2)}(\alpha)K_0\left (2z\cosh\frac{\alpha}{2}\right)
+ \int_{2\ln(1/z)}^{\infty} \frac{d\alpha}{2\pi^2} g^{(2)}(\alpha)K_0\left (2z\cosh\frac{\alpha}{2}\right)=\mathcal{F}^{(2)}_1 + \mathcal{F}^{(2)}_2 \, .
\ee
When $z\rightarrow 0$, $\mathcal{F}^{(2)}_2$ goes to zero because $K_0$ is bounded within the integration support and the function $g^{(2)}$ behaves as $C\alpha^{-2}+O(\alpha ^{-4})$ for large rapidity, resulting in an $O(1/\ln z)$ contribution.
As far as $\mathcal{F}^{(2)}_1$ is concerned, 
in order to estimate the diverging and the finite contributions for $z\rightarrow 0$, we are allowed to expand $K_0(2z \cosh \frac{\alpha}{2})$ for small argument (\ref{Bessel}). 
Renaming the function $h(\alpha)\equiv \frac{1}{2\pi^2}g^{(2)}(\alpha)$, we get
\ba\label{leading}
\mathcal{F}^{(2)}&=&\ln\frac{1}{z} \int _{0}^{2\ln(1/z)} d\alpha h(\alpha)- \int _{0}^{2\ln(1/z)} d\alpha h(\alpha) \ln \left (\cosh \frac{\alpha}{2} \right ) - \gamma \int _{0}^{2\ln(1/z)} d\alpha h(\alpha) +  O\left(\frac{1}{\ln z}\right) \nonumber \\
&=& J^{(2)}\ln\frac{1}{z} - \int _{0}^{2\ln(1/z)} d\alpha h(\alpha) \ln \left (\cosh \frac{\alpha}{2} \right ) - J^{(2)}\gamma - \ln\frac{1}{z} \int _{2\ln(1/z)}^{\infty} d\alpha h(\alpha) + O\left(\frac{1}{\ln z}\right) \nn\\
\ea
where $\displaystyle J^{(2)}\equiv\int_{0}^{\infty} d\alpha h(\alpha)$ is the leading term of the series (\ref{F2}).
The second term in (\ref{leading}) is of order $\ln\ln(1/z)$ since the integrand behaves like $\sim\frac{1}{\alpha}$, while the remaining ones are finite, given that
\be
-\ln(1/z) \int _{2\ln(1/z)}^{\infty} d\alpha h(\alpha) \simeq -\frac{C}{2\pi^2}\ln(1/z)\int _{2\ln(1/z)}^{\infty}\frac{d\alpha}{\alpha^2}=-\frac{C}{(2\pi)^2} \, .
\ee
In order to disentangle the $O(\ln\ln(1/z))$ contribution from the constant ones in
\be
- \int _{0}^{2\ln(1/z)} d\alpha h(\alpha) \ln \left (\cosh \frac{\alpha}{2} \right ) \, ,
\ee
we split the integration domain into two intervals
\be
- \int _{0}^{1} d\alpha h(\alpha) \ln \left (\cosh \frac{\alpha}{2} \right )- \int _{1}^{2\ln(1/z)} d\alpha h(\alpha) \ln \left (\cosh \frac{\alpha}{2} \right ) \, ,
\ee
the latter housing the divergence $\ln\ln(1/z)$: to extract it, we add and subtract a counterterm
\be
- \int_{1}^{2\ln(1/z)} d\alpha\left[ h(\alpha) \ln \left (\cosh \frac{\alpha}{2} \right )-\frac{C}{(2\pi)^2\alpha}\right]-\frac{C}{(2\pi)^2}\int_{1}^{2\ln(1/z)}\frac{d\alpha}{\alpha} \, .
\ee
The piece stays finite in the limit $2\ln(1/z)\rightarrow\infty$ while the second yields the subleading $\ln\ln(1/z)$, up to an additive constant
\be
-\frac{C}{(2\pi)^2}\int_{1}^{2\ln(1/z)}\frac{d\alpha}{\alpha}=-\frac{C}{(2\pi)^2}\ln \ln (1/z)-\frac{C}{(2\pi)^2}\ln 2 \, .
\ee
Keeping track of all the divergent and finite pieces, we obtain (\ref{F2}) with:
\be
J^{(2)}=\int  _{0}^{+\infty} d\alpha h(\alpha )= \frac{1}{2\pi^2}\int _{0}^{+\infty} d\alpha g^{(2)}(\alpha ) \simeq 0.03109
\ee
\be
s^{(2)}=-\frac{C}{(2\pi)^2}=- \frac{3}{2\pi } \frac{\Gamma ^2(3/4)}{\Gamma ^2(1/4)}\simeq -0.05454
\ee
\ba
t^{(2)} &=& - J^{(2)}\gamma - \frac{C}{(2\pi)^2}\left(1+\ln 2\right) - \frac{1}{2\pi^2}\int_{0}^{1}d\alpha g^{(2)}(\alpha) \ln \left (\cosh \frac{\alpha}{2} \right) + \nn\\
&+& \frac{1}{2\pi^2}\int_{1}^{\infty}d\alpha \left[\frac{C}{2\alpha}-g^{(2)}(\alpha) \ln \left ( \cosh \frac{\alpha}{2} \right ) \right ] \, .
\ea
Our numerical estimate for $t^{(2)}$ amounts to $t^{(2)}\simeq -0.00819$, in agreement with the Montecarlo evaluation by \cite{BSV4,BEL}.

\medskip
\noindent\textbf{$\bullet $\ Four scalars:}\\

As far as the leading order $J\ln(1/z)$ is concerned, it is not difficult to evaluate the correction $\delta J^{(4)}$ coming from the explicit expression of the four scalar connected function $g^{(4)}$. Specializing (\ref{fin-scal}) for $n=2$ we get
\be
\delta J^{(4)}=\frac{1}{12(2\pi)^4}\int d\alpha_1 d\alpha_2 d\alpha_3 g^{(4)}(\alpha_1,\alpha_2,\alpha_3) \, .
\ee
We simply integrate it with \texttt{Mathematica}\textsuperscript{\textregistered} and obtain a correction to $J$ by an amount $\delta J^{(4)}= (-3.44\pm 0.01)\cdot 10^{-3}$, {\it i.e.} $J^{(2)}+\delta J^{(4)}\equiv J^{(4)}\simeq 0.02765$: this value differs from the 2D-CFT prediction $J=\frac{1}{36}=0.02\bar{7}$ \cite{BSV4,BEL} by just $0.5\%$.

The correction $\delta s^{(4)}$ to the subleading coefficient $s$ is more involved, as it  depends on the asymptotic behaviour of the connected function $g^{(4)}$ and there are many different regions to take into account. We remind from formula (\ref {I2n-quater}) that the divergence $\ln\ln(1/z)$ comes from the combined action of the cutoff $z\xi<1$ and the piece $g^{(4)}\ln\xi$ due to the expansion of $K_0(z\xi)$. More precisely, it is contained in the integral
\be\label{deltas4int}
\delta s^{(4)}\ln\ln(1/z)=-\frac{1}{12(2\pi)^4}\int_{z\xi<1} d\alpha_1 d\alpha_2 d\alpha_3 g^{(4)}(\alpha_1,\alpha_2,\alpha_3) \ln\xi + O(1) \, .
\ee
When one or more variables are large we have, from (\ref{xi}), $\ln\xi\simeq\frac{|\alpha_i|}{2}$, where $\alpha_i$ is the largest of them, and the cutoff condition translates into $|\alpha_i|<2\ln(1/z)$. The linearity in $\alpha_i$ tells us that the only region where the integral becomes divergent corresponds to the split $4\to 3+1$,  where $g^{(4)}$ goes to zero with the minimum power required by convergence, see Appendix \ref{int-con}. This region has multiplicity four\footnote{Three are explicit in the representation with the $\alpha_i$ variables, while in the other one we send all of them to infinity which means that the rapidity $\theta_1$ is split far away.} and they are all physically equivalent, so we choose to send $\alpha_1\to\infty$ and keep the other variables finite.
We define the asymptotic function $g^{(4)}_{as}$
\be
\lim_{\alpha_1\to\pm\infty}\alpha_1^2 g^{(4)}(\alpha_1,\alpha_2,\alpha_3)= g^{(4)}_{as}(\alpha_2,\alpha_3) \, .
\ee
Then, the contribution from the regions $4\to 3+1$ follows from
\be
-\frac{2}{3(2\pi)^4}\int^{2\ln(1/z)}d\alpha_1\frac{1}{\alpha_1^2}\frac{\alpha_1}{2}\int d\alpha_2 d\alpha_3 g^{(4)}_{as}(\alpha_2,\alpha_3) \, ,
\ee
where we considered only the upper integration limit, which contains the divergent part. In addition, a factor $4\cdot 2$ due to the number of regions (a particle can be sent either to $+\infty$ or $-\infty$) shows up.
The coefficient in (\ref{deltas4int}) is then
\be\label{deltas4}
\delta s^{(4)}=-\frac{1}{3(2\pi)^4}\int d\alpha_2 d\alpha_3 g^{(4)}_{as}(\alpha_2,\alpha_3) \, ,
\ee
with
\ba
 &&g^{(4)}_{as}(\alpha_2,\alpha_3)=-6\mu^2\left[g^{(2)}(\alpha_2-\alpha_3) + g^{(2)}(\alpha_2)+g^{(2)}(\alpha_3)\right] +\mu^4 \left(\frac{2}{\pi}\right)^2 36\Pi(u_3)\Pi(u_4)\Pi(u_{34}) \cdot \nn\\
&\cdot & \frac{(u_3^2+4)(u_4^2+4) + (u_3^2+4)(u_{34}^2+4)+(u_4^2+4)(u_{34}^2+4) + \frac{3}{2}(u_3^2+u_4^2+u_{34}^2 +24)}{(u_3^2+1)(u_3^2+4)(u_4^2+1)(u_4^2+4)(u_{34}^2+1)(u_{34}^2+4)} \, ,
\ea
where in the last piece we used the variables $u_{3,4}=\frac{2}{\pi}\alpha_{2,3}$ for brevity.
The numerical integration yields $\delta s^{(4)}\simeq 0.017650$ and the four particles prediction sums up to $s^{(4)}=s^{(2)} + \delta s^{(4)}\simeq -0.036894$: the discrepancy with respect to the expected value \cite{BSV4} $s=-1/24=-0.041\bar{6}$ is about $11\%$.

\vspace{0.3cm}

The finite part $\delta t^{(4)}$ of the four particles integral has three different contributions $\delta t^{(4)}=\delta t^{(4)}_1+\delta t^{(4)}_2+\delta t^{(4)}_3$. They can be obtained from
\be
\mathcal{F}^{(4)}=\frac{1}{12(2\pi)^4}\int_{z\xi<1} d\alpha_1 d\alpha_2 d\alpha_3 g^{(4)}(\alpha_1,\alpha_2,\alpha_3)K_0(z\xi)
+ O\left(\frac{1}{\ln z}\right) \, ,
\ee
once we subtract both the divergent terms, $\delta J^{(4)}\ln(1/z)$ and $\delta s^{(4)}\ln\ln(1/z)$, previously analysed.
Referring to formula (\ref {I2n-quater}), we immediately see that a finite contribution comes from the constant term in the expansion of the Bessel function $\ln 2-\gamma$, then
\be
\delta t^{(4)}_1=\frac{(\ln 2-\gamma)}{12(2\pi)^4}\int d\alpha_1 d\alpha_2 d\alpha_3 g^{(4)}(\alpha_1,\alpha_2,\alpha_3)=(\ln 2-\gamma)\delta J^{(4)} \, .
\ee
Another one appears when we remove the cutoff $z\xi < 1$ (see the first line of (\ref {I2n-quater})) in the computation of $\delta J^{(4)}$
\be
\delta t^{(4)}_2=\lim_{z\to 0}\left[-\frac{\ln(1/z)}{12(2\pi)^4}\int_{z\xi>1}d\alpha_1 d\alpha_2 d\alpha_3 g^{(4)}(\alpha_1,\alpha_2,\alpha_3)\right] \, .
\ee
Repeating the same argument as for the subleading $\delta s^{(4)}$, only the regions $4\to 3+1$ matter and their contribution is exactly the same
\be
\delta t^{(4)}_2=-\frac{2\ln(1/z)}{3(2\pi)^4}\int_{2\ln(1/z)}^{\infty}\frac{d\alpha_1}{\alpha_1^2}\int d\alpha_2 d\alpha_3 g_{as}^{(4)}(\alpha_2,\alpha_3) =\delta s^{(4)} \, .
\ee
The last one, $\delta t^{(4)}_3$, must be extracted from the integral (\ref{deltas4int})
\be
-\frac{1}{12(2\pi)^4}\int_{z\xi<1} d\alpha_1 d\alpha_2 d\alpha_3 g^{(4)}(\alpha_1,\alpha_2,\alpha_3) \ln\xi \simeq \delta s^{(4)}\ln\ln(1/z) + \delta t^{(4)}_3 ,
\ee
which, besides the $\ln\ln(1/z)$ contribution obtained in (\ref{deltas4}), contains also a finite piece.
As in the two particle case, we regulate the infinity by subtracting the asymptotic behaviours and get the finite integral (it is allowed to remove the cutoff $z\xi<1$)
\ba
&& -\frac{1}{12(2\pi)^4}\int d\alpha_1d\alpha_2d\alpha_3 \Bigl[g^{(4)}(\alpha_1,\alpha_2,\alpha_3)\ln\xi-\frac{g^{(4)}_{as}(\alpha_2,\alpha_3)}{2(|\alpha_1|+ a)}-\frac{g^{(4)}_{as}(\alpha_1,\alpha_3)}{2(|\alpha_2|+a)}-\nn\\
&&-\frac{g^{(4)}_{as}(\alpha_1,\alpha_2)}{2(|\alpha_3|+a)} -
 \frac{g^{(4)}_{as}(\alpha_2 -\alpha_1,\alpha_3 -\alpha_1)}{2(|\alpha_1| + a)}\Bigr] \, ,
\ea
where we introduced $a>0$ to prevent the singularities on the axes $\alpha_i=0$. This parameter does not affect the large $\alpha_i$ limit and we can take any finite value we want. On the contrary of the two particle case we do not split the integration in parts, for it turns out to be rather involved; the insertion of a parameter $a$ which avoid the singularity in $\alpha_i=0$ is more effective.\footnote{Alternatively, we could have used this procedure also for the $n=1$ case.}
The divergence $\delta s^{(4)}\ln\ln(1/z)$ is isolated in
\ba
&& -\frac{1}{12(2\pi)^4}\int_{z\xi<1}d\alpha_1d\alpha_2d\alpha_3 \Bigl[\frac{1}{2(|\alpha_1|+ a)}g^{(4)}_{as}(\alpha_2,\alpha_3)+\frac{1}{2(|\alpha_2|+a)}g^{(4)}_{as}(\alpha_1,\alpha_3)+\nn\\
&&+ \frac{1}{2(|\alpha_3|+a)}g^{(4)}_{as}(\alpha_1,\alpha_2) +
 \frac{1}{2(|\alpha_1| + a)}g^{(4)}_{as}(\alpha_2 -\alpha_1,\alpha_3 -\alpha_1)\Bigr] \, ,
\ea
which also contains $\delta t_3^{(4)}$.
The four terms contribute the same thanks to the invariance of the cutoff $z\xi <1$ under permutation of variables and $(\alpha_1,\alpha_2,\alpha_3)\to (-\alpha_1,\alpha_2-\alpha_1,\alpha_3-\alpha_1)$, therefore we are left with
\be
-\frac{1}{6(2\pi)^4}\int_{z\xi <1}d\alpha_1d\alpha_2d\alpha_3 \frac{1}{|\alpha_1|+ a}\,g^{(4)}_{as}(\alpha_2,\alpha_3) \, .
\ee
Disregarding the vanishing terms, the integral simplifies to
\be
-\frac{1}{3(2\pi)^4}\int_{0}^{2\ln(1/z)}d\alpha_1 \frac{1}{\alpha_1+ a} \int_{\mathbb{R}^2}d\alpha_2d\alpha_3\, g^{(4)}_{as}(\alpha_2,\alpha_3) \, ,
\ee
as the divergence appears only where $|\alpha_1|$ is large and we can safely remove the cutoff in the other directions.
The integral over $\alpha_1$ yields
\be
-\frac{1}{3(2\pi)^4}\left[\ln\ln(1/z) + \ln\frac{2}{a}\right]\int_{\mathbb{R}^2}d\alpha_2d\alpha_3 \,g^{(4)}_{as}(\alpha_2,\alpha_3) 
\, ,
\ee
which reproduces (\ref{deltas4}) plus a finite correction proportional to $\delta s^{(4)}$ and eventually we get
\ba
&&\delta t^{(4)}_3 = -\frac{1}{12(2\pi)^4}\int d\alpha_1d\alpha_2d\alpha_3 \Bigl[g^{(4)}(\alpha_1,\alpha_2,\alpha_3)\ln\xi-\frac{1}{2(|\alpha_1|+ a)}g^{(4)}_{as}(\alpha_2,\alpha_3)-\nn\\
&&-\frac{1}{2(|\alpha_2|+a)}g^{(4)}_{as}(\alpha_1,\alpha_3)-
 \frac{1}{2(|\alpha_3|+a)}g^{(4)}_{as}(\alpha_1,\alpha_2) -
 \frac{1}{2(|\alpha_1| + a)}g^{(4)}_{as}(\alpha_2 -\alpha_1,\alpha_3 -\alpha_1)\Bigr] + \nn\\
 && + \delta s^{(4)}\ln\frac{2}{a} ,
\ea
where the dependence on $a$ drops out thanks to $\int_{0}^{\infty}d\alpha\left(\frac{1}{\alpha + a}-\frac{1}{\alpha + a'}\right)=\ln\frac{a'}{a}$. To simplify the result we choose $a=2$ and sum up everything to get the final answer
\ba
&&\delta t^{(4)} = -\frac{1}{12(2\pi)^4}\int d\alpha_1d\alpha_2d\alpha_3 \Bigl[g^{(4)}(\alpha_1,\alpha_2,\alpha_3)\ln\xi-\frac{1}{2(|\alpha_1|+ 2)}g^{(4)}_{as}(\alpha_2,\alpha_3)-\nn\\
&&-\frac{1}{2(|\alpha_2|+2)}g^{(4)}_{as}(\alpha_1,\alpha_3)-
 \frac{1}{2(|\alpha_3|+2)}g^{(4)}_{as}(\alpha_1,\alpha_2) -
 \frac{1}{2(|\alpha_1| + 2)}g^{(4)}_{as}(\alpha_2 -\alpha_1,\alpha_3 -\alpha_1)\Bigr] + \nn\\
 && + (\ln 2 -\gamma)\delta J^{(4)} + \delta s^{(4)} \, .
\ea
A numerical estimate returns the value $\delta t^{(4)}\simeq -0.006133$,
which added to the two-particle contribution yields $t^{(4)}=t^{(2)} + \delta t^{(4)}\simeq -0.01432$. Differently from $\delta J^{(4)}$ and $\delta s^{(4)}$, $\delta t^{(4)}$ is almost as large as the previous approximation $t^{(2)}\simeq -0.00819$. This suggests that we might need a larger $n$ to obtain a better evaluation of this coefficient. However, an accurate estimate of $t$ is still missing, as the (Montecarlo) numerical evaluations by \cite{BSV4,BEL} furnish $t\simeq -0.01$ with one significant digit (compatible, though, with our $t^{(4)}$).

\medskip
\noindent\textbf{$\bullet $\ $2n$ scalars:}\\
Referring to the notation
\be
\ln W = \mathcal{F}\simeq J \ln (1/z) + s \ln \ln (1/z) + t
\ee
and recalling the expansion (\ref{Jst2n}), we get the following expressions of the $2n$ particle contributions to $J$, $s$ and $t$: the leading divergence $J$ gets corrected by
\be
\delta J^{(2n)}=-\frac{2}{(2n)! (2\pi)^{2n}} \int \prod _{i=1}^{2n-1} d\alpha _i g^{(2n)}(\alpha _1,\ldots , \alpha _{2n-1}) \, ,
\ee
while the subleading $\delta s^{(2n)}$ is contained in the integral
\be\label{deltas2n}
-\frac{2}{(2n)! (2\pi)^{2n}} \int _{z\xi <1}\prod _{i=1}^{2n-1} d\alpha _i g^{(2n)}(\alpha _1,\ldots , \alpha _{2n-1})
\ln \xi \simeq \delta s^{(2n)} \ln \ln (1/z) + \delta t^{(2n)}_3 \, ,
\ee
which also yields the finite piece $\delta t^{(2n)}_3$. As for $t$, we have three contributions $\delta t^{(2n)}=\delta t^{(2n)}_1+\delta t^{(2n)}_2+\delta t^{(2n)}_3$, where the first is simply due to the constant term in the expansion of $K_0$
\be
\delta t^{(2n)}_1=(\ln 2 - \gamma)\delta J^{(2n)} \, ,
\ee
while the second comes from the removal of the cutoff in the computation of $\delta J^{(2n)}$ and reads
\be
\delta t^{(2n)}_2=\lim_{z\to 0}\left[-\frac{2\ln(1/z)}{(2n)!(2\pi)^{2n}}\int_{z\xi >1}\prod _{i=1}^{2n-1} d\alpha _i g^{(2n)}(\alpha _1,\ldots , \alpha _{2n-1})\right] \, .
\ee
As in the $n=1,2$ cases, it can be shown to equal\footnote{The contributing regions are the same, in which the decay is just enough for the function $g^{(2n)}$ to be $L^{1}(\mathbb{R}^{2n-1})$.} $\delta s^{(2n)}$. Collecting all the contributions, we get
\be
\delta t^{(2n)}=(\ln 2 - \gamma)\delta J^{(2n)} + \delta s^{(2n)} + \delta t^{(2n)}_3 \, .
\ee
In summary, we provided many explicit formul{\ae} for the coefficients $J$, $s$ and $t$, which parametrise the small $z$ limit of $W$ (\ref{conf-corr}). Thanks to the expansion of $\ln W$, we have been able to represent them as a series (\ref{Jst2n}), which is a very effective procedure, as their contributions $\delta J^{(2n)}$, $\delta s^{(2n)}$ and $\delta t^{(2n)}$ can be easily extracted from the integral (\ref{I2n-quater}), in most cases analytically.  Already for $n=2$, the expected values \cite{BSV4} for $J$ and $s$ are reproduced with a good accuracy. A deeper numerical analysis of (\ref{I2n-quater}) could confirm their values with even more precision, and would allow to compute the constant contribution $t$ very precisely. On the other hand, by means of (\ref{Wlambda}), the coefficients $J$, $s$ and $t$ can be used to parametrise the scalar contribution $W_{O(6)}$ in terms of the coupling constant $\lambda$.

\section{Conclusions and perspectives}
\label {conc}

For our purposes, the strong coupling behaviour of the quantum GKP dynamics shows at least two different regimes depending on the value of the rapidities. In the first, the strong coupling non-perturbative regime (fixed rapidity), the scalars of the hexagonal WL (OPE) series yield a dominant contribution, $W_{O(6)}$, and decouple from the other particles (gluons and fermions) whose effect is negligible. Yet, in the perturbative regime (rapidity scaling like $\sim \sqrt{\lambda}$), only gluons and fermions yield a contribution, $W_{AdS}$, and it is nothing but that of the classical $AdS$ string \cite{TBuA, FPR2}. The two contributions are comparable and compete one with the other depending on the values of the cross ratios. Naturally, this scenario admits a generalisation for the other polygons, and the analysis has been restricted here only to the simplest polygon for sake of simplicity.

In fact, the matrix factors appearing in the hexagonal WL are the simplest ones and they have been recast into a shape recalling the Nekrasov instanton partition function of ${\cal N}=2$ SYM. Thus, they have allowed us an efficient and elegant treatment through Young tableaux, which culminated in the calculation (\ref{MatYoung}) in terms of rational functions. Since the starting formula inherits its structure from the $SO(6)$ symmetry of the scalars, we would like to think that the entire procedure may be generalised to the more complicated matrix parts appearing in the other polygons. To support this idea and elucidate the method, we have explicitly performed the computations for two and four scalars and eventually obtained final elegant expressions. In fact, this rather holds in general as we have provided explicit closed expressions for the polynomials $\delta_{2n}$ entering the integral expression for the matrix factor, (\ref{Nekr-scal}), and we have introduced and partially disentangled the polynomials $P_{2n}$ (\ref {P2n}), which completely define the explicit expressions of the matrix factors. Eventually, this Young tableaux approach naturally applies as well to the $SU(4)$ matrix part of the fermion/anti-fermion contributions, which give rise at strong coupling to the so-called meson excitation \cite{FPR2, BFPR1}. This is actually the topic of an upcoming paper \cite{BFPR3}.

On the contrary, it is suitable that we recall why the dynamical parts of the pentagonal transitions are more explicit: they are always products of two-body components. Besides, they become relativistic when the 't Hooft coupling grows.

The product of the dynamical and matrix parts makes the (modulus) square of the full pentagonal transition, $G^{(2n)}$ (for $2n$ particles). This enters the OPE series (multiplied by the exponential of the free propagation) and, for large rapidities, enjoys a factorisation property in terms of products of squared transitions $G^{(2m)}$ with less particles, $m<n$, up to corrections decreasing as inverse powers (of the rapidities). This feature constitutes a crucial issue of our paper in itself and for future studies, but also because it has led us to interesting achievements. For instance, the structure of the matrix factor, {\it i.e.} formula (\ref {P2n}): the poles of the matrix factor have been completely identified and all the unknowns are the polynomials $P_{2n}$, whose structure is studied and discussed in Appendix \ref {sca-pol}. Another consequence is the power-like decay of the connected counterparts $g^{(2n)}$, which in their turn characterise the series expansion of the logarithm, $\ln W$. This passage has allowed us the strong coupling expansion despite the ambiguous behaviour caused by an exponentially small mass gap $m_{gap}\sim e^{-\frac{\sqrt{\lambda}}{4}}$. In perspective, this manoeuvre should be very efficient and fruitful for future studies. For instance for performing a numerical summation of the OPE series, also for other polygons. This seems quite evident if we look at the expressions for the $2n$ scalar leading terms at the end of Subsection \ref{smallmass}. We have proved explicitly the two and four particles contributions to $J$, $s$ and $t$, which give an estimate for $A$, $B$ and $C(\tau,\sigma)$, parameters of the strong coupling behaviour of the scalar contribution (\ref{Wlambda}). This line of research should allow us a precise numerical determination of the coefficients $A,B,C$ of the aforementioned formul\ae \, from the OPE series. On the contrary, all this does not apply to the series for $W$, because of the single term behaviour $W^{(2n)} \sim (1/z)^n$ at LO. Physically, the fundamental difference is due to the asymptotic behaviour of the connected functions $g^{(2n)}$ with respect to the $G^{(2n)}$.

Enlarging our point of view, if, as written above, it is natural to apply these ideas to arbitrary null polygonal Wilson loops with $n$ sides, it is also very important to understand how our approach compares with FFs for twist fields \cite{CCAD} and how much of it survives for other operators in $SO(6)$ ($SU(4)$) symmetric (relativistic) theories.

\medskip
{\bf Acknowledgements}
We thank Z. Bajnok, J. Balog, A. Belitsky, J.-E. Bourgine, O. A. Castro-Alvaredo, B. Doyon, M. de Leeuw, K. Ito, Y. Satoh, H. Shu,  and especially Ivan Kostov and Didina Serban for discussions. This project was partially supported by the grants: GAST (INFN), UniTo-SanPaolo Nr TO-Call3-2012-0088, the ESF Network HoloGrav (09-RNP-092 (PESC)), Grant-in-Aid for JSPS Fellows 16F16735, the MPNS--COST Action MP1210 and the EC Network Gatis. DF thanks the Galileo Galilei Institute (GGI) for Theoretical Physics for the hospitality, INFN and Organisers for financial support during the completion of this work, within the program \textit{New Developments in AdS3/CFT2 Holography}. SP is an Overseas researcher under Postdoctoral Fellowship of Japan Society for the Promotion of Science (JSPS). AB acknowledges the CEA and the Institut de Physique Th\'eorique (IPhT) for partial financial support and hospitality.

\appendix

\section{Properties of the $\delta_{2n}$ polynomials} \label {delta}
\setcounter{equation}{0}

In this Appendix we list some properties of the polynomials $\delta_{2n}(b_1,\ldots,b_{2n})$ (\ref{def-delta}) which appear in the integrand (\ref{Nekr-scal}), giving the matrix factor $\Pi_{mat}^{(2n)}(u_1,\ldots,u_{2n})$.\\
For convenience, we recall their expression as sums over partitions
\ba
\delta _{2n}(b_1,\ldots,b_{2n})&\equiv &\frac{n!}{2n} \sum _{\alpha _1<\alpha _2< \ldots  < \alpha _{n}=1}^{2n} \left (  \prod _{\stackrel {i\in S_{\vec{\alpha}},
j\in S_{\vec{\alpha}}, i<j} {i\in \bar S_{\vec{\alpha}},
j\in \bar S_{\vec{\alpha}}, i<j}} [ (b_i-b_j)^2+1] \right ) \cdot \nonumber \\
&\cdot & \prod _{k=1}^n \prod _{\beta \in  \bar S_{\vec{\alpha}}} \frac{b_{\alpha _k}-b_{\beta}-i}{b_{\alpha _k}-b_{\beta}} \, .
\label{def-deltaA}
\ea
In the first place, the function $\delta_{2n}(b_1,\ldots,b_{2n})$ is invariant under the exchange of its arguments, {\it i.e.}
\be
\delta_{2n}(b_1,\ldots,b_i,b_{i+1},\ldots,b_{2n})=\delta_{2n}(b_1,\ldots,b_{i+1},b_i,\ldots,b_{2n})
\ee
and vanishes whenever three or more variables lie aligned ({\it i.e.} spaced by $i$) in the complex plane
\be
\delta_{2n}(b_1,b_1+i,b_1+2i,b_4,\ldots,b_{2n})=0 \ .
\ee
From (\ref {def-deltaA}) a more compact representation to $\delta_{2n}$ can be obtained, by borrowing some results from the Quantum Hall effect: indeed, one can recognise the Moore-Read wave function \cite{Mooread,Pasq} in the highest degree $2n(n-1)$ of the $\delta$-polynomials
\be
\delta^{(0)} _{2n}(b_1,\ldots,b_{2n})\equiv\frac{n!}{2n} \sum _{\alpha _1<\alpha _2< \ldots < \alpha _{n}=1}^{2n}   \prod _{\stackrel {i\in S_{\vec{\alpha}},
j\in S_{\vec{\alpha}}, i<j} {i\in \bar S_{\vec{\alpha}},
j\in \bar S_{\vec{\alpha}}, i<j}}  (b_i-b_j)^2 \ ,
\ee
so that a more elegant expression in terms of a Pfaffian\footnote{Of course, the diagonal terms of the matrix $1/b_{ij}$, often called Trummer's matrix, are defined to be zero.} follows
\be\label{delta0det}
\delta^{(0)} _{2n}(b_1,\ldots,b_{2n})=\frac{n!}{2n}2^n\displaystyle\prod_{i<j}b_{ij}\textrm{Pf}\left(\frac{1}{b_{ij}}\right) \, .
\ee
We can extend (\ref{delta0det}) to the full $\delta_{2n}$ by means of the substitution\footnote{This is not granted a priori, since the equivalence depends on the particular form of the matrix elements $1/b_{ij}$; however, in our case the procedure can be employed harmlessly.} $b_{ij}\to \frac{b_{ij}^2+1}{b_{ij}}$, to find the compact formula \cite{DID}
\be\label{deltadet}
\delta_{2n}(b_1,\ldots ,b_{2n})=
\frac{n!}{2n}2^n\displaystyle\prod_{i<j}\frac{b_{ij}^2+1}{b_{ij}}\textrm{Pf} \, D \, , \quad D_{ij}=\left(\frac{b_{ij}}{b_{ij}^2+1}\right) \, ,
\ee
in terms of the Pfaffian of the $2n \times 2n$ matrix $D$. The Pfaffian representation (\ref{deltadet}) also allows for a recursive description of the $\delta$-polynomials:
\ba
\delta_2(b_1,b_2) &=& 1 \nn\\
\delta_{2n}(b_1,\ldots,b_{2n}) &=& 2(n-1)\sum_{\stackrel{l=1}{l\neq k}}^{2n}\prod_{\stackrel{i=1}{i\neq k,l}}^{2n}
\frac{b_{ik}^2+1}{b_{ik}}\frac{b_{il}^2+1}{b_{il}}\,\delta_{2n-2}(b_1,\ldots,\underline{b_k},\ldots,\underline{b_l}\ldots,b_{2n})
\label {delta-rec} \, ,
\ea
(for any arbitrarily chosen $k\in\{1,\ldots,2n\}$), where the notation $\underline{b_k}$ means that $b_k$ does not appear as a variable of the function $\delta_{2n-2}$\,.\\
Referring to (\ref{diagr}), the functions $\delta_{2n}(b_1,\ldots,b_{2n})$ take simple forms under some specific configurations for their arguments, resulting from the residue computations described in Section \ref{young}. We make use of the shorthand notation introduced in Section \ref{young}, as for instance $\delta_{2n}(Y)$ to indicate that the variables of $\delta_{2n}$ are computed on the residue configuration $Y=(l_1,\ldots,l_{2n})$\,.\\
When two variables are displaced by $i$ we have the recursion relation \cite{DID}
\ba
\delta_{2n}(2,0,1,\ldots, 1)&\equiv &\delta_{2n}(u_1, u_1+i, u_3,\ldots , u_{2n})= \nn\\
&=& 2(n-1)\displaystyle\prod_{j=3}^{2n}(u_1-u_j-i)(u_1-u_j+2i)\delta_{2n-2}(u_3,\ldots , u_{2n}) \, ,
\ea
which can be iterated to get the most general one, with $0\leq k \leq n-2$:
\ba
&&\delta_{2n}(2,0,..,2,0_{2k+2},1,\ldots ,1)\equiv\delta_{2n}(u_1, u_1+i,u_3, u_3+i,.., u_{2k+1}, u_{2k+1}+i, u_{2k+3},\ldots , u_{2n})= \nn\\
&& =2^{k+1}\frac{(n-1)!}{(n-2-k)!} \displaystyle\prod_{i<j=0}^{k}[(u_{2i+1}-u_{2j+1})^2+1][(u_{2i+1}-u_{2j+1})^2+4]\cdot \nn\\ &&\cdot\displaystyle\prod_{j=2k+3}^{2n}\displaystyle\prod_{l=0}^{k}(u_{1+2l}-u_j-i)(u_{1+2l}-u_j+2i)\delta_{2n-2-2k}(u_{2k+3}, \ldots , u_{2n}) \, .
\ea
The relation above allows us to express any $\delta_{2n}(Y)$ in terms of the fundamental one $\delta_{2k}(1,\ldots ,1)$, with a fewer number of particles; if we consider the particular case $k=n-2$ and choose $u_{2n}=u_{2n-1}+i$ we get
\ba\label{delta20}
\delta_{2n}(2,0,\ldots ,2,0)&\equiv &\delta_{2n}(u_1, u_1+i, \ldots , u_{2n-1}, u_{2n-1}+i)= \nn\\
&=& 2^{n-1}(n-1)!\displaystyle\prod_{i<j=0}^{n-1}[(u_{2i+1}-u_{2j+1})^2+1][(u_{2i+1}-u_{2j+1})^2+4] \, .
\ea
Combining the last two equations we are able to express $\delta_{2n} (Y=(Y_1,Y_2))$, where $Y_1=(2,0,\cdots ,2,0)$ and $Y_2=(1,\cdots ,1)$, in terms of the product $\delta_{2k+2}(Y_1)\delta_{2n-2k-2}(Y_2)$ times a mixing part, through
\ba\label{delta-fact}
&&\delta_{2n}(2,0,\ldots ,2,0_{2k+2},1,\ldots ,1)=2\displaystyle\prod_{j=2k+3}^{2n}\displaystyle\prod_{l=0}^{k}(u_{1+2l}-u_j-i)(u_{1+2l}-u_j+2i)\cdot \nn\\
&& \cdot \frac{(n-1)!}{(n-2-k)!k!}\delta_{2k+2}(2,0,\ldots ,2,0)\delta_{2n-2-2k}(1,1,\ldots ,1,1) \, ,
\ea
which holds for $0\leq k \leq n-2$. In (\ref{delta20}) we can move all the columns to the left to obtain
\ba
\delta_{2n}(2,2,\ldots ,0,0)&\equiv &\delta_{2n}(u_1, u_1+i, u_2, u_2 +i, \ldots , u_{n}, u_{n}+i)= \nn\\
&=& 2^{n-1}(n-1)!\displaystyle\prod_{i<j}^{n}[(u_{i}-u_{j})^2+1][(u_{i}-u_{j})^2+4] \, ,
\ea
which is the configuration considered in the main text.

When a set of $2k$ particle rapidities is boosted to infinity, the polynomials $\delta_{2n}$ enjoy a factorisation similar to the one occurring to the functions $G^{(2n)}$: from the Pfaffian representation (\ref{deltadet}) we get
\be
\delta_{2n}(u_1+\Lambda,\cdots ,u_{2k}+\Lambda, u_{2k+1},\cdots ,u_{2n})= \Lambda^{4k(n-k)}\frac{(n-1)!}{(k-1)!(n-k-1)!}\delta_{2k}\delta_{2n-2k}\left[1+O(\Lambda^{-1})\right] \, .
\ee

We end this Appendix by giving some other explicit expressions in particular cases for the lower polynomial $n=2$:
\ba\label{pdelta4}
&& \delta_4(a,a,b,b) = 2(7+(4+(a-b)^2)(a-b)^2) \nn \, , \\
&& \delta_4(a,a,a,b) = 2(3(a-b)^2+7) \nn \, , \\
&& \delta_4\left(a,b,\frac{a+b}{2} +\frac{i}{\sqrt{3}}\frac{\sqrt{(a-b)^2+4}}{2},\frac{a+b}{2} +i\sqrt{3}\frac{\sqrt{(a-b)^2+4}}{2}\right)=0 \nn \, , \\
&& \delta_4\left(a,b,\frac{a+b}{2} -\frac{i}{\sqrt{3}}\frac{\sqrt{(a-b)^2+4}}{2},\frac{a+b}{2} -i\sqrt{3}\frac{\sqrt{(a-b)^2+4}}{2}\right)=0 \, ;
\ea

\section{Factorisations}
\setcounter{equation}{0}

In this Appendix, we report the main calculations which prove the asymptotic factorisation of the $2n$-point function
when some of the rapidities $u_i$ get large. Following the main text (Section \ref {asy}), we first discuss the four point functions, then the general $2n$-point case.

\subsection{Four point functions}
\label {calc-4}

Starting from the integral representation (\ref {Pi_mat}) for $\Pi_{mat}^{(4)}(u_1+\Lambda_1,u_2+\Lambda_2,u_3,u_4)$ and performing the shifts in the isotopic variables described in the main text, we get
\ba
&&\Pi_{mat}^{(4)}(u_1+\Lambda_1,u_2+\Lambda_2,u_3,u_4)=  \nonumber \\
&=&\frac{1}{4}\sum _{\textrm{shifts}}\int_{-\infty}^{+\infty}\frac{da_1db_1db_2dc_1}{(2\pi)^4}\frac{g(b_1-b_2+\Lambda _{1}^{b}-\Lambda _{2}^{b})}{\displaystyle\prod_{i=1}^2f(a_1-b_i+\Lambda _{1}^{a}-\Lambda _i^{b})f(c_1-b_i+\Lambda _1^{c}-\Lambda _{i}^{b})\prod_{i,j=1}^{2}f(u_i-b_j+\Lambda _{i}-\Lambda _j^{b})} \times \nn\\
&& \times \int_{-\infty}^{+\infty}\frac{da_2db_3db_4dc_2}{(2\pi)^4}\frac{g(b_3-b_4)}{\displaystyle\prod_{i=3}^4f(a_2-b_i)f(c_2-b_i)\prod_{i,j=3}^{4}f(u_i-b_j)}
\,\m{R}^{(4,2)}(a_1,a_2,b_1,\ldots ,b_4,c_1,c_2;\Lambda_1,\Lambda _2) \, ,\nn\\ \label {Pi_mat2}
\ea
where we defined
\ba
&& \m{R}^{(4,2)}(a_1,a_2,b_1,\ldots ,b_4,c_1,c_2;\Lambda_1,\Lambda_2) =
\frac{\displaystyle\prod_{i=1}^2\prod_{j=3}^{4}g(b_i-b_j+\Lambda_i^{b})}{\displaystyle\prod_{i=1}^{2}\prod_{j=3}^{4}f(u_i-b_j+\Lambda_i)f(u_j-b_i-\Lambda_i^{b})} \times\nn\\
&& \times \frac{g(a_1-a_2+\Lambda_1^{a})g(c_1-c_2+\Lambda_1^{c})} {\displaystyle\prod_{i=3}^{4}f(a_1-b_i+\Lambda_1^{a})f(c_1-b_i+\Lambda_1^{c})\prod_{i=1}^{2}f(a_2-b_i-\Lambda_i^{b})f(c_2-b_i-\Lambda_i^{b})} \ . \nn
\ea
The reason to shift the isotopic variables is that we get an expression, (\ref{Pi_mat2}), in which one is allowed to perform the limit $\Lambda_i\to\infty$ inside the integrals. We have
\ba
&& \m{R}^{(4,2)}(a_1,a_2,b_1,\ldots ,b_4,c_1,c_2;\Lambda_1,\Lambda_2) \rightarrow
\Lambda_1^{-4}\Lambda _2^{-4}  \Bigl [ 1+2(u_3+u_4)\left (\frac{1}{\Lambda _1^b}+\frac{1}{\Lambda _2^b} \right )-4\left (\frac{u_1}{\Lambda _1}+\frac{u_2}{\Lambda _2} \right ) + \nonumber \\
&+& 2(b_3+b_4) \left ( \frac{1}{\Lambda _1}+\frac{1}{\Lambda _2}-\frac{2}{\Lambda _1^b}-\frac{2}{\Lambda _2^b}+\frac{1}{\Lambda _1^a}+\frac{1}{\Lambda _1^c}\right ) + 2a_2\left ( \frac{1}{\Lambda _1^b}+\frac{1}{\Lambda _2^b}-\frac{2}{\Lambda _1^a} \right )+2c_2 \left(\frac{1}{\Lambda _1^b}+\frac{1}{\Lambda _2^b}-\frac{2}{\Lambda _1^c}\right )+ \nn \\
&+& O(R^{-2}) \Bigr ] \, . \nn
\ea
The possible values for the string of shifts $\{ \Lambda _1^b,\Lambda _2^b,\Lambda _1^a,\Lambda _1^c\} $ are ten:
\ba
\Lambda _1^b &=& \Lambda _1 \ \ \Lambda _1 \ \ \Lambda _1 \ \ \Lambda _1 \ \ \Lambda _2 \ \ \Lambda _2 \ \ \Lambda _2 \ \ \Lambda _2 \ \ \Lambda _1 \ \ \Lambda _2  \nn \\
\Lambda _2^b &=& \Lambda _2 \ \ \Lambda _2 \ \ \Lambda _2 \ \ \Lambda _2 \ \ \Lambda _1 \ \ \Lambda _1 \ \ \Lambda _1 \ \ \Lambda _1 \ \ \Lambda _1 \ \ \Lambda _2 \  \nn \\
\Lambda _1^a &=& \Lambda _1 \ \ \Lambda _2 \ \ \Lambda _1 \ \ \Lambda _2 \ \ \Lambda _1 \ \ \Lambda _2 \ \ \Lambda _1 \ \ \Lambda _2 \ \ \Lambda _1 \ \ \Lambda _2 \  \nn \\
\Lambda _1^c &=& \Lambda _1 \ \ \Lambda _2 \ \ \Lambda _2 \ \ \Lambda _1 \ \ \Lambda _1 \ \ \Lambda _2 \ \ \Lambda _2 \ \ \Lambda _1 \ \ \Lambda _1 \ \ \Lambda _2 \  \nn \\
\ea
Summing over the ten possible shifts, we eventually obtain
\ba
&&\Pi_{mat}^{(4)}(u_1+\Lambda_1,u_2+\Lambda_2,u_3,u_4)\simeq \Lambda_1^{-4}\Lambda _2^{-4} \frac{1}{4}\frac{1}{\Lambda_{12}^{4}}\int\frac{da_1db_1db_2dc_1}{(2\pi)^4}\frac{1}{\displaystyle f(a_1-b_1)f(c_1-b_1)\prod_{i=1}^{2}f(u_i-b_i)}\times \nn\\
&& \times \int\frac{da_2db_3db_4dc_2}{(2\pi)^4}\frac{g(b_3-b_4)}{\displaystyle\prod_{i=3}^4f(a_2-b_i)f(c_2-b_i)\prod_{i,j=3}^{4}f(u_i-b_j)}= \label{pimat-final}\\
&=&
\Lambda_1^{-4}\Lambda _2^{-4}\left [1+2(u_3+u_4)\left (\frac{1}{\Lambda _1}+\frac{1}{\Lambda _2} \right )-4\left (\frac{u_1}{\Lambda _1}+\frac{u_2}{\Lambda _2} \right ) +O(R^{-2}) \right ]  \Pi_{mat}^{(2)}(u_1+\Lambda _1,u_2+\Lambda _2)\Pi_{mat}^{(2)}(u_3,u_4) \ , \nn
\ea
where $\Pi_{mat}^{(2)}(u_1+\Lambda _1,u_2+\Lambda _2) =6/\Lambda _{12}^4 + \ldots $. In order to evaluate the subleading terms in the square bracket of (\ref {pimat-final}), we use the following formul{\ae}:
\ba
&& \int \frac{dadcdb_1db_2}{(2\pi)^4} \frac{1}{f(u_1-b_1)f(u_2-b_2)f(a-b_1)f(c-b_1)}=1 \nn \\
&& \int \frac{dadcdb_1db_2}{(2\pi)^4} \frac{g(b_1-b_2)}{f(u_1-b_1)f(u_1-b_2)f(a-b_1)f(a-b_2)f(c-b_1)f(c-b_2)}=2 \, .
\label {use}
\ea
Formul{\ae} (\ref {use}) are necessary to reconstruct the term $6/\Lambda _{12}^4$, the leading one
in $ \Pi_{mat}^{(2)}(u_1+\Lambda _1,u_2+\Lambda _2)$, after summing over the ten possible shifts.

\subsection{$2n$ point functions}
\label{calc-2n}

Starting from the integral representation (\ref {Pi_mat}) for $\Pi_{mat}^{(2n)}(u_1+\Lambda_1,\ldots, u_m+\Lambda_m,u_{m+1},\ldots, u_{2n})$ and performing the shifts in the isotopic variables described in the main text, we get
\ba \label {Intpimat}
&&\Pi_{mat}^{(2n)}(u_1+\Lambda _1,\ldots,u_{m}+\Lambda _m,u_{m+1},\ldots,u_{2n})=  \frac{1}{(2n)!(n!)^2} \binom{2n}{m} \binom{n}{k}^{2} \sum _{\textrm{shifts}} \\
&& \int\prod_{i=1}^k\frac{da_i dc_i}{(2\pi)^2}\prod_{i=1}^m\frac{db_i}{2\pi}\,
\frac{\displaystyle\prod_{i<j,\,i=1}^{k}\left[g(a_i-a_j+\Lambda _{ij}^a)g(c_i-c_j+\Lambda _{ij}^c)\right]\displaystyle\prod_{i<j,\,i=1}^{m}g(b_i-b_j+\Lambda _{ij}^b)}
{\displaystyle\prod_{j=1}^{m}\left[\prod_{i=1}^{k}f(a_i-b_j+\Lambda _{ij}^{ab})f(c_i-b_j+\Lambda _{ij}^{cb})\prod_{l=1}^{m}f(u_l-b_j+\Lambda _{l}-\Lambda_j^b)\right]}\times \nn\\
&& \times \int\prod_{i=k+1}^n\frac{da_i dc_i}{(2\pi)^2}\prod_{i=m+1}^{2n}\frac{db_i}{2\pi}\,
\frac{\displaystyle\prod_{i<j,\,i=k+1}^{n}\left[g(a_i-a_j)g(c_i-c_j)\right]\displaystyle\prod_{i<j,\,i=m+1}^{2n}g(b_i-b_j)}
{\displaystyle\prod_{j=m+1}^{2n}\left[\prod_{i=k+1}^{n}f(a_i-b_j)f(c_i-b_j)\prod_{l=m+1}^{2n}f(u_l-b_j)\right]}
\,\m{R}^{(2n,m)}(a_1,\ldots , c_{2n};\Lambda) \nn
\ea
where
\ba
&& \m{R}^{(2n,m)}(a_1,\ldots , c_{2n};\Lambda)= \frac{\displaystyle\prod_{i=1}^{m}\prod_{j=m+1}^{2n}g(b_i-b_j+\Lambda _i^b)}{\displaystyle\prod_{i=1}^{m}\prod_{j=m+1}^{2n}f(u_j-b_i-\Lambda_i^b)f(u_i-b_j+\Lambda_i)} \times \nn\\
&& \times
\frac{\displaystyle\prod_{i=1}^{k}\prod_{j=k+1}^{n}g(a_i-a_j+\Lambda_i^a)g(c_i-c_j+\Lambda_i^c)}
{\displaystyle\prod_{j=1}^{m}\prod_{i=k+1}^{n}f(a_i-b_j-\Lambda_j^b)f(c_i-b_j-\Lambda_j^b)
\prod_{j=m+1}^{2n}\prod_{i=1}^{k}f(a_i-b_j+\Lambda_i^a)f(c_i-b_j+\Lambda_i^c)} \, , \label {r-int}
\ea
where we used the shorthand notations $\Lambda _{ij}^{a}=\Lambda _i^a -\Lambda _j^a$, $\Lambda _{ij}^{ab}=\Lambda _i^a -\Lambda _j^b$ and where we multiplied the previous expression by a
combinatorial factor $\binom{2n}{m} \binom{n}{k}^{2}$ which takes into account the different choices of isotopic roots that after shifting give the same result. We are now allowed to send $\Lambda _i \rightarrow +\infty$ inside the integrals in (\ref  {Intpimat}).
We have
\ba
&& \m{R}^{(2n,m)} \rightarrow  \frac{1}{\left ( \prod \limits _{i=1}^m \Lambda _i \right )^{4n-2m}}\Bigl \{ 1+2 \sum _{i=1}^{m}\frac{1}{\Lambda _i^b}\sum _{j=m+1}^{2n}u_j-2(2n-m)\sum _{i=1}^{m}\frac{u_i}{\Lambda _i} + \nonumber \\
&& + \sum _{j=m+1}^{2n} 2b_j \left [ \sum _{i=1}^{m}\frac{1}{\Lambda _i}-2\sum _{i=1}^{m} \frac{1}{\Lambda _i^b} + \sum _{i=1}^k\left (\frac{1}{\Lambda _i^a}+\frac{1}{\Lambda _i^c} \right ) \right ] + \nn \\
&& + \sum _{j=k+1}^{n} 2a_j \left ( \sum _{i=1}^{m}\frac{1}{\Lambda _i^b}-2 \sum _{i=1}^{k}\frac{1}{\Lambda _i^a} \right ) +  \sum _{j=k+1}^{n} 2c_j \left ( \sum _{i=1}^{m}\frac{1}{\Lambda _i^b}-2 \sum _{i=1}^{k}\frac{1}{\Lambda _i^c} \right ) + O\left (R^{-2} \right ) \Bigr \} \, . \label {R-int}
\ea
Summing over all the choices for the shifts on the auxiliary variables, the second and the third line of (\ref {R-int})
cancel and we are left with
\ba
&& \Pi_{mat}^{(2n)}(u_1+\Lambda _1,\ldots,u_{2k}+\Lambda _{2k},u_{2k+1},\ldots,u_{2n}) \rightarrow
\frac{1}{\left ( \prod \limits _{i=1}^m \Lambda _i \right )^{4n-4k}}\Bigl [1+2 \sum _{i=1}^{2k}\frac{1}{\Lambda _i}\sum _{j=2k+1}^{2n}u_j- \nonumber \\
&-& 2(2n-2k)\sum _{i=1}^{2k}\frac{u_i}{\Lambda _i} + O\left (R^{-2} \right ) \Bigr ] \Pi_{mat}^{(2k)}(u_1+\Lambda _1,\ldots,u_{2k}+\Lambda_{2k})\Pi_{mat}^{(2n-2k)}(u_{2k+1},\ldots,u_{2n})
\ea
where, however, $\Pi_{mat}^{(2k)}(u_1+\Lambda _1,\ldots,u_{2k}+\Lambda_{2k})$ goes to zero like
\be
\Pi_{mat}^{(2k)}(u_1+\Lambda _1,\ldots,u_{2k}+\Lambda_{2k})\sim
\prod _{i<j=1}^{2k} (\Lambda _i-\Lambda _j)^{-2} \left[ \Lambda _{12}^{-2}\Lambda _{34}^{-2}\ldots \Lambda _{2k-1,2k}^{-2}+pairings \right] \ .
\ee

\section{The polynomials $P_{2n}$}
\label{sca-pol}
\setcounter{equation}{0}

In Section \ref{asy} we proved the polar structure (\ref {P2n}),
\be
\label{P2nbis}
\Pi_{mat}^{(2n)}(u_1,\ldots ,u_{2n})=\frac{P_{2n}(u_1,\ldots,u_{2n})}{\displaystyle\prod_{i<j}^{2n}(u_{ij}^2+1)(u_{ij}^2+4)} \, ,
\ee
where $P_{2n}$ is a polynomial of degree $4 n(n-1)$, symmetric under permutations of its variables. In this Appendix we list some properties of these functions and give explicit expressions for $P_2$ and $P_4$. In addition, by means of the factorisation, we derive a simple formula for the highest degree $P_{2n}^{(0)}$ for any $n$ and relate it to that of $\delta_{2n}$.

To start with, we express the full polynomial $P_{2n}$ computed in specific values in terms of smaller polynomials $P_{2k}$, with $k<n$. This follows from the residue formula of the matrix part (\ref{ResPimat}), which implies a (restricted) ''recursion'' relation for the polynomials
\be\label{RecSca}
P_{2n}(u_1-i,u_1+i,u_3,\ldots ,u_{2n})=6P_{2n-2}(u_3,\ldots ,u_{2n})\displaystyle\prod_{j=3}^{2n}(u_{1j}^2+4)(u_{1j}^2+9)\, ,
\ee
that gives, upon iteration, the most general one valid for $k=0,\ldots, n$
\ba
&&P_{2n}(u_1-i,u_1+i,\ldots ,u_k-i,u_k+i,u_{2k+1},\ldots, u_{2n})=6^k P_{2n-2k}(u_{2k+1},\ldots ,u_{2n})\cdot \nn\\
&& \cdot\displaystyle\prod_{i<j}^k (u_{ij}^2+1)(u_{ij}^2+4)(u_{ij}^2+9)(u_{ij}^2+16)\displaystyle\prod_{i=1}^k\displaystyle\prod_{j=2k+1}^{2n}(u_{ij}^2+4)(u_{ij}^2+9) \, .
\ea
The complete iteration $k=n$ yields the exact expression of the polynomials $P_{2n}$, evaluated in a specific configuration:
\be
P_{2n}(u_1-i,u_1 +i,u_2-i,u_2 + i,\ldots ,u_{n}-i,u_{n}+i)= 6^n \displaystyle\prod_{i<j}^n(u_{ij}^2+1)(u_{ij}^2+4)(u_{ij}^2+9)(u_{ij}^2+16) \, .
\ee
In addition, the recursion formula (\ref{RecSca}) tells us that the polynomial vanishes in some particular configurations
\ba
&& P_{2n}(u_1,u_1+i,u_1+3i,u_4,\ldots ,u_{2n})=0 \, , \nn\\
&& P_{2n}(u_1,u_1+2i,u_1+4i,u_4,\ldots ,u_{2n})=0 \, . \label {van-conf}
\ea

\subsection{Explicit expressions}
\label{exp-P2n}

We now provide the polynomials appearing in (\ref {P2n}) in the cases $n=1, n=2$:
\be
P_2(u_1,u_2) =6
\ee
\begin{align}\label{P4}
P_4(u_1,u_2,u_3,u_4) &= 36 \left[9((u_1-u_2)^2+4)((u_3-u_4)^2+4)+9((u_1-u_3)^2+4)((u_2-u_4)^2+4)+\right. \nn\\
&+ 9((u_1-u_4)^2+4)((u_2-u_3)^2+4) +\nn\\
&+ ((u_1-u_3)^2+4)((u_1-u_4)^2+4)((u_2-u_3)^2+4)((u_2-u_4)^2+4)+\nn\\
&+ ((u_1-u_2)^2+4)((u_1-u_4)^2+4)((u_3-u_2)^2+4)((u_3-u_4)^2+4)+\nn\\
&+ ((u_1-u_2)^2+4)((u_1-u_3)^2+4)((u_4-u_2)^2+4)((u_4-u_3)^2+4)+\nn\\
&+ \frac{3}{2}((u_1 - u_2)^2+4)((u_2 - u_3)^2+4)((u_1- u_4)^2+4) +\nn\\
&+ \frac{3}{2}((u_1 - u_3)^2+4)((u_2-u_3)^2+4)((u_1 - u_4)^2+4) + \nn\\
&+ \frac{3}{2}((u_1 - u_2)^2+4)((u_1-u_3)^2+4)((u_2 - u_4)^2+4) + \nn\\
&+ \frac{3}{2}((u_1 - u_3)^2+4) ((u_2 - u_3)^2+4) ((u_2 - u_4)^2+4) + \nn\\
&+ \frac{3}{2}((u_1 - u_3)^2+4) ((u_1 - u_4)^2+4) ((u_2 - u_4)^2+4) + \nn\\
&+ \frac{3}{2}((u_2 - u_3)^2+4) ((u_1 - u_4)^2+4) ((u_2 - u_4)^2+4) + \nn\\
&+ \frac{3}{2}((u_1 - u_2)^2+4) ((u_1 - u_3)^2+4) ((u_3 - u_4)^2+4) + \nn\\
&+ \frac{3}{2}((u_1 - u_2)^2+4) ((u_2 - u_3)^2+4) ((u_3 - u_4)^2+4) + \nn\\
&+ \frac{3}{2}((u_1 - u_2)^2+4) ((u_1 - u_4)^2+4) ((u_3 - u_4)^2+4) + \nn\\
&+ \frac{3}{2}((u_2 - u_3)^2+4) ((u_1 - u_4)^2+4) ((u_3 - u_4)^2+4) + \nn\\
&+ \frac{3}{2}((u_1 - u_2)^2+4) ((u_2 - u_4)^2+4) ((u_3 - u_4)^2+4) + \nn\\
&+ \frac{3}{2}((u_1 - u_3)^2+4) ((u_2 - u_4)^2+4) ((u_3 - u_4)^2+4)-\nn\\
&- \frac{3}{2}((u_1 - u_2)^2(u_2 - u_3)^2(u_1 - u_3)^2 + (u_1 - u_2)^2(u_2 - u_4)^2(u_1-u_4)^2 +\nn\\
&+ (u_1 - u_4)^2(u_4 - u_3)^2(u_1 - u_3)^2 +(u_4 - u_2)^2(u_2 - u_3)^2(u_4 -u_3)^2)+ \nn\\
&+ 48(u_1 - u_2)^2 + 48(u_1 - u_3)^2 + 48(u_1 - u_4)^2 + 48(u_3 - u_2)^2 + 48(u_4 - u_2)^2 +\nn\\
&+ 48(u_3 -u_4)^2 +\frac{3}{2}((u_1 - u_2)^2(u_3 - u_4)^2 +(u_1 - u_3)^2(u_2 - u_4)^2 +\nn\\
&+ (u_1 - u_4)^2(u_3 - u_2)^2) + \left. 1152 \right]
\end{align}
From the expression of $P_4$ and the factorisation property (\ref{factP}), we can guess
the highest degree term, {\it i.e.} the term of degree (in $u$) $4n(n-1)$. For $n=2$ the exact formula (\ref{P4}) yields
\be
P_{4}^{(0)}(u_1,u_2,u_3,u_4)=6^2\displaystyle\prod_{i<j}^4u_{ij}^2\left(\frac{1}{u_{12}^2 u_{34}^2} + \frac{1}{u_{13}^2 u_{24}^2} + \frac{1}{u_{14}^2 u_{23}^2}\right) \, .
\ee
This formula has a nice interpretation as a sum over the pairings, resembling the Wick theorem for bosons: except for the common prefactor, we can think of $P^{(0)}_4$ as the four point function of a free boson with propagator $u^{-2}_{ij}$. The generalization of this formula to the $2n$ goes through an expression that, in the factorisation limit, reproduces exactly the property (\ref{factP}) for $P_{2n}^{(0)}$, with any $n,k$.
We thus conjecture
\be\label{P0}
P_{2n}^{(0)}(u_1,\ldots ,u_{2n})=6^n\displaystyle\prod_{i<j}^{2n} u_{ij}^2 \sum'_p\displaystyle\prod_{i=1}^{n}\frac{1}{(u_{p(2i-1)} - u_{p(2i)})^2}  \, ,
\ee
where the sum is restricted\footnote{Alternatively, we can write the unconstrained sum over all the permutations, with its specific prefactor to account for the overcounting, $\sum'_p=\frac{1}{2^nn!}\sum_p$.} over the (inequivalent) pairings, such that the total number of terms is $(2n-1)!!$, as in the Wick expansion.
A careful analysis shows that (\ref{P0}) is the only polynomial solution satisfying factorisation (\ref{factP}) and the required symmetry under $u_i \leftrightarrow u_j$. Formula (\ref{P0}) is confirmed for $n=3$, directly from the sum over Young tableaux (\ref{MatYoung}), by taking the leading order in the large rapidities limit of the general contribution (\ref{general}).

As anticipated in the main text, there is an interesting link with the polynomials $\delta_{2n}$: we use the identity\footnote{From the physical point of view, this identity is a sort of bosonisation, as the LHS can be thought as a correlator of a free fermion with propagator $u^{-1}_{ij}$.} for the special $2n\times 2n$ antisymmetric matrix
\be\label{Pf-Wick}
\textrm{Det} \left(\frac{1}{u_{ij}}\right)=\left[\textit{Pf}\left(\frac{1}{u_{ij}}\right)\right]^2 = \sum'_p\displaystyle\prod_{i=1}^{n}\frac{1}{(u_{p(2i-1)} - u_{p(2i)})^2} \, ,
\ee
to relate the highest degrees of the polynomials $P_{2n}$ and $\delta_{2n}$. Combining (\ref{delta0det}),(\ref{P0}) and (\ref{Pf-Wick}) we can express the highest degree in terms of a determinant as
\be
P^{(0)}_{2n}(u_1,\ldots ,u_{2n})=6^n\displaystyle\prod_{i<j}^{2n} u_{ij}^2\textrm{Det}\left(\frac{1}{u_{ij}}\right)=\frac{6^n4n^2}{4^n(n!)^2}\left[\delta^{(0)}_{2n}(u_1,\ldots ,u_{2n})\right]^2 \, .
\ee
This remarkable equality does not survive when we consider the full polynomial $P_{2n}$, as we can verify for $n=2$ with the explicit formula (\ref{P4}).  We do not know if a determinant representation of the full $P_{2n}$ exists: however, it is an interesting idea to pursue since it would allow to find a nice representation of $W$.

To close this appendix, we notice that
in the large (differences of) rapidities limit the Young diagram $(1,1,\ldots ,1,1)_{2n}$, see (\ref{1111}), enjoys the same structure as the whole matrix part $\Pi_{mat}^{(2n)}$; more precisely, through a rescaling of rapidities, we get
\be\label{LargeDelta}
\lim_{\Delta\to\infty} \frac{(1,1,\ldots ,1,1)_{2n}}{\Pi^{(2n)}_{mat}}=\left(\frac{2}{3}\right)^n , \quad u_{i}\to \Delta u_i \, ,
\ee
$\delta^2_{2n}$ being contained, also for finite rapidities, in $(1,1,\ldots ,1,1)_{2n}$, see formula (\ref{1111}).
It is not a trivial fact that in the large rapidities limit the other Young diagrams - which for finite rapidities do not contain $\delta^2_{2n}$ - reproduce the same structure as $(1,1,\ldots ,1,1)_{2n}$.

\subsection{Factorisation of $P_{2n}$}
\label{fact-p2n}

The factorisability of the functions $G^{(2n)}$ also affects the behaviour of the polynomials $P_{2n}$. 
Indeed, by requiring the factorisation property when a set of $2k$ rapidities is sent to infinity, we find
\ba\label{factP}
&& P_{2n} (u_1 + \Lambda,\ldots,u_{2k}+\Lambda,u_{2k+1},\ldots,u_{2n})= \Lambda ^{8(n-k)k}
P_{2k}(u_1,\ldots\,u_{2k})P_{2n-2k}(u_{2k+1},\ldots\,u_{2n})\cdot \nonumber \\
&\cdot & \left [ 1+2\Lambda^{-1} \sum _{i=1}^{2k} \sum _{j=2k+1}^{2n} (u_i-u_j)  + \Delta_{2n,2k}^{(2)}(u_1,\ldots ,u_{2n})\Lambda^{-2} + O(\Lambda^{-3})\right] \, ,
\ea
where we encoded the quadratic subleading in the function $\Delta_{2n,2k}^{(2)}$.
On the other hand, by shifting an odd number of particles, we get instead the power behaviour
\be
P_{2n} (u_1 + \Lambda,\ldots,u_{2k+1}+\Lambda,u_{2k+2},\ldots,u_{2n})= O(\Lambda ^{2(2k+1)(2n-2k-1)-2}) \, ,
\ee

\section{Connected functions}
\label{int-con}
\setcounter{equation}{0}

This Appendix focuses on the connected functions $g^{(2n)}$. We first analyse the relation with the 'Green' functions $G^{(2n)}$, that we sketched for the first few cases $n=2,3$ in the main text (Section \ref{conn-func}). We also mentioned the importance of the property $g^{(2n)}\in L^1(\mathbb{R}^{2n-1})$: here we tackle this issue and give evidence of its validity. Sometimes, concerning the asymptotic behaviour of $G^{(2n)}$, we will refer to the results obtained in Section \ref{asy}.

The expansion of $G^{(2n)}$ 
in terms of the connected parts is well-known, it involves a sum over all the possible arrangements of $2n$ particles in subgroups of even particles\footnote{For the sake of compactness, we omitted the dependence on the rapidities.}
\be\label{direct}
G^{(2n)}=\sum_{\left\{m\right\}}\sum_{\textit{pair.}}\displaystyle\prod_{k=1}^{n}\underbrace{g^{(2k)}\ldots g^{(2k)}\,}_\text{$m_k$ terms} \, ,
\ee
where $\left\{m\right\}$ represents the set of integers $m_k$ with $k=1,\ldots,n$, identifying a specific cluster configuration for the $2n$ particles and fulfilling the constraint $\displaystyle\sum_{k=1}^{n}2k \, m_k=2n$\footnote{A similar formula holds in general, when also odd numbers of particles are allowed.}. For any definite set $\{m\}$, the number of non equivalent ways of clustering is given by $\left(\prod_{k=1}^{n}\frac{1}{m_k!}\right)\frac{(2n)!}{\prod_{k=1}^{n}((2k)!)^{m_k}}$.

The inverse relation can be obtained, resulting in the general expansion
\be\label{inverse}
g^{(2n)}=\sum_{\left\{m\right\}}f(\left\{m\right\})\sum_{\textit{pair.}}\displaystyle\prod_{k=1}^{n}\underbrace{G^{(2k)}\ldots G^{(2k)}\,}_\text{$m_k$ terms} \, ,
\ee
where, in contrast to (\ref {direct}), the products of functions are weighted by a prefactor
\be
f(\left\{m\right\})=(-1)^{p}p! \, , \quad p\equiv\sum_{k=1}^{n}m_k - 1 \, ,
\ee
containing also an oscillating sign.
In an equivalent manner, it is possible to sum over all the permutations and account for the overcounting with the specific prefactor, and rewrite (\ref{direct}) and (\ref{inverse}) as \cite{Smirnov}
\ba
&&G^{(n)}(u_1,\ldots,u_n)=\sum_{q=1}^{n}\frac{1}{q!}\sum_{k_1+\ldots+k_q=n}\frac{1}{k_1!\ldots k_q!} \cdot \nn\\
&&\cdot\sum_P g^{(k_1)}(u_{P_1},\ldots,u_{P_{k_1}})\ldots g^{(k_q)}(u_{P_{n-k_q+1}}\ldots,u_{P_{n}}) \, ,
\ea
\ba
&&g^{(n)}(u_1,\ldots,u_n)=\sum_{q=1}^{n}\frac{(-1)^{q-1}}{q}\sum_{k_1+\ldots+k_q=n}\frac{1}{k_1!\ldots k_q!}\cdot \nn\\
&&\cdot\sum_P G^{(k_1)}(u_{P_1},\ldots,u_{P_{k_1}})\ldots G^{(k_q)}(u_{P_{n-k_q+1}}\ldots,u_{P_{n}}) \, ,
\ea
which actually holds also for $n$ odd.

\vspace{0.3cm}

Now we turn our attention to the asymptotic properties of $g^{(2n)}$. In Section \ref{conn-func} we stated that the connected functions must be integrable over the $2n-1$ variables $\alpha_i$ on which they depend: a sufficient condition for that is $g^{(2n)}\in L^1(\mathbb{R}^{2n-1})$, which involves the integral of $|g^{(2n)}|$, being then a stronger requirement.  The general condition to be fulfilled is (\ref{convg2n}), which covers all the possible asymptotic regions in the integration space. Here we will address some of them for any $n$, giving hints for $g^{(2n)}\in L^1(\mathbb{R}^{2n-1})$. Moreover, we perform a complete study of the functions $g^{(4)}$ and $g^{(6)}$ and show that they belong respectively to $L^1(\mathbb{R}^3)$ and $L^1(\mathbb{R}^5)$.

Let us start with a general $n$ analysis: consider a set of asymptotic regions, by shifting a subset of the entries of $G^{(2n)}(u_1+\Lambda _1,u_2+\Lambda_2, \ldots,u_{m}+\Lambda_{m},u_{m+1},\ldots,u_{2n})$ by amounts $\Lambda_i$, $i=1,\ldots ,m$, $1\leq m \leq 2n$, all of order $R\gg 1$: as usual, we suppose that $\Lambda _i =c_i R +O(R^0)$, with $c_i$ constants and $R\rightarrow +\infty$ \footnote {Due to the fact that $G^{(2n)}$ depends only on differences of rapidities, this way of shifting covers also the case $G^{(2n)}(u_1+\Lambda _1,u_2+\Lambda_2, \ldots,u_{2n}+\Lambda _{2n})$, with $\Lambda _{m+1}=\ldots =\Lambda _{2n}$.}.
Recalling Section \ref {asy}, we observe that the function $G^{(2n)}$ behaves as
\be
G^{(2n)}(u_1+\Lambda _1,u_2+\Lambda_2, \ldots,u_{m}+\Lambda_{m},u_{m+1},\ldots,u_{2n}) \sim R^{-2k+2\left [ \frac{p}{2}\right ]} \label{Goddeven} \, ,
\ee
where $2k=m$ for even $m$, either $2k=m+1$ for $m$ odd and $p\leq m$ is the number of $\Lambda _i$ which mutually coincide ({\it i.e.} we have $\Lambda _1=\Lambda _2=\ldots =\Lambda _p \not = \Lambda _{p+1} \ldots \not= \Lambda _{m}$). As a direct consequence of (\ref{direct}), along with the factorisation of $G^{(2n)}$, the connected functions exhibit instead a different asymptotic behaviour. In fact, when $m$ is odd, {\it i.e.} $m=2k-1$, we find
\footnote {The different asymptotic form with respect to (\ref {Goddeven}) comes from the fact that for even $p$ we have factorisation of $G^{(2n)}$ and then an extra $R^{-2}$ in the behaviour of $g^{(2n)}$.}
\be\label{gRodd}
g^{(2n)}(u_1+\Lambda _1,u_2+\Lambda_2,\ldots,u_{2k-1}+\Lambda_{2k-1},u_{2k},\ldots,u_{2n}) \sim R^{-2k+2\left [ \frac{p-1}{2}\right ]} \,;
\ee
otherwise for even $m=2k$, with $1<m\leq 2n-2$, a faster decay shows up
\be\label{gReven}
g^{(2n)}(u_1+\Lambda _1,u_2+\Lambda_2,\ldots,u_{2k}+\Lambda_{2k},u_{2k+1},\ldots,u_{2n}) \sim R^{-2k-2+2\left [ \frac{p}{2}\right ]} \,.
\ee
So far, see (\ref{gRodd}) and (\ref{gReven}), we showed that if we send $m$ (even or odd) particles to infinity shifting them or by $m$ different quantities $\Lambda_i$, $i=1,\ldots ,m$, or by $m-p+1$ different quantities, which means that $\Lambda _1=\Lambda _2=\ldots =\Lambda _p \not = \Lambda _{p+1} \ldots \not= \Lambda _{m}$, the function $g^{(2n)}$ decays fast enough to be integrable. This asymptotic region corresponds to the split $2n\to p + m' +1+ \ldots + 1$, where $m'\equiv 2n-m$. In contrast, relation (\ref {Goddeven}) shows that the integral of $G^{(2n)}$ is divergent. If $p=m$ we find the result reported in \cite {BFPR2}, {\it i.e.} that if $m$ rapidities are shifted by the same amount, $g^{(2n)}$ decays as $R^{-2}$, while $G^{(2n)}$ stays constant for even $m$ as it factorises. However, the cases $2n\to p + m' +1 +\ldots + 1$ form only a subset of all the different ways of grouping particles. The most general case, depicted in (\ref{convg2n}), concerns the split $2n\to k_1 + \cdots + k_{l+1}$. It turns out that a general proof is very complicated, as the number of asymptotic regions grows very fast with $n$. Nevertheless, a thorough analysis of the simplest cases $g^{(4)}$, $g^{(6)}$ strongly hints that $g^{(2n)}$ belongs to $L^1(\mathbb{R}^{2n-1})$, for any $n$.

\vspace{0.3cm}

We are going to analyse in detail $g^{(4)}$ and $g^{(6)}$, to show that they do belong to $L^1(\mathbb{R}^{2n-1})$. 
To start with, we discuss the four point function $g^{(4)}$, for which the conditions above are sufficient to guarantee $g^{(4)}\in L^1(\mathbb{R}^3)$, as all the asymptotic regions are of the type $2n\to p+m'+1+\ldots + 1$.
We can push further the analysis thanks to its explicit expression as a rational function, see the polynomial (\ref{P4}). This allows us to find the actual decay
of the connected function $g^{(4)}$ in the different regimes.
We use the variables $\alpha_i\equiv \theta_{i+1}-\theta_1$, thus the invariance under exchange of rapidities, in addition to the symmetry under permutation of the $\alpha_i$, implies
\be
g^{(4)}(\alpha_1,\alpha_2,\alpha_3)=g^{(4)}(-\alpha_1,\alpha_2-\alpha_1,\alpha_3-\alpha_1)=g^{(4)}(\alpha_1-\alpha_2,-\alpha_2,\alpha_3-\alpha_2)=g^{(4)}(\alpha_1-\alpha_3,\alpha_2-\alpha_3,-\alpha_3) \,.
\ee
The polar expression (\ref{P2n}), combined with (\ref{P4}), provides a compact formula for $g^{(4)}$.
When one of the $\alpha_i$ grows to infinity, which corresponds to the split $4\to 3+1$,
we have only one shift $\Lambda$ and obtain
\be\label{1p-3p}
g^{(4)}(\alpha_1 +\Lambda,\alpha_2,\alpha_3) = \frac{g^{(4)}_{as}(\alpha_2,\alpha_3) }{\Lambda^2} + O(\Lambda^{-3})\, ,
\ee
being the minimum decay assuring integrability at infinity. We defined the asymptotic function $g^{(4)}_{as}$ according to the main text, where it has been used to compute the subleading contribution.
A physically different limit occurs when we consider the split $4\to 2+2$, realized by sending two $\alpha_i$ to infinity together and resulting in a faster decay\footnote {The result above means that the correction to the factorisation (\ref{4-fact}) are in fact of order $O(\Lambda^{-4})$, if we consider the case $\Lambda_1=\Lambda_2$.} than the expected $O(\Lambda^{-2})$
\be\label{g4faster}
g^{(4)}(\alpha_1 +\Lambda,\alpha_2 +\Lambda,\alpha_3) = O(\Lambda^{-4}) \,.
\ee
Now we deal with different shifts, taking all the $\Lambda_i$ and their differences $\Lambda_{ij}$ to be large and of order $R$.
We consider $g^{(4)}(\alpha_1 + \Lambda_1, \alpha_2+\Lambda_2,\alpha_3)$, $g^{(4)}(\alpha_1 + \Lambda_1, \alpha_2+\Lambda_1,\alpha_3 +\Lambda_2)$ along with all the possible permutations (corresponding to a multiplicity six in the integration domain): this is the split $4\to 2+1+1$, where the function behaves as
\be
g^{(4)}(\alpha_1 +\Lambda_1,\alpha_2 +\Lambda_2,\alpha_3) = O(R^{-4}) \, ,
\ee
which turns out to be faster than the required $O(R^{-3})$.
The last limit to analyse is $4\to 1+1+1+1$, where our function decays as
\be
g^{(4)}(\alpha_1 +\Lambda_1,\alpha_2 +\Lambda_2,\alpha_3+\Lambda_3) =
O(R^{-6}) \, ,
\ee
which is, again, faster than the minimum $O(R^{-4})$. Summarising, in all regions except $4\to 3+1$, the connected function $g^{(4)}$ goes to zero faster than the minimum required by integrability. This fact has important effects on the computation of the subleading term $\ln\ln(1/z)$, as clarified in Section \ref{smallmass}. In particular, the function $g^{(4)}_{as}(\alpha_2,\alpha_3)$ belongs to $L^1(\mathbb{R}^2)$, as it satisfies
\be
g^{(4)}_{as}(\alpha_2+\Lambda,\alpha_3) = O(\Lambda^{-2}), \quad g^{(4)}_{as}(\alpha_2+\Lambda_1,\alpha_3+\Lambda_2) = O(R^{-4}) \, .
\ee

\vspace{0.3cm}

The six scalar case, $n=3$, is more involved due to the presence of many different asymptotic regions. However, all of them but one are included in the subset $2n\to m + p + 1 + \cdots + 1$ analysed before.
Let us recall the connected function $g^{(6)}$, in the shorthand notation
\be\label{inverse-bis}
g_{123456}=G_{123456}-(G_{12}G_{3456} + 14 \textit{ d.e.}) + 2(G_{12}G_{34}G_{56} + 14 \textit{ d.e.}) \, .
\ee
The only process we need to address is $6\to 2+2+2$, which is not trivial as it involves only groups composed by an even number of particles, making the RHS of (\ref{inverse-bis}) of order $O(1)$. Therefore, in addition to the finite part $O(1)$, a refined cancellation of the subleading terms $O(R^{-2})$ needs to occur, a fact that not guaranteed by (\ref{2n-fact}) itself.
We choose to group the particles according to $(12 \quad 34 \quad 56)$, thus we are left with
\be\label{g6left}
g_{123456} = G_{123456} - G_{12}G_{3456} - G_{34}G_{1256} - G_{56}G_{1234} + 2G_{12}G_{34}G_{56} + O(R^{-4}) \, ,
\ee
Thanks to (\ref{4-fact}) and (\ref{g4faster}), the condition (\ref{convg2n}) becomes
\be\label{cond6222}
G^{(6)}(u_1+\Lambda_1, u_2+\Lambda_1, u_3+\Lambda_2, u_4+\Lambda_2, u_5,u_6)= G^{(2)}(u_1,u_2)G^{(2)}(u_3,u_4)G^{(2)}(u_5,u_6) + O(R^{-3}) \, ,
\ee
which is not a straightforward extension of the results in Section \ref{asy} and, as we are going to show, represents a sort of multiple factorisation.\\
To shed light on (\ref{cond6222}), we define the corrections to the single factorisation process $2n\to 2k + 2(n-k)$ through
\be
G^{(2n)}\to G^{(2k)}G^{(2n-2k)} + \sum_l \Lambda^{-l}S_{2n,2k}^{(l)}(u_1, \ldots , u_{2n}) \, .
\ee
We neglect for a moment that the quadratic subleading $S^{(2)}_{(4,2)}$ is vanishing, the cancellation of the terms $O(R^{-2})$ in (\ref{g6left}) occurs if
\ba\label{mult6222}
&& G^{(6)}(u_1 + \Lambda_1 , u_2 + \Lambda_1 , u_3 + \Lambda_2, u_4 + \Lambda_2 , u_5, u_6)= G^{(2)}(u_1, u_2)G^{(2)}(u_3, u_4)G^{(2)}(u_5, u_6) + \nn\\
&& + \Lambda_1^{-2}G^{(2)}(u_3, u_4)S^{(2)}_{4,2}(u_1, u_2, u_5, u_6) + \Lambda_2^{-2}G^{(2)}(u_1, u_2)S^{(2)}_{4,2}(u_3, u_4, u_5, u_6) + \nn\\
&& + \Lambda_{12}^{-2}G^{(2)}(u_5, u_6)S^{(2)}_{4,2}(u_1, u_2, u_3, u_4) + O(R^{-3})
\ea
If we use $S^{(2)}_{(4,2)}=0$, we recover the previous formula (\ref{cond6222}): nevertheless, (\ref{mult6222}) is interesting by itself for its clear physical meaning and it may be easily generalized to any process of the type $2n\to 2k_1 + \cdots + 2k_l$.
The formula (\ref{mult6222}) represents a relation among the subleading corrections of different factorisation processes: to put it simply, the multi-factorisation process gets corrected by all the subleading terms associated to the various sub-factorisation processes involved, which are three in the case $6\to 2+2+2$.
The constraint (\ref{mult6222}) can be translated into the following condition on the polynomial $P_6$
\ba\label{sublP6}
&& P_6(12_{\Lambda_1} \quad 34_{\Lambda_2} \quad 56)=\Lambda_1^{8}\Lambda_2^{8}\Lambda_{12}^{8}P_2P_2P_2 \biggl[ 1 + 2(u_{13}+u_{14}+u_{23}+u_{24})\Lambda_{12}^{-1} + \nn\\
&+& 2(u_{15}+u_{16}+u_{25}+u_{26})\Lambda_{1}^{-1} + 2(u_{35}+u_{36}+u_{45}+u_{46})\Lambda_{2}^{-1}  + \nn\\
&+& 4(u_{13} + u_{14} +u_{23} +u_{24} )(u_{15} + u_{16} +u_{25} +u_{26} )\Lambda_{12}^{-1}\Lambda_1^{-1}  + \nn\\
&+& 4(u_{13} + u_{14} +u_{23} +u_{24} )(u_{35} + u_{36} +u_{45} +u_{46} )\Lambda_{12}^{-1}\Lambda_2^{-1}  + \nn\\
&+& 4(u_{15} + u_{16} +u_{25} +u_{26} )(u_{35} + u_{36} +u_{45} +u_{46} )\Lambda_{2}^{-1}\Lambda_1^{-1}  + \nn\\
&+& \Delta_{4,2}^{(2)}(u_1,u_2,u_3,u_4)\Lambda_{12}^{-2} + \Delta_{4,2}^{(2)}(u_1,u_2,u_5,u_6)\Lambda_{1}^{-2} + \Delta_{4,2}^{(2)}(u_3,u_4,u_5,u_6)\Lambda_{2}^{-2} +  O(R^{-3}) \biggr] \nn \, , \\
\ea
where an obvious shorthand notation has been introduced. In plain words, the quadratic corrections to the multifactorisation $6 \to 2+2+2$ of $P_6$ shall be fixed by the quantity $\Delta_{4,2}^{(2)}$, which parametrises the correction to the factorization $4 \to 2+2$, $cfr.$ (\ref{factP}). It is easy to generalize (\ref{sublP6}) for the process $2n\to 2k_1 + \cdots 2k_l$ and check that the highest degree $P_{2n}^{(0)}$ satisfies these constraints. To prove the general integrability condition (\ref{convg2n}) for any $n$ and any split, the generalizations of (\ref{cond6222}), (\ref{mult6222}) and (\ref{sublP6}) must hold.
A deeper analysis of $g^{(6)}$, employing the sum over Young tableaux (\ref{PiMat-Young}), confirms that in the limit $6 \to 2+2+2$ the function $g^{(6)}$ behaves like $O(R^{-4})$ and thus it is of class $L^1(\mathbb{R}^5)$, which also means that the property (\ref{mult6222}) is valid.

In conclusion, despite the lack of a general proof, we collected much evidence for $g^{(2n)}\in L^1(\mathbb{R}^{2n-1})$. First, both functions $g^{(4)}$ and $g^{(6)}$ satisfy the requirement and their decay is even faster in some regions. The extension of the factorisation analysed in Section \ref{asy} to the region $6\to 2+2+2$ (\ref{mult6222}) is needed for $g^{(6)}$, which actually holds. In addition, the constraint (\ref{mult6222}) is very physical and we can imagine it holds true, conveniently extended, for any splitting of the type $2n\to 2k_1 + \cdots + 2k_l$, which would guarantee the integrability of our connected functions.


\begin{thebibliography}{xx}




\bibitem{MGKPW1}
J. Maldacena, {\sl The large N limit of superconformal field theories and supergravity},
Adv. Theor. Math. Phys. {\bf 2} (1998) 231 and hep-th/9711200;

\bibitem{MGKPW2}
S. Gubser, I. Klebanov, A. Polyakov, {\sl Gauge theory
correlators from non-critical string theory},
Phys. Lett. {\bf B428} (1998) 105 and hep-th/9802109;

\bibitem{MGKPW3}
E. Witten, {\sl Anti-de Sitter space and holography},
Adv. Theor. Math. Phys. {\bf 2} (1998) 253 and hep-th/9802150;

\bibitem{MZ}
J. Minahan, K. Zarembo,
{\sl The Bethe Ansatz for ${\cal N}=4$ Super Yang-Mills},
JHEP{\bf 03} (2003) 013 and hep-th/0212208;

\bibitem{BS1}
N. Beisert, M. Staudacher,
{\sl The ${\cal N}=4$ SYM integrable super spin chain},
Nucl. Phys. {\bf B670} (2003) 439 and hep-th/0307042;

\bibitem{BS2}
V. Kazakov, A. Marshakov, J. Minahan, K. Zarembo,
{\sl Classical/quantum integrability in AdS/CFT},
JHEP{\bf 05} (2004) 024 and hep-th/0402207;

\bibitem{BS3}
N. Beisert, C. Kristjansen and M. Staudacher,
{\sl The Dilatation operator of conformal N=4 superYang-Mills theory},
Nucl. Phys. {\bf B664} (2003) 131 and hep-th/0303060;

\bibitem{BS4}
N. Beisert, V. Kazakov, K. Sakai, K. Zarembo,
{\sl The Algebraic curve of classical superstrings on AdS(5) x S**5},
Commun. Math. Phys. {\bf 263} (2006) 659 and hep-th/0502226;

\bibitem{BS5}
N. Beisert, M. Staudacher,
{\sl Long-range $PSU(2,2|4)$ Bethe Ansatz for gauge theory and strings},
Nucl. Phys. {\bf B727} (2005) 1 and hep-th/0504190;

\bibitem{BES}
N. Beisert, B. Eden, M. Staudacher, {\sl Transcendentality and
crossing}, J. Stat. Mech. {\bf 07} (2007) P01021 and hep-th/0610251;

\bibitem{TBA1}
D. Bombardelli, D. Fioravanti and R. Tateo,
{\sl Thermodynamic Bethe Ansatz for planar AdS/CFT: a proposal},
J. Phys. {\bf A42} (2009) 375401 and arXiv:0902.3930 [hep-th];

\bibitem{TBA2}
N. Gromov, V. Kazakov, A. Kozak and P. Vieira,
{\sl Exact Spectrum of Anomalous Dimensions of Planar N = 4 Supersymmetric Yang-Mills Theory: TBA and excited states},
Lett. Math. Phys. {\bf 91} (2010) 265 and arXiv:0902.4458 [hep-th];

\bibitem{TBA3}
G. Arutyunov and S. Frolov,
{\sl Thermodynamic Bethe Ansatz for the $AdS_5 \times  S^5$ Mirror Model},
JHEP{\bf 05} (2009) 068 and arXiv:0903.0141 [hep-th];

\bibitem{QSC1}
N. Gromov, V. Kazakov, S. Leurent, D. Volin,
{\sl Quantum Spectral Curve for Planar N=4 Super-Yang-Mills Theory},
Phys. Rev. Lett. {\bf 112}  (2014) 1, 011602 and arXiv:1305.1939 [hep-th];

\bibitem{QSC2}
A. Cavagli\`a, D. Fioravanti, N. Gromov and R. Tateo,
{\sl Quantum Spectral Curve of the $\mathcal N=$ 6 Supersymmetric Chern-Simons Theory},
Phys. Rev. Lett. {\bf 113} (2014) 2 021601 and arXiv:1403.1859 [hep-th];

\bibitem{AM-amp}
L. Alday and J. Maldacena, {\sl  Gluon scattering amplitudes at strong coupling},
JHEP{\bf 06} (2007) 064 and arXiv:0705.0303 [hep-th];

\bibitem{DKS}
J. Drummond, G. Korchemsky and E. Sokatchev,
{\sl  Conformal properties of four-gluon planar amplitudes and Wilson loops},
Nucl. Phys. {\bf B795} (2008) 385 and arXiv:0707.0243 [hep-th];

\bibitem{BHT}
A. Brandhuber, P. Heslop and G. Travaglini, {\sl MHV amplitudes in N=4 super Yang-Mills and Wilson loops},
Nucl. Phys. {\bf B794} (2008) 231 and arXiv:0707.1153 [hep-th];

\bibitem{Anope}
L. Alday, D. Gaiotto, J. Maldacena, A. Sever, P. Vieira,
{\sl An Operator Product Expansion for Polygonal null Wilson Loops},
JHEP{\bf 04} (2011) 088 and arXiv:1006.2788 [hep-th];

\bibitem{Knizhnik:1987xp}
V. G. Knizhnik,
{\sl Analytic Fields on Riemann Surfaces. 2}, Commun.\ Math.\ Phys.\  {\bf 112} (1987) 567;

\bibitem{CCAD}
J. L. Cardy, O. A. Castro-Alvaredo and B. Doyon,
{\sl Form factors of branch-point twist fields in quantum integrable models and entanglement entropy}, J.\ Statist.\ Phys.\  {\bf 130} (2008) 129 and arXiv:0706.3384 [hep-th].

\bibitem{BSV1}
B. Basso, A. Sever, P. Vieira,
{\sl Space-time S-matrix and Flux-tube S-matrix at Finite Coupling},
Phys. Rev. Lett. {\bf 111} (2013) 091602 and arXiv:1303.1396 [hep-th];


\bibitem{GKP}
S. Gubser, I. Klebanov, A. Polyakov,
{\sl  A Semiclassical limit of the gauge / string correspondence},
Nucl. Phys. {\bf B636 } (2002)  99  and hep-th/0204051;

\bibitem{Bel-Ope}
A.V.Belitsky,
{\sl OPE for null Wilson loops and open spin chains},
Phys. Lett. {\bf B709} (2012) 280 and arXiv:1110.1063 [hep-th];

\bibitem{Bel-Qua}
A.V.Belitsky, S.E.Derkachov, A.N.Manashov,
{\sl  Quantum mechanics of null polygonal Wilson loops},
Nucl. Phys. {\bf B882} (2014) 303 and arXiv:1401.7307 [hep-th];



\bibitem{BSV2}
B. Basso, A. Sever, P. Vieira,
{\sl  Space-time S-matrix and Flux tube S-matrix II. Extracting and Matching Data},
JHEP{\bf 01} (2014) 008 and arXiv:1306.2058 [hep-th];

\bibitem{BSV3}
B. Basso, A. Sever, P. Vieira,
{\sl  Space-time S-matrix and Flux tube S-matrix III. The two-particle contributions},
JHEP{\bf 08} (2014) 085 and arXiv:1402.3307 [hep-th];

\bibitem{BSV5}
B. Basso, A. Sever, P. Vieira,
{\sl  Space-time S-matrix and Flux tube S-matrix IV. Gluons and fusion},
JHEP{\bf 09} (2014) 149 and arXiv:1407.1736 [hep-th];

\bibitem{BCCSV1}
B. Basso, J. Caetano, L. Cordova, A. Sever, P. Vieira,
{\sl OPE for all helicity amplitudes},
JHEP{\bf 08} (2015) 018 and arXiv:1412.1132 [hep-th];

\bibitem{FPR2}
D. Fioravanti, S. Piscaglia, M. Rossi,
{\sl Asymptotic Bethe Ansatz on the GKP vacuum as a defect spin chain: scattering, particles and minimal area Wilson loops},
Nucl. Phys. {\bf B898} (2015) 301 and arXiv:1503.08795 [hep-th];

\bibitem{BFPR1}
A. Bonini, D. Fioravanti, S. Piscaglia, M. Rossi, {\sl Strong Wilson polygons from the lodge of free and bound mesons},
 JHEP{\bf 04} (2016) 029 and
arXiv:1511.05851 [hep-th];

\bibitem{BFPR2}
A. Bonini, D. Fioravanti, S. Piscaglia, M. Rossi,
{\sl The contribution of scalars to ${\cal N}=4$ SYM amplitudes},
Phys. Rev. {\bf D95} (2017) no. 4, 041902 and arXiv:1607.02084 [hep-th];

\bibitem{Cordova}
L. C\'ordova, {\sl Hexagon POPE: effective particles and tree level resummation}, JHEP{\bf 1701} (2017) 051  and
arXiv:1606.00423 [hep-th];

\bibitem{DP}
J. Drummond, G. Papathanasiou,
{\sl Hexagon OPE Resummation and Multi-Regge Kinematics},
JHEP{\bf 02} (2016) 185 and arXiv:1507.08982 [hep-th];

\bibitem{Dixon}
L. J. Dixon, J. M. Drummond, J. M. Henn,
{\sl Analytic result for the two-loop six-point NMHV amplitude in N=4 super Yang-Mills theory},
JHEP{\bf 01} (2012) 024 and arXiv:1111.1704 [hep-th];



\bibitem{Basso}
B. Basso, {\sl Exciting the GKP String at Any Coupling}, Nucl. Phys. {\bf B857} (2012) 254 and
arXiv:1010.5237 [hep-th];

\bibitem{FPR1}
D. Fioravanti, S. Piscaglia, M. Rossi,
{\sl On the scattering over the GKP vacuum},
Phys. Lett. {\bf B728} (2014) 288 and arXiv:1306.2292 [hep-th];

\bibitem{FRO6}
D. Fioravanti, M. Rossi,
{\sl TBA-like equations and Casimir effect in (non-)perturbative AdS/CFT},
JHEP{\bf 12} (2012) 013 and arXiv:1112.5668 [hep-th];

\bibitem{Basso-Rej}
B. Basso and A. Rej,
{\sl Bethe Ans\"{a}tze for GKP strings},
Nucl. Phys. {\bf B879} (2014) 162 and arXiv:1306.1741 [hep-th];

\bibitem{P1}
G. Papathanasiou,
{\sl Evaluating the six-point remainder function near the collinear limit},
Int. J. Mod. Phys. {\bf A29} (2014) 27, 1450154 and arXiv:1406.1123 [hep-th];

\bibitem{P2}
G. Papathanasiou,
{\sl Hexagon Wilson Loop OPE and Harmonic Polylogarithms},
JHEP{\bf 11} (2013) 150 and arXiv:1310.5735 [hep-th];


\bibitem{BSV4}
B. Basso, A. Sever, P. Vieira,
{\sl  Collinear limit of scattering amplitudes at strong coupling},
Phys. Rev. Lett. {\bf 113} (2014) 26, 261604 and arXiv:1405.6350 [hep-th];

\bibitem{Bel1509}
A.V. Belitsky, {\sl Towards NMHV amplitudes at strong coupling},
Nucl.Phys. {\bf B911} (2016) 517 and arXiv:1509.06054 [hep-th];

\bibitem{TBuA}
L. Alday, D. Gaiotto, J. Maldacena,
{\sl Thermodynamic Bubble Ansatz},
JHEP{\bf 09} (2011) 032 and arXiv:0911.4708 [hep-th];

\bibitem{YSA}
L. Alday, J. Maldacena, A. Sever, P. Vieira,
{\sl Y-system for Scattering Amplitudes},
J. Phys. {\bf A43} 485401 (2010) and arXiv:1002.2459 [hep-th];

\bibitem{Hatsuda:2010cc}
Y. Hatsuda, K. Ito, K. Sakai and Y. Satoh,
{\sl Thermodynamic Bethe Ansatz Equations for Minimal Surfaces in $AdS_3$},
JHEP{\bf 04} (2010) 108 and arXiv:1002.2941 [hep-th];

\bibitem{AM}
L. Alday, J. Maldacena,
{\sl Comments on operators with large spin},
JHEP{\bf 11} (2007) 019 and arXiv:0708.0672 [hep-th];

\bibitem{FGR}
D. Fioravanti, P. Grinza, M. Rossi,
{\sl Strong coupling for planar N=4 SYM theory: An All-order result},
Nucl. Phys. {\bf B810} (2009) 563 and arXiv:0804.2893 [hep-th];

\bibitem{BK}
B. Basso, G. Korchemsky,
{\sl Embedding nonlinear $O(6)$ sigma model into $N=4$ super-Yang-Mills theory},
Nucl. Phys. {\bf B807} (2009) 397 and arXiv:0805.4194 [hep-th];

\bibitem{FGR1}
D. Fioravanti, P. Grinza, M. Rossi,
{\sl The generalised scaling function: a note},
Nucl. Phys. {\bf B827} (2010) 359 and arXiv:0805.4407 [hep-th];

\bibitem{FB}
F. Buccheri and D. Fioravanti, {\sl The Integrable O(6) model and the correspondence: Checks and predictions}, arXiv:0805.4410 [hep-th];





\bibitem{BEL}
A.V. Belitsky, 	
{\sl Nonperturbative enhancement of superloop at strong coupling},
Nucl. Phys. {\bf B911} (2016) 425  and arXiv:1512.00555 [hep-th];

\bibitem{BEL-last}
A.V. Belitsky, 	
{\sl Twisted perturbed parafermions},
Phys. Lett. {\bf B770} (2017) 35 and
arXiv:1701.08914 [hep-th];

\bibitem{BoF}
J.-E. Bourgine, D. Fioravanti,
{\sl Finite $\varepsilon _2$-corrections to the N=2 SYM prepotential},
Phys. Lett. {\bf B750} (2015) 139
and arXiv:1506.01340 [hep-th];

\bibitem{BoF2}
J.-E. Bourgine, D. Fioravanti,
{\sl Mayer expansion of the Nekrasov prepotential: the subleading $\epsilon_2$-order}, Nucl. Phys. {\bf B906} (2016) 408 and arXiv:1511.02672 [hep-th];

\bibitem{Smirnov}
F.A. Smirnov, {\sl Reductions of the Sine-Gordon as a perturbation of minimal models of conformal field theory},
Nucl. Phys {\bf B337} (1990) 156;

\bibitem{NEK}
N. Nekrasov,
{\sl Seiberg-Witten prepotential from instanton counting},
Adv. Theor. Math. Phys. {\bf 7} (2003) 831 and hep-th/0206161;

\bibitem{NOK}
N. Nekrasov, A. Okounkov,
{\sl Seiberg-Witten theory and random partitions},
Prog. Math.  {\bf 244} (2006) 525 and  hep-th/0306238;

\bibitem{DID}
I.Kostov and D. Serban, private communication;

\bibitem{Bel1501}
A.V. Belitsky, 	
{\sl On factorization of multiparticle pentagons},
Nucl. Phys. {\bf B897} (2015) 346 and arXiv:1501.06860 [hep-th];

\bibitem{Bel1607}
A.V. Belitsky, 	
{\sl Matrix pentagons},  arXiv:1607.06555 [hep-th];





\bibitem{BFPR3}
A. Bonini, D. Fioravanti, S. Piscaglia, M. Rossi,
{\sl On the fermion contribution to amplitudes},
to appear;

\bibitem{Mooread}
N. Read, G. Moore, {\sl Fractional quantum hall effect and nonabelian statistics}, Prog. Theor. Phys. Suppl. {\bf 107} (1992) 157 and hep-th/9202001;

\bibitem{Pasq}
V. Pasquier, {\sl Incompressible Representations of the Birman-Wenzl-Murakami Algebra}, Annales Henri Poincar\'e {\bf 7} (2006) 603 and math/0507364.


\end{thebibliography}
\end{document}